\documentclass[12pt]{article}
\pdfoutput=1
\usepackage{putex}
\usepackage{graphicx}
\usepackage{epstopdf}
\usepackage{enumerate}
\usepackage[nosort]{cite}
\usepackage{tensor}
\usepackage{slashed}
\usepackage{feynmf}
\usepackage{hyperref}
\usepackage{float}
\usepackage[bottom]{footmisc}
\usepackage[utf8]{inputenc}
\usepackage{amsmath}
\usepackage{xfrac}
\usepackage{extarrows}
\usepackage[usenames,dvipsnames,svgnames,table]{xcolor}
\usepackage[bf,hang,footnotesize]{caption}
\usepackage{bbold}

\usepackage{graphicx,wrapfig,lipsum,float,subcaption,caption,sidecap}
\usepackage{hyperref,enumerate,array,multirow,tabularx,bm,soul,multicol}
\usepackage[toc,page]{appendix}
\usepackage{tikz}
\usetikzlibrary{intersections,shapes.geometric}
\usepackage{tkz-euclide}
\usepackage{pgfplots}
\usepgfplotslibrary{fillbetween}

\numberwithin{equation}{section}
\hypersetup{
	pdfstartview={FitH},    % fits the width of the page to the window
	pdftitle={Title},    % title
	pdfauthor={Kim, Kraus, Sun},     % authors
	colorlinks=true,       % false: boxed links; true: colored links
	linkcolor=blue,          % color of internal links
	citecolor=blue!59!black! ,        % color of links to bibliography
	filecolor=magenta,      % color of file links
	urlcolor=blue!75!black,           % color of external links
	linktoc=all
}

\makeatletter
\g@addto@macro\bfseries{\boldmath}
\makeatother

\numberwithin{equation}{section}

\newcommand{\eq}[2]{\begin{align}\label{#1}#2\end{align}}

\def\Lc{{\cal L}}

\newcommand {\be} {\begin {equation}}
\newcommand {\ee} {\end {equation}}

\newcommand{\p}{\partial}

\def\Tr{\mathop{\rm Tr}}

\newcommand\rt{{\rightarrow}}
\def\eps{\epsilon}

\newcommand{\rf}[1]{(\ref{#1})}
\newcommand{\rff}[1]{\ref{#1}}

\newcommand{\zb}{\overline{z}}

\newcommand{\arrow}{\rightarrow}

\definecolor{greenC}{rgb}{0.0, 0.38, 0.18}

\newcommand{\phib}{\bar{\phi}}

\newcommand{\dr}{d}

\newcommand{\Gh}{\hat{G}}

\newcommand{\ct}{\tilde{c}}
\newcommand{\Lambdah}{\hat{\Lambda}} 
\newcommand{\Lamh}{\hat{\Lambda}} 
\newcommand{\phid}{\dot{\phi}} 
\newcommand{\phit}{\tilde{\phi}}
\newcommand{\Oc}{{\cal O}}

\newcommand{\RN}[1]{%
	\textup{\uppercase\expandafter{\romannumeral#1}}%
}

\usepackage[framemethod=TikZ]{mdframed} % fancy boxes
\usepackage{tikz-cd} %for complicated commutative diagrams
\usetikzlibrary{arrows,snakes,shapes.arrows,decorations.markings}
     \tikzset{>=triangle 90}
     \tikzstyle{bbc}=[draw,circle,fill=black,scale=.75]
     \tikzstyle{rc}=[circle,fill=red,scale=.6]
     \tikzstyle{wc}=[draw,circle,scale=.75]

\usetikzlibrary{cd}
\tikzset{snake it/.style={decorate, decoration=snake}}

\tikzset{
	on each segment/.style={
		decorate,
		decoration={
			show path construction,
			moveto code={},
			lineto code={
				\path [#1]
				(\tikzinputsegmentfirst) -- (\tikzinputsegmentlast);
			},
			curveto code={
				\path [#1] (\tikzinputsegmentfirst)
				.. controls
				(\tikzinputsegmentsupporta) and (\tikzinputsegmentsupportb)
				..
				(\tikzinputsegmentlast);
			},
			closepath code={
				\path [#1]
				(\tikzinputsegmentfirst) -- (\tikzinputsegmentlast);
			},
		},
	},
	mid arrow/.style={postaction={decorate,decoration={
				markings,
				mark=at position .5 with {\arrow[#1]{stealth}}
	}}},
}

\usetikzlibrary{decorations.markings}

\tikzset{line/.style={line width=0.25mm},
curve/.style={line,smooth,tension=1},
->-/.style={decoration={
  markings,
  mark=at position #1 with {\arrow[>=stealth]{>}}},postaction={decorate}},
-<-/.style={decoration={
  markings,
  mark=at position #1 with {\arrow[>=stealth]{<}}},postaction={decorate}},
}

\tikzset{bg/.style={opacity=.5}}

\tikzset{
    partial ellipse/.style args={#1:#2:#3}{
        insert path={+ (#1:#3) arc (#1:#2:#3)}
    }
}

\definecolor{azure}{rgb}{0.0, 0.5, 1.0}
\definecolor{darkblue}{rgb}{0.15,0.35,0.7}
\definecolor{reddish}{rgb}{0.65, 0.2, 0.2}
\definecolor{brandeisblue}{rgb}{0.0, 0.44, 1.0}
\definecolor{ceruleanblue}{rgb}{0.16, 0.32, 0.75}
\definecolor{indigo(dye)}{rgb}{0.0, 0.25, 0.42}
\definecolor{dgrey}{rgb}{0.3,0.3,0.3}
\definecolor{grey}{rgb}{0.9,0.9,0.9}

\newcommand{\Zd}{{Z_d}} %This is the partition function computed in the defect Hilbert space channel
\newcommand{\Zo}{{Z_o}} %This is the partition function computed in the operator channel

\begin{document}

\institution{UCLA}{ \quad\quad\quad\quad\quad\quad\quad\ ~ \, $^{1}$Mani L. Bhaumik Institute for Theoretical Physics
		\cr Department of Physics \& Astronomy,\,University of California,\,Los Angeles,\,CA\,90095,\,USA}

\title{Codimension one defects in \\ free scalar field theory
}

\authors{Seolhwa Kim$^1$, Per Kraus$^{1}$, Zhengdi Sun$^{1}$}
	
\abstract{  We study various aspects of codimension one defects in free scalar field theory, with particular emphasis on line defects in two-dimensions.  These defects are generically non-conformal, but include conformal and topological defects as special cases.   Our analysis is based on the interplay between two complementary descriptions, the first involving matching conditions imposed on fields and their derivatives across the defect, and the second on the resummation of  perturbation theory in terms of renormalized defect couplings.  Using either description as appropriate we compute a variety of observables: correlators of fields in the presence of such defects; the defect anomalous dimension; multiple defects and their fusion; canonical quantization and instabilities; ring shaped defects with application to the g-theorem and the entanglement entropy of accelerating defects;  defects on the torus and Cardy formulas for the asymptotic density of states of the defect Hilbert space; and quenches produced by spacelike defects.   The simplicity of the model allows for explicit computation of all these quantities, and provides a starting point for more complicated theories involving interactions.       }
	
	\date{}
	
	\maketitle
	\setcounter{tocdepth}{2}
	\begingroup
	\hypersetup{linkcolor=black}
	\tableofcontents
	\endgroup
	
%%%%%%%%%%%%%%%%%%%%%%%%%%%%%%%%%%%%%%%%%%%%%%%%%%%

\section{Introduction and Summary}

Defects in quantum field theory arise in a wide variety of physical contexts, and often elucidate features of the ambient theory in which they are embedded.   By a ``defect" we mean a lower dimensional submanifold that hosts localized interactions and/or degrees of freedom.  In this paper we consider a simple example of such a defect: a codimension-one defect in free scalar field theory.  We will  be interested mainly in non-conformal defects embedded in a theory that is conformal in the bulk.\footnote{Non-conformal defects in free field theory were considered in a similar spirit to ours  in \cite{Goldberger:2001tn,Michel:2014lva}. The folding trick relates defects to boundaries, and  string theory applications arise in the context of boundary  string field theory \cite{Witten:1992cr,Kutasov:2000qp}.  More generally, boundary RG flows have been considered in many works, including \cite{Affleck:1991tk,Friedan:2003yc}, along with other references noted below.  }    As we will see, this simple example  illustrates a variety of phenomena that are also present in more elaborate contexts. 

\subsection{Two formulations} 

As we will discuss in detail, there are two distinct formulations of this problem.  In the first description, which we refer to as the ``perturbative formulation" since it will be based on resumming perturbation theory, we start from a Lagrangian comprising bulk and defect contributions,
\eq{a1}{ S &=  \int\! d^{d-1}y dx\big(  \Lc_{\rm b}(\phi) +   \Lc_{\rm d}(\phi) \delta(x) \big)}
where we have taken the defect to lie at $x=0$.       The bulk theory is a free scalar field\footnote{We mostly work in Euclidean signature and usually set the bulk mass to zero.}
\eq{a2}{ \Lc_{\rm b}  = {1\over 2} \big( (\p \phi)^2+m^2 \phi^2\big) }
while the defect Lagrangian contains quadratic terms\footnote{The restriction to quadratic terms is made for simplicity but can also be viewed as a starting point for an interacting theory, where the interactions are treated perturbatively in the exactly soluble quadratic theory.}    with any number of derivatives transverse or parallel to the defect, compatible with translation and rotation invariance on the defect, written schematically as
\eq{a3}{ \Lc_{\rm d}  =  \sum_{m,n,p,q} \alpha_{mnpq } (\p_y)^m (\p_x)^n \phi  (\p_y)^p (\p_x)^q \phi ~.}
However, a field redefinition may be used to eliminate all terms with $n$ or $q$ greater than $1$.  We may also integrate by parts along the defect to write all terms with derivatives along the defect in the form $\phi  (\nabla^2_y)^n \phi$.   It will therefore suffice to consider 
\eq{a4}{ \Lc_d = c_0^B  + {1\over 2} c_1^B \phi^2 + c_2^B \phi \phi' + {1\over 2}c_3^B (\phi')^2 }
where $\phi' = \p_x \phi$.  Here, the dependence on $\nabla_y^2$ has been absorbed into the couplings $c^B_{1,2,3}$; in fact, since we'll work in momentum space along the defect this dependence plays no significant role, and we'll take $c^B_{1,2,3}$ to be constants, noting that the dependence on $\nabla^2_y$ can easily be incorporated by allowing these couplings to depend on the momentum along the defect.  The $B$ superscript on the couplings indicates that these are bare couplings; computation of physical quantities in this description will be UV divergent, necessitating a UV cutoff $\Lambda$ and the introduction of renormalized couplings $c_{1,2,3}$ that are finite as $\Lambda \rt \infty$.  Choosing a sharp momentum cutoff, $|p|< \Lambda$, we will find 
\eq{a3a}{ c^B_1 & = c_1 +{ c_2^2 \Lamh \over 1- c_3 \Lamh} \cr
c^B_2 & = {c_2  \over 1- c_3 \Lamh} \cr
c^B_3 & = {c_3  \over 1- c_3 \Lamh} }
with $\Lamh = { \Lambda \over \pi}$. 
By the same token, we have included a defect cosmological term $c_0^B$ to cancel divergences arising from vacuum diagrams.   This term depends on the defect dimension and the couplings $c_{1,2,3}$.    For  example, in the case of  a line defect embedded in a two-dimensional bulk, and setting $c_2=c_3=0$, we have  
\eq{a4a}{ c_0^B = -{c_1 \over 4\pi} \ln\left( {\Lambda \over \mu}\right) }
where $\mu$ is an RG scale.  Computations in this description, such as correlators involving bulk and boundary fields,  are obtained by employing perturbation theory in the couplings $c_i$ and then resumming.   The expressions \rf{a3a} are obtained by demanding finiteness at each order of perturbation theory.   

For general couplings the defect is non-scale invariant and so describes an RG flow, typically between a Neumann boundary condition in the UV and a Dirichlet boundary condition in the IR.   For special values of the couplings the defect is conformal or topological, as we discuss in the next section.  

In the second description,   we use that the theory is quadratic to encode the defect couplings in terms of matching conditions across the defect; this approached was employed in \cite{Bachas:2001vj} for conformal defects.   Using $(L,R)$ to denote values immediately to the left or right of the defect, we impose\footnote{Note that this is different than the matching conditions in  \cite{Bachas:2001vj}  since we are not imposing conformal invariance.}
\eq{a5}{ \left(
           \begin{array}{c}
             \phi_R \\
             \phi'_R \\
           \end{array}
         \right)  =  \left(
                       \begin{array}{cc}
                         a & b \\
                         c & d \\
                       \end{array}
                     \right)
           \left(
           \begin{array}{c}
             \phi_L \\
             \phi'_L \\
           \end{array}
         \right) }
for some constants $(a,b,c,d)$.
As we discuss in the next section, time translation invariance imposes the condition $ad-bc=1$.  The remaining three independent couplings are related to the renormalized couplings defined above as 
\eq{a6}{ \left(
                       \begin{array}{cc}
                         a & b \\
                         c & d \\
                       \end{array}
                     \right)
=\left(\begin{array}{cc}
\frac{\left(2-c_2\right)^2-c_1 c_3}{4-c_2^2+c_1 c_3} & \frac{-4 c_3}{4-c_2^2+c_1 c_3} \\
\frac{4 c_1}{4-c_2^2+c_1 c_3} & \frac{\left(2+c_2\right)^2-c_1 c_3}{4-c_2^2+c_1 c_3}
\end{array}\right) ~. }
These relations are fixed by matching observables computed in the two descriptions.   Alternatively, we can write down an action in the matching approach involving $(\phi_{L,R}, \phi'_{L,R})$ treated as independent such that the variational principle yields the matching conditions \rf{a6}.  As we'll see, the couplings in this description may be identified with the renormalized couplings in the perturbative description.  Note that the action in the perturbative description involves just $(\phi,\phi')$, and hence would be ambiguous if we attempted to incorporate field discontinuities  in the Lagrangian.   Physical discontinuities arise in the perturbative approach via the discontinuities of correlators computed after perturbative resummation. 

Either of the two descriptions may be more convenient for different purposes.   For example, the $\mu$ dependence of the $c_0^B$ term in the perturbative description makes it clear that the defect, regarded as an operator, is scale dependent and so acquires an ``anomalous dimension", which should be taken into account when  considering  the modular properties of the  torus  partition function with a defect insertion.  On the other hand, the matching description makes it clear how to understand the fusion of two parallel defects, as it simply involves the multiplication  of two matching matrices in \rf{a6}, and is also suitable for canonical quantization via  mode expansions and computing partition sums.   We can pass back and forth between the two descriptions using the dictionary \rf{a6}. 

\subsection{Computations}

In the remainder of the introduction we summarize the computations we have carried out in this theory.   The two formulations discussed above are laid out in detail in section \rff{general}.  

In section \rff{twopoint}  we compute the 2-point function of bulk fields in the presence of the defect using the perturbative description.   The computation is carried out by summing up loop diagrams in momentum space; this is possible due to the iterative structure of the loop diagrams.   This determines the relation between the bare and renormalized couplings, and we also use the renormalized correlator  to read off the effective matching conditions on the field and its normal derivative   across the defect.    We then discuss various special cases: the UV and IR limits, the conformal defect, and the topological defect, the latter implementing a sign flip of $\phi$ across the defect. 

In section \rff{vacuum} we compute the sum of vacuum diagrams, which determines the defect anomalous dimension\footnote{We use the term ``anomalous dimension" to characterize the $\mu$ dependence, though the defect in general has no well defined scaling dimension since the couplings change under a scale transformation.  } via the logarithmically divergent part of $c_0^B$. In particular, the renormalized defect operator ${\cal O}_{\rm d} = e^{-S_{\rm d}}$ obeys the RG equation 
\eq{a7}{ \mu {d\over d\mu}{\cal O}_{\rm d}  = \gamma_{\rm d}  {\cal O}_{\rm d}  }
A closed form expression for $ \gamma_{\rm d}$ may be obtained when $c_3=0$, 
\eq{a8}{  \gamma_{\rm d} = -{1\over 4\pi}  {c_1 L_d \over 1+ {1\over 4} c_2^2}~,}
where $L_d$ is the length of the defect. For $c_3$ nonzero  $ \gamma_{\rm d}$  is given by a divergent series in the couplings $c_{1,2,3}$. 

Section \rff{fusion_sec} considers the fusion of two parallel defects.  We verify that perturbative computation of the renormalized two-point function yields a fusion rule result consistent with the multiplication of matching matrices \rf{a6}.   We also work out the leading dependence of the partition function on the distance  $L$ between the two defects, which for $c_2=c_3=0$  can be described in terms of a defect OPE as 
\eq{a9}{ \Oc_{\rm d}^{(1)} (L) \Oc_{\rm d}^{(2)} (0) \approx L^{{c_1^{(1)} c_1^{(2)} L L_d \over 4\pi} } \Oc_{\rm d}^{(12)}~,\quad L\rt 0} 

In section \rff{mode} we canonically quantize the  $d=1+1$ scalar field in the presence of the defect by working out a complete set of mode solutions.  Depending on the values of the couplings $c_i$, these can include bound state solutions, as well as exponentially growing (in real time) solutions that signal an instability.   The instabilities, when present, can be stabilized by giving the scalar field a bulk mass term.  The bound state solutions are identified with the poles of the two-point function computed in an earlier section. 

In section \rff{ring} we replace  the  line defect (with $c_2=c_3=0$)  with a circular ring defect.  The partition function $Z_{\rm ring}$ depends on the dimensionless combination $c_1 \beta$, where $\beta$ is the circumference of the ring.   We find
\eq{a10}{ \log Z_{\rm ring} (c_1 \beta)  ={1\over 2} \log{ \beta c_1 \over 4\pi}  + \log \Gamma\left({\beta c_1 \over 4\pi}\right) ~,}
up to an additive constant.      We can equivalently think of this as the partition function of a ring placed on a cone at a fixed distance from the tip, with the length of the ring $\beta$ controlled by the opening angle of the cone.   We can  then compute the ``entropy" as  
\eq{a11}{  s &= (1-\beta \p_\beta) \log Z_{\rm ring} (c_1\beta) \cr
& ={1\over 2} \log{ \beta c_1 \over 4\pi}  + \log \Gamma\left({\beta c_1 \over 4\pi}\right) -{\beta c_1 \over 4\pi} \Psi\left({\beta c_1 \over 4\pi}\right) -{1\over 2}  }
where $\Psi(x) = \Gamma'(x)/\Gamma(x)$.  Using the folding trick to convert this to the partition function of a CFT on a disk with a $c_1\phi^2$ boundary interaction, the entropy $s$ corresponds to the $g$-function \cite{Affleck:1991tk,Kutasov:2000qp,Friedan:2003yc}.    The $g$-function in \rf{a11} is monotonically decreasing in $c_1 \beta$, in accord with the $g$-theorem \cite{Friedan:2003yc}.   Alternatively, the quantity $s$ may be related to the entanglement entropy of a half-space in the presence of two accelerating defects.  This interpretation arises from analytic continuation of the Euclidean ring configuration to Lorentzian space time.

In section \rff{torusZ} we consider the torus partition function $Z(L_d,L)$ with the $c_1$ defect wrapped around a  cycle of a square torus with length $L_d$, the length of the other cycle being $L$.     The torus partition function depends on the two dimensionless variables $(c_1 L, c_1L_d)$ as well as the RG scale $\mu$ that is part of the definition of the renormalized defect.    The torus partition function may be computed as a sum over states in either of two channels, depending on which direction we view as time.   If we take time to run along the defect direction, then the computation involves summing over modes with discontinuities given by the matching conditions.  The result in this channel receives a nontrivial contribution from the zero point energy.  Alternatively, taking time to run orthogonally to the defect, the computation corresponds to computing matrix elements of the defect operator between excited states of the scalar field.   The relation between these two descriptions is a bit subtle due to the fact that the 
partition function depends on a renormalization scale $\mu$, as in \rf{a7}, which must be chosen to agree in the two descriptions.

We can use these results to give a ``Cardy formula" for the asymptotic density of states of the defect Hilbert space; i.e. of the Hilbert space of the defect located at a point on a spatial circle of size $L$.  The result depends on relative sizes of the dimensionless combinations of $EL$ and $c_1 L$.  For $c_1 L = O(1)$   the asymptotic density of states as a function of energy works out to be 
  \eq{a13}{ \rho(E) \approx   {1\over \sqrt{c_1 L} }   \left({ 6 EL\over \pi }\right)^{{1\over 8} \sqrt{c_1^2 L \over 6 \pi E}   }   \left( { L/2\pi \over 48 E^3}\right)^{1/4}  e^{2\pi \sqrt{  EL/2\pi \over 3} } \left(1 + O\left(\frac{1}{\sqrt{EL}}\right)\right)~,\quad EL \gg 1 ~.}
The divergence as $c_1\rt 0$ corresponds to integration over the scalar field zero mode, which acquires vanishing action in this limit.      This same result also applies to the regime  $EL \gg (c_1L)^2 \gg 1$ except the leading correction is $O\left({c_1L \over \sqrt{EL} }  \right)$.   Another regime of interest is $c_1 L \gg EL \gg 1$, for which the partition function becomes that of a Dirichlet defect after carefully taking into account the zero point energy.

In section \ref{quench}   we consider the  defect as an operator inserted at a fixed Lorentzian  time $t=0$.  The scalar field is taken to be in the vacuum state for $t<0$.  As in the quench scenario studied in \cite{Calabrese:2006rx}, the insertion of the defect creates an excited state at $t>0$, whose explicit form we work out by computing the Bogoliubov coefficients relating the natural in and out modes.    We also compute the expectation value of the stress tensor for $t>0$, finding it to be UV divergent on account of the sharply localized (in time) nature of the defect.  The introduction of a UV regulator $\Lambda$ yields a result logarithmically divergent in $\Lambda$, corresponding to smearing the defect operator over a time window $\Delta t \sim \Lambda^{-1}$.  

We conclude by commenting on some natural extensions of these results in section  \rff{comments}.

A series of appendices fill in some technical details, and outline how adding bulk interactions can stabilize the defect when the couplings  $c_i$ are such that unstable modes are present. 

Other references on defects relevant to what follows include \cite{Raviv-Moshe:2023yvq,Giombi:2023dqs,Trepanier:2023tvb,Bashmakov:2024suh,Casini:2022bsu,Cuomo:2021rkm,Diatlyk:2024zkk,Kravchuk:2024qoh,Diatlyk:2024qpr,Shachar:2024ubf,Cuomo:2024psk,Shachar:2024cwk,Fredenhagen:2006dn,Pannell:2023pwz,Brunner:2007ur,Gaiotto:2012np,Brunner:2015vva,Shimamori:2024yms,Ge:2024hei,Giombi:2022vnz,Barrat:2023ivo,Konechny:2015qla,Lauria:2020emq,DiPietro:2020fya,Behan:2020nsf,Behan:2021tcn,Castiglioni:2022yes,Castiglioni:2023uus,DiPietro:2023gzi,Nagar:2024mjz,Sun:2025ihw}.

\section{General aspects }
\label{general}

In this section we describe our two approaches to the defect problem --- the matching approach and the perturbative approach --- and their relation.  We describe the general space of defect couplings along with special cases that lead to conformal or topological defects.

\subsection{Matching conditions}

We consider a noncompact free massless scalar field $\phi$ on the Euclidean plane with coordinates $(x,t)$.
We place an infinite line defect at $x=0$, extending in the $t$ direction, and write the action as 

\eq{b1}{ S = {1\over 2} \int_{x<0}\! d^2x (\p \phi)^2+{1\over 2} \int_{x>0}\! d^2x (\p \phi)^2+ S_{\rm d}~,}
where $S_d$ is the action associated with the defect.  One approach to this system consists of imposing matching conditions across the  defect while  ignoring direct reference to $S_d$. We will consider linear matching conditions of the form 
\eq{b2}{ \left(
           \begin{array}{c}
             \phi_R \\
             \phi'_R \\
           \end{array}
         \right)  =  \left(
                       \begin{array}{cc}
                         a & b \\
                         c & d \\
                       \end{array}
                     \right)
           \left(
           \begin{array}{c}
             \phi_L \\
             \phi'_L \\
           \end{array}
         \right) }
where $\phi_{L,R}$ are to be evaluated just to the left or right of the defect,  and $\phi' = {\p \phi\over \p x}$.

\begin{figure}[H]
    \centering
    \begin{tikzpicture}[baseline={([yshift=-1ex]current bounding box.center)},vertex/.style={anchor=base,circle,fill=black!25,minimum size=18pt,inner sep=2pt},scale=0.7]
    \draw[line width = 0.8mm, blue] (0,-3) -- (0,3);
    \draw[line width = 0.5mm, red, ->-=0.2] (-1.7,0.3) circle (1.2);
    \draw[line width = 0.5mm, red, ->-=0.2] (2.3,0.3) circle (1.3);
    \filldraw[black] (-1.8,1) circle (2pt);
    \node[right] at (-1.8,1) {\footnotesize $\Oc_1$};
    \filldraw[black] (-1.7,-0.3) circle (2pt);
    \node[right] at (-1.7,-0.3) {\footnotesize $\Oc_2$};
    \filldraw[black] (1.8,1) circle (2pt);
    \node[right] at (1.8,1) {\footnotesize $\Oc_{n-1}$};
    \filldraw[black] (1.7,-0.4) circle (2pt);
    \node[right] at (1.7,-0.4) {\footnotesize $\Oc_n$};
\end{tikzpicture}
    \caption{ A  two component contour $C$ (red)  surrounding the operators $\Oc_i$ on each side of the defect (blue).  The current $j_\mu$ is integrated over the contour. }
    \label{con1}
\end{figure}
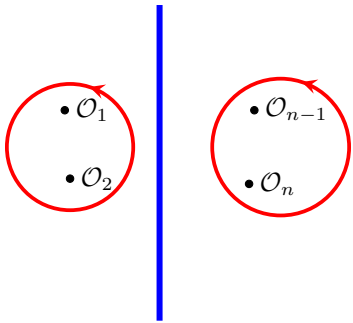

Imposing symmetries  constrains the values of $(a,b,c,d)$.  Consider a correlator in the presence of the defect, $ \langle {\cal O}_1(x_1,t_1)\ldots  {\cal O}_n(x_n,t_n)\rangle_d$.  Given a current $j^\mu(x,t)$  that is conserved away from the defect, we may attempt to derive a Ward identity in the standard fashion by inserting an integral of $j^\mu$ along a contour $C$ that surrounds all the operators, as shown in Figure \ref{con1}. 

\begin{figure}[H]
\centering
\begin{tikzpicture}[baseline={([yshift=-1ex]current bounding box.center)},vertex/.style={anchor=base,circle,fill=black!25,minimum size=18pt,inner sep=2pt},scale=0.7]
    \draw[line width = 0.8mm, blue] (0,-3) -- (0,3);
    \draw[line width = 0.5mm, red, -<-=0.5] (0.5,-2.5) -- (0.5,2.5);
    \draw[line width = 0.5mm, red, ->-=0.5] (0.5,-2.5) arc (-90:90:3 and 2.5);
    \draw[line width = 0.5mm, red, ->-=0.5] (-0.5,-2.5) -- (-0.5,2.5);
    \draw[line width = 0.5mm, red, -<-=0.5] (-0.5,-2.5) arc (-90:-270:3 and 2.5);
    \filldraw[black] (-1.8,1) circle (2pt);
    \node[right] at (-1.8,1) {\footnotesize $\Oc_1$};
    \filldraw[black] (-1.7,-0.3) circle (2pt);
    \node[right] at (-1.7,-0.3) {\footnotesize $\Oc_2$};
    \filldraw[black] (1.8,1) circle (2pt);
    \node[right] at (1.8,1) {\footnotesize $\Oc_{n-1}$};
    \filldraw[black] (1.7,-0.4) circle (2pt);
    \node[right] at (1.7,-0.4) {\footnotesize $\Oc_n$};
\end{tikzpicture}
\caption{ Deformation of the contour $C$ into two arbitrarily large semi-circles and flat sides running along the defect. }
\label{con2}
\end{figure}
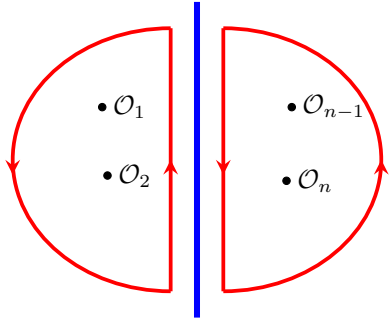

That is, we insert $\oint_C j_\mu dx^\mu$.  On one hand, we can deform the contour into a sum of contours surrounding each operator and then use the $j^\mu {\cal O}$ OPE to evaluate the integrals, yielding
\eq{b3}{ &\langle \oint_C j_\mu(x) dx^\mu  {\cal O}_1(x_1,t_1)\ldots  {\cal O}_n(x_n,t_n)\rangle_d\cr& \quad\quad\quad =  \langle \delta {\cal O}_1(x_1,t_1)\ldots  {\cal O}_n(x_n,t_n) + \ldots +  {\cal O}_1(x_1,t_1)\ldots \delta {\cal O}_n(x_n,t_n)\rangle_d ~.}
In the absence of a defect, we alternatively expand the contour to an arbitrarily large circle, which yields zero due to field falloffs at infinity. The Ward identity is then the statement that the right hand side of \rf{b3} vanishes.  In the presence of the defect we can instead first expand two  contours that surround the operators on either side of the defect into  arbitrarily large semi-circles, with the flat sides lying along the defect, as illustrated in Figure \ref{con2}.

In order to get zero in the large radius limit, we need that the two segments lying along the defect cancel. That is, the Ward identity (i.e the vanishing of the rhs of \rf{b3}) is only valid provided
\eq{b4}{ \int_{-\infty}^\infty\! j_t(\phi_L)dt  =  \int_{-\infty}^\infty\! j_t(\phi_R)dt ~.}
Demanding that the defect preserves a symmetry amounts to the condition that \rf{b4} be obeyed for the Noether current $j^\mu$ corresponding to that symmetry. 

As a first condition, we require time translation invariance, meaning invariance under a common time shift $t_i \rt t_i + \delta t_i$ of all operator insertions.  The corresponding current is 
\eq{b5}{ j_z= {1\over 2\pi} T_{zz}~,\quad j_{\zb} = {1\over 2\pi}T_{\zb\zb} }
where we use complex coordinates $z=x+it$ and $\zb=x-it$, and the stress tensor has nonzero components 
\eq{b6}{ T_{zz}=-2\pi (\p_z \phi)^2~,\quad  T_{\zb\zb}=-2\pi (\p_{\zb}  \phi)^2 ~.}
This gives
\eq{b7}{ j_t = {1\over 2\pi} T_{xt} = -\phi' \dot{\phi} ~.}
Time translation invariance then requires
\eq{b8}{ \phi_R'\dot{\phi}_R -  \phi_L'\dot{\phi}_L = {dX\over dt} }
for some local operator $X$.  Using \rf{b2} it's easy to see that this requires 
\eq{b8a}{ ad-bc=1}
in which case
\eq{b9}{ X = {1\over 2} \big( ac \phi_L^2+ bd {\phi'_L}^2+ bc (\phi_L^2)'\big)~.}
We will therefore restrict henceforth to matching conditions obeying \rf{b8a}. 

Next we consider the restrictions imposed by demanding invariance of the correlation functions under scale transformations $x_i^\mu \rt \lambda x_i^\mu$.  The current is now 
\eq{b10}{ j_z= {1\over 2\pi i} zT_{zz}~,\quad j_{\zb} = {-1\over 2\pi i}\zb T_{\zb\zb} }
and the condition \rf{b4} becomes
\eq{b11}{  \int_{-\infty}^\infty\! \phi'_L \dot{\phi}_L t dt  =  \int_{-\infty}^\infty\! \phi'_R \dot{\phi}_R t dt ~.}
This now requires equality at the integrand level, $ \phi'_L \dot{\phi}_L =  \phi'_R \dot{\phi}_R$, which requires that $X=0$ in \rf{b9}. Along with \rf{b8a} this  imposes  $d=1/a$ with $b=c=0$ as the condition of scale invariance.  Assuming just scale invariance we automatically have conformal invariance since the currents
\eq{b12}{ j^{(n)}_z= {1\over 2\pi i} z^n T_{zz}~,\quad j^{(n)}_{\zb} = {(-1)^n\over 2\pi i}\zb^n T_{\zb\zb}~,\quad n=0,1,2, \ldots  }
automatically obey \rf{b4}, corresponding to a single copy of Virasoro currents.

An even more restrictive condition is to allow symmetries under both the holomorphic and anti-holomorphic Virasoro currents.   This impose the additional condition $ \dot{\phi}_L^2- (\phi'_L)^2 = \dot{\phi}_R^2- (\phi'_R)^2$, which requires $a=d=\pm 1$ with $b=c=0$.  This defines a topological (or transparent) interface.   In particular, we have the nontrivial case of $a=d=-1$ for which $\phi$ flips sign across the defect.  

Given the matching conditions \rf{b2} it is natural to ask for an action that reproduces these.  In this section we look for an action whose variational principle implies the matching conditions.  Later we develop a second approach in which the defect action is treated diagrammatically by summing up Feynman diagrams.   

In considering the form of a defect action an immediate issue is that the field and its first derivative are discontinuous across the defect, rendering ambiguous the meaning of the ``field on the defect".    To avoid this issue we work with two fields $(\phi_L,\phi_R)$ defined at $(x=0^-,x=0^+)$, as already anticipated in \rf{b1}.  The appropriate form of $S_d$ appearing in \rf{b1} turns out to take the form 
\eq{b11a}{ S_d = \int d t  \big[\left(\phi_R-\phi_L\right)\phi'_A+\frac{1}{2} \ct_1 \phi_A^2+ \ct_2 \phi_A \phi_A^{\prime}+\frac{1}{2} \ct_3\left(\phi_A^{\prime}\right)^2 \big] }
where  we have defined the averaged field 
\eq{b12a}{  \phi_A =  {\phi_L+\phi_R\over 2}~.}
The tildes on $\ct_i$ are there to distinguish these couplings from  analogous couplings that will appear in an alternative formulation.   To read off the boundary conditions we compute the on-shell variation of the action, which works out to be 
\eq{b13}{ \delta S =-\int d t\Big(\left[\phi_R^{\prime}-\phi_L^{\prime}-\ct_1 \phi_A-\ct_2 \phi_A^{\prime}\right] \delta \phi_A+\left[\phi_L-\phi_R-\ct_2 \phi_A-\ct_3 \phi_A^{\prime}\right] \delta \phi_A^{\prime} \Big)~. } 
A good variational principle is thus achieved by imposing the following conditions at the defect
\eq{b14}{ \phi_R^{\prime}-\phi_L^{\prime}-\ct_1 \phi_A-\ct_2 \phi_A^{\prime} & = 0 ~, \cr
\phi_L-\phi_R-\ct_2 \phi_A-\ct_3 \phi_A^{\prime} & =0 ~.}
These conditions may be put in the form \rf{b2} with 
\eq{b15}{ \left(
                       \begin{array}{cc}
                         a & b \\
                         c & d \\
                       \end{array}
                     \right)
=\left(\begin{array}{cc}
\frac{\left(2-\ct_2\right)^2-\ct_1 \ct_3}{4-\ct_2^2+\ct_1 \ct_3} & \frac{-4 \ct_3}{4-\ct_2^2+\ct_1 \ct_3} \\
\frac{4 \ct_1}{4-\ct_2^2+\ct_1 \ct_3} & \frac{\left(2+\ct_2\right)^2-\ct_1 \ct_3}{4-\ct_2^2+\ct_1 \ct_3}
\end{array}\right) ~. }
This obeys $ad-bc=1$ as expected.  Inverting gives
\eq{b16}{\ct_1 ={4ac\over (a+1)^2+bc}~,\quad \ct_2 = {2a(d-a)\over (a+1)^2+bc}~,\quad \ct_3={-4ab\over (a+1)^2+bc}~.}
The condition of conformal invariance, $a=1/d$ and $b=c=0$ translates to $\ct_1=\ct_3=0$, giving $a= {2-\ct_2\over 2+\ct_2}$. Further taking $\ct_2\rt \pm \infty$ yields the nontrivial topological defect with $b=c=0$ and $a=d=-1$.

Although we will not consider it further in this paper, in order to make contact with previous work~\cite{Bachas:2001vj}  we now comment on how to incorporate matching conditions of the form 
\eq{b15a}{ \left(
           \begin{array}{c}
            \dot{ \phi}_R \\
             \phi'_R \\
           \end{array}
         \right)  =  \left(
                       \begin{array}{cc}
                         a & b \\
                         c & d \\
                       \end{array}
                     \right)
           \left(
           \begin{array}{c}
            \dot{  \phi}_L \\
             \phi'_L \\
           \end{array}
         \right) }
in this framework.   For example, the case $a=d=0$,  $b=1/c$ yields a conformal defect not included in our analysis above.   To incorporate this example we consider the action \rf{b1} and now choose  
\eq{b15b}{ S_d =  c \int\! dt \dot{\phi}_L \phi_R }
so that 
\eq{b15c}{ \delta S = \int\! dt \big[(\phi'_L-c\dot{\phi}_R)\delta \phi_L - (\phi'_R-c\dot{\phi}_L)\delta \phi_R   \big], }
which implies the stated matching conditions.  

\subsection{Perturbative action approach}
\label{pertaction}

Another approach is to consider the action 
\eq{b15d}{ S = \int\! d^2x (\p \phi)^2  + S_d }
with 
\eq{b15e}{ S_d =   \int\! dt  \left(  c^B_0+{1\over 2} c^B_1 \phi^2+ c^B_2 \phi \phi' +{1\over 2} c^B_3 {\phi'}^2 \right) }
and to compute correlators and other quantities via perturbative expansion in the couplings $c_i^B$, as in \cite{Goldberger:2001tn,Michel:2014lva}.   To regulate the ultraviolet divergences that appear in perturbation theory we introduce a cutoff on the momentum transverse to the defect, $|p|<\Lambda$, and express the bare couplings $c_i^B$ in terms of renormalized couplings $c_i$ as in \rf{a3a}.  This will be illustrated in the next section when we compute the scalar 2-point function in the presence of the defect by summing up the perturbative expansion.

In this approach $\phi$ and $\phi'$ are treated as being continuous across the defect at each order in perturbation theory.\footnote{Since we impose a momentum cutoff $\Lambda$ the defect is effectively smeared out, and the fields are continuous within this smeared region, which shrinks to zero as the cutoff is removed.}  Indeed, the defect action \rf{b15e} would be ambiguous if discontinuities were allowed.  Nonetheless the resummed two-point correlation exhibits discontinuities that satisfy the matching conditions \rf{b2}.  So we recover the same correlators as in the $(\phi_L,\phi_R)$ approach, and the dictionary equating the two turns out to be extremely simple, $\tilde{c}_i =c_i $. 

We now explain the sense in which the action \rf{b15e} represents what is effectively the most general local action quadratic in fields.    First consider the inclusion of terms with time derivatives, ${1\over 2} \dot{\phi}^2$ etc.  In momentum space this becomes ${1\over 2}\omega^2  \tilde{\phi}^2$, where $\omega$ represents the momentum along the time direction.  Since the action is quadratic and  $\omega$ is conserved by time translation invariance, there is no mixing among different values of $\omega$.   Hence the effect of a ${1\over 2} \dot{\phi}^2$ term is equivalent to redefining $c_1^B \rt c_1^B + \omega^2$.  More generally, all terms involving time derivatives can be included by allowing $c_{1,2,3}^B$ to be arbitrary functions of $\omega$.  Computing momentum space correlation functions in the presence of $\omega $ dependent couplings is the same as for constant couplings since the value of $\omega$ is conserved in any connected correlator.  For this reason we will not explicitly consider terms with time derivatives, noting that they are simple to include if desired. 

We can also add terms in the defect action involving $\phi''$ and higher transverse derivatives.   However,  by using the field equations --- equivalently, the equations of motion --- such terms can always be traded for terms involving only $\phi$ and $\phi'$ and time derivatives thereof,\footnote{We illustrate this with an example in appendix \rff{higher}.}  and we already discussed the incorporation of such time derivative terms in the previous paragraph. 
   
Finally, it is worth commenting that while we focus on $d=1+1$ examples with a line defect, what really matters as far as most of our computations go is the defect codimension.   A codimension-1 planar defect in $d$ dimension can be treated uniformly for any $d$, as follows from the fact that the $d-1$ momenta parallel to the defect are conserved in any connected diagram contributing to a momentum space correlator.  On the other hand, extending our computations to  codimension-$q$ defects is nontrivial, since the 1-dimensional loop integrals we encounter turn into $q$-dimensional loop integrals, which of course changes the divergence structure among other things.   Other codimensions were considered in \cite{Goldberger:2001tn,Michel:2014lva}

\section{Scalar two-point function in presence of defect}
\label{twopoint}
In this section we use the perturbative approach described in section \rff{pertaction} to compute the scalar two-point function in the presence of a defect.  The defect is placed at $x=0$ and  the Euclidean action is
\eq{c1}{  S &=  {1\over 2} \int\! d^2x ( \phid^2 + \phi'^2) +\int\! dt \big( {1\over 2} c^B_1 \phi^2+ c^B_2 \phi \phi' +{1\over 2} c^B_3 {\phi'}^2 \big)\cr
& =  {1\over 2} \int \! {d\omega dp\over (2\pi)^2} \phit(-\omega,-p)(\omega^2+p^2) \phit(\omega,p) \cr
& \quad  +{1\over 2} \int\! {d\omega dp_1 dp_2\over (2\pi)^3} \phit(-\omega,p_2) \big( c^B_1  +ic^B_2 (p_1+p_2) -c^B_3 p_1p_2  \big) \phit(\omega,p_1)  }
where we Fourier transformed as 
\eq{c2}{ \phi(x,t) = \int\! {d\omega dp \over (2\pi)^2} \phit(\omega ,p) e^{i\omega t+ipx} ~.}
The $c_i^B$ denote bare couplings that depend on a UV cutoff $\Lambda$ to be introduced below.  These will eventually be re-expressed in terms of renormalized couplings $c_i$ that are finite as $\Lambda \rt \infty$.   In the above we have omitted the defect cosmological constant term $\int\! dt c_0^B$ since it doesn't affect the connected diagrams contributing to the 2-point function. The full correlator is the product of the connected part and vacuum diagram contribution; the latter will be included in section~\ref{sec: vacuum}, while in this section we consider only the connected part. 

\subsection{Computation} 

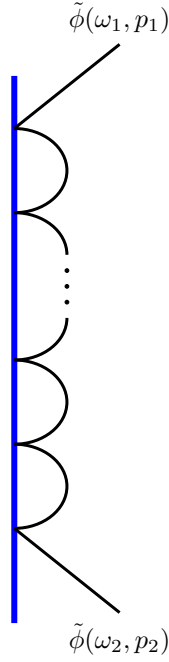
\begin{figure}[H]
\centering
\begin{tikzpicture}[baseline={([yshift=-1ex]current bounding box.center)},vertex/.style={anchor=base,circle,fill=black!25,minimum size=18pt,inner sep=2pt},scale=0.7]
    \draw[line width = 0.8mm, blue] (0,-6.4) -- (0,4);
    \draw[line width = 0.4mm] (0,-4.6) arc(-90:90:1 and 0.8);
    \draw[line width = 0.4mm] (0,-3) arc(-90:90:1 and 0.8);
    \draw[line width = 0.4mm] (0,-1.4) arc(-90:0:1 and 0.8);
    \draw[line width = 0.4mm] (0,1.4) arc(90:0:1 and 0.8);
    \draw[line width = 0.4mm] (0,3) arc(90:-90:1 and 0.8);
    \draw[line width = 0.4mm] (0,-4.6) -- (2,-6.2);
    \node[below] at (2,-6.2) {\footnotesize $\tilde{\phi}(\omega_2,p_2)$};
    \draw[line width = 0.4mm] (0,3) -- (2,4.6);
    \node[above] at (2,4.6) {\footnotesize $\tilde{\phi}(\omega_1,p_1)$};
    \filldraw[black] (1,0.3) circle (1pt);
    \filldraw[black] (1,0) circle (1pt);
    \filldraw[black] (1,-0.3) circle (1pt);
\end{tikzpicture}
\caption{The defect (blue) contributes the quadratic interaction vertices to the connected two point correlation function (shown in momentum space). We sum up these contributions. }
\label{con3}
\end{figure}

Our strategy in this section is to treat the defect action as a perturbation and then sum up contributions to all orders.  We will compute the sum of connected diagrams in momentum space (thus omitting the defect self-energy vacuum diagrams, which will be discussed in section~\ref{sec: vacuum}).   This is shown in Figure \ref{con3}.
 
The two-point function takes the form 
\eq{c3}{ \langle \phit(\omega_1,p_1)\phit(\omega_2,p_2)\rangle = \langle \phit(\omega_1,p_1)\phit(\omega_2,p_2)\rangle_0 + \langle \phit(\omega_1,p_1)\phit(\omega_2,p_2)\rangle_d }
where the correlator in the absence of the defect is 
\eq{c4}{  \langle \phit(\omega_1,p_1)\phit(\omega_2,p_2)\rangle_0 = {1\over \omega_1^2+p_2^2}(2\pi)^2 \delta(\omega_1+\omega_2)\delta(p_1+p_2) }
while the contribution from the defect takes the form 
\eq{c5}{ \langle \phit(\omega_1,p_1)\phit(\omega_2,p_2)\rangle_d = {2\pi \delta(\omega_1+\omega_2) \over (\omega_1^2+p_1^2)(\omega_2^2+p_2^2)} X(\omega_1,p_1,p_2)~. }
where the to-be-computed object $X$ may be expanded in powers of the bare couplings,
\eq{c6}{ X(\omega_1,p_1,p_2) = \sum_{n=0}^\infty I_n(\omega_1,p_1,-p_2)~,\quad I_n \sim (c_i^B)^{n+1}~.}
 An elementary computation yields
\eq{c7}{ I_0(\omega_1,p_1,p_2) =- c_1^B+ic_2^B(p_1-p_2) -c_3^B p_1 p_2~.}
The  $I_{n>0}$ are then determined recursively via 
\eq{c8}{ I_{n+1}(\omega_1,p_1,p_2) & = \int_{-\Lambda}^\Lambda  {dq_1 \over 2\pi} \ldots {dq_{n+1} \over 2\pi} {I_0(\omega_1,p_1,q_1)I_0(\omega_1,q_1,q_2) \ldots I_0(\omega_1,q_{n+1},p_2) \over (\omega_1^2+ q_1^2) \ldots (\omega_1^2+ q_{n+1}^2)} \cr
& = \int_{-\Lambda}^\Lambda \! {dq_{n+1} \over 2\pi} {I_n(\omega_1,p_1, q_{n+1}) I_0(\omega_1,q_{n+1},p_2) \over \omega_1^2+ q_{n+1}^2}   }
where we have introduced the UV cutoff $\Lambda$.  Evaluating for $n=0$ gives a linearly divergent part and a finite part, 
\eq{c9}{ I_1(\omega_1,p_1,p_2) & =\Lambdah (c^B_2-ic^B_3 p_1)(c^B_2+i c^B_3 p_2)  \cr
&- \big[(c^B_1-ic^B_2p_1)(c^B_1+i c^B_2 p_2)-\omega_1^2(c^B_2-ic^B_3 p_1)(c^B_2+i c^B_3 p_2) \big] f_\Lambda(\omega_1)    }
where 
\eq{c10}{ \Lambdah \equiv {\Lambda \over \pi} }
and
\eq{c11}{ f_\Lambda(\omega_1) ={1\over 2\pi } \int_{-\Lambda}^\Lambda {dq \over q^2 +\omega_1^2}= {1\over \pi \omega_1} \tan^{-1}{\Lambda \over \omega_1}~. }
To solve the recursion relation it's useful write out the $p_2$ dependence of $I_n(\omega_1,p_1,p_2)$ as
\eq{c12}{ I_n(\omega_1,p_1,p_2) = a_n + b_n p_2~.}
From \rf{c7} we have 
\eq{c13}{ a_0 & = -c^B_1 +ic_2^B p_1 ~,\cr
b_0 & = -i c^B_2 -c^B_3p_1~. }
Then the recursion relation \rf{c8} becomes
\eq{c14}{\left(
                \begin{array}{c}
                  a_{n+1} \\
                  b_{n+1}   \\
                \end{array}
              \right)  =\left(
                \begin{array}{cc}
                  M_{11} & M_{12} \\
                  M_{21} & M_{22}  \\
                \end{array}
              \right)
              \left( \begin{array}{c}
                  a_{n} \\
                  b_{n}   \\
                \end{array}
              \right)
}
with 
\eq{c15}{
 M_{11}& = -c^B_1 f_\Lambda(\omega_1)\quad\quad M_{12} = ic^B_2 \left( \Lambdah -\omega_1^2 f_\Lambda(\omega_1) \right) \cr
M_{21}& =- ic^B_2 f_\Lambda(\omega_1) \quad\quad M_{22} = -c^B_3 \left( \Lambdah -\omega_1^2 f_\Lambda(\omega_1)  \right)~.   }
This allows us to compute $X$ as 
\eq{c16}{ X(\omega_1,p_1,p_2)  &= \sum_{n=0}^\infty I_n(\omega_1,p_1,-p_2)  \cr
& =  \sum_{n=0}^\infty   \big( a_n  - b_n p_2\big)\cr
& = \sum_{n=0}^\infty \left( \begin{array}{cc}  1 & -p_2 \end{array} \right)  \left( \begin{array}{c}  a_{n} \\ b_{n}  \end{array}\right)      \cr
& =  \sum_{n=0}^\infty \left( \begin{array}{cc}  1 & -p_2 \end{array} \right)  M^n  \left( \begin{array}{c}  a_{0} \\ b_{0}  \end{array}\right) \cr
& = \left( \begin{array}{cc}  1 & -p_2 \end{array} \right)  (I-M)^{-1}   \left( \begin{array}{c}  a_{0} \\ b_{0}  \end{array}\right) ~. }
Evaluating gives
\eq{c17}{ &X(\omega_1,p_1,p_2) =\cr&  {-c^B_1+(c^B_1 c^B_3-(c^B_2)^2)\left[ \Lambdah -\omega_1^2 f_\Lambda(\omega_1) \right] +ic^B_2(p_1+p_2) + \left[ (c^B_1 c^B_3-(c^B_2)^2) f_\Lambda(\omega_1) +c^B_3 \right]p_1 p_2    \over \det(I-M) }\cr }
with
\eq{c18}{ \det(I-M)& = 1+c^B_1 f_\Lambda(\omega_1) +\Big[ (c^B_1 c^B_3-(c^B_2)^2)f_\Lambda(\omega_1)+c^B_3\Big] \left[ \Lambdah-\omega_1^2 f_\Lambda(\omega_1) \right]~. }
Now we wish to remove the cutoff $\Lambda$.  To do so, we expand order by order in the $c^B_i$, and then rewrite them in terms of $\Lambda$ and renormalized couplings $c_i$ such that $\Lambda \rt \infty $ limit at fixed $c_i$ is well defined.  This procedure results in 
\eq{c19}{ c^B_1 & = c_1 +{ c_2^2 \Lamh \over 1- c_3 \Lamh} ~, \cr
c^B_2 & = {c_2  \over 1- c_3 \Lamh} ~,\cr
c^B_3 & = {c_3  \over 1- c_3 \Lamh} ~.}
Taking the large $\Lambda$ limit, using 
$ \lim_{\Lambda \rt \infty} f_\Lambda(\omega_1)  = {1 \over 2|\omega_1| } $, then yields our final result for the renormalized correlator
\eq{c20}{  X(\omega_1,p_1,p_2) = {-c_1 +{|\omega_1| \over 2} (c_1 c_3-c_2^2) +ic_2(p_1+p_2) +\big( {1 \over 2|\omega_1| } (c_1 c_3-c_2^2)  +c_3\big)p_1p_2   \over 1+  {1 \over 2|\omega_1| } (c_1- c_3 \omega_1^2)   -{1\over 4} (c_1 c_3-c_2^2) }~.}

\subsection{Matching conditions from the correlator}

The result  \rf{c20}  can be used to infer matching conditions \rf{b2} across the defect.  We first Fourier transform to position space in the direction transverse to the defect by defining
\eq{c21}{ G(x_1,x_2)&=  \int\! {dp_1 \over 2\pi} {dp_2 \over 2\pi}  \langle \phit(\omega_1,p_1)\phit(\omega_2,p_2)\rangle  e^{ip_1 x_1 + ip_2 x_2} \cr
& = \Gh(x_1,x_2) 2\pi \delta(\omega_1+\omega_2) }
with
\eq{c22}{ \Gh(x_1,x_2) =  \int\! {dp_1 \over 2\pi} {e^{ip_1(x_1-x_2)} \over p_1^2 +\omega_1^2}  +   \int\! {dp_1 \over 2\pi} {dp_2 \over 2\pi}  {X(\omega_1,p_1,p_2) \over  (\omega_1^2 +p_1^2)  (\omega_2^2 +p_2^2) } e^{ip_1 x_1 + ip_2 x_2} ~.  }
Note that in $G$ and $\hat{G}$  we have suppressed the dependence on the frequencies $\omega_{1,2}$.
Now suppose a function takes the form 
\eq{c23}{ f(x) =\int_{-\infty}^\infty \!{dp \over 2\pi} {\alpha+ \beta p \over \omega^2+p^2}e^{ipx}~.}
Using elementary contour integration, the values of the function on either side of the origin are 
\eq{c24}{ f(\pm \eps)  =    {\alpha \over 2|\omega|} \pm {i \beta\over 2} ~,\quad f'(\pm \eps) =    \mp  {\alpha \over 2} - {i\beta |\omega| \over 2}~. }
If we apply these formulas to \rf{c21} and \rf{c20} for $x_1 = \pm \eps$ we find 
\eq{c25}{ \left(
           \begin{array}{c}
             \Gh(\eps,x_2) \\
             \Gh'(\eps,x_2)  \\
           \end{array}
         \right)  = \left(
                       \begin{array}{cc}
                         a & b \\
                         c & d \\
                       \end{array}
                     \right)
           \left(
           \begin{array}{c}
             \Gh(-\eps,x_2) \\
             \Gh'(-\eps,x_2) \\
           \end{array}
         \right)}
with
\eq{c25a}{ \left(
                       \begin{array}{cc}
                         a & b \\
                         c & d \\
                       \end{array}
                     \right)  = \left(\begin{array}{cc}
\frac{\left(2-c_2\right)^2-c_1 c_3}{4-c_2^2+c_1 c_3} & \frac{-4 c_3}{4-c_2^2+c_1 c_3} \\
\frac{4 c_1}{4-c_2^2+c_1 c_3} & \frac{\left(2+c_2\right)^2-c_1 c_3}{4-c_2^2+c_1 c_3}
\end{array}\right) ~.}
We recognize the matrix as being the same as appeared in \rf{b15}, except that the couplings are now the renormalized couplings that appear in the perturbative formulation.   Hence we identify $\tilde{c}_i = c_i$.   The equality between these two sets of couplings is not obvious a priori.  

\subsection{Special cases}

We now consider several special cases of interest.

\subsubsection{IR/UV limit}

To extract the IR/UV limits we rescale the momenta as $(\omega,p)\rt  (\lambda\omega, \lambda p)$.  From \rf{c20} we find 
\eq{c31}{  X(\omega,p_1,p_2)  \rt   \begin{cases}
			-2|\omega |   &   (\omega,k)\rt 0\\
            -{2p_1 p_2  \over |\omega|  } &    (\omega,k)\rt \infty
		 \end{cases} ~.  }
This implies (after a bit of rewriting) the following result for the IR/UV limit of the correlator
\eq{m12}{  \langle \phi(t_1,x_1)\phi(t_2,x_2)\rangle= \int_{-\infty}^\infty\! {d\omega\over 2\pi}  { e^{i\omega (t_1-t_2)} \over 2|\omega|} \left[  e^{-|\omega||x_1-x_2|}  \pm e^{-|\omega| ( x_1 +x_2) }  \right] }
where we have taken $x_{1,2}>0$, and the relative sign is $+$ in the UV and $-$ in the IR.   The first term on the right hand side is the correlator in the absence of the defect, and the second term represents the image charge contribution appropriate for a Neumann ($+$ sign) or Dirichlet ($-$ sign) boundary condition.   This aligns with the fact that the $c_3 (\phi')^2 $ term in the defect action dominates in the UV, setting $\phi'=0$ on the defect; while $c_1 \phi^2$ dominates in the IR, setting $\phi=0$ on the defect. 

\subsubsection{Conformal defect}

Here we set the dimensionful couplings to zero:  $c_1=c_3=0$.  
We  take $x_1>0$ but allow either sign for $x_2$, yielding the defect contribution
\eq{m12a}{ \langle \phi(t_1,x_1)\phi(t_2,x_2)\rangle_d &=  \int_{-\infty}^\infty\! {d\omega\over 2\pi}  { e^{i\omega (t_1-t_2)} \over 2|\omega|}\left\{  \begin{array}{cc}  -{2c_2^2 \over 4+c_2^2} e^{-|\omega|(x_1-x_2)}     &  x_2<0 \cr  -{4c_2 \over 4+c_2^2}  e^{-|\omega|(x_1+x_2)} &  x_2>0
   \end{array} \right\} ~.}

\subsubsection{Topological defect}

For this we take the $c_2\rt \infty $ limit of the conformal case.   The total correlator in this limit, including the free part, is 
\eq{m12b}{ \langle \phi(t_1,x_1)\phi(t_2,x_2)\rangle &=  \int_{-\infty}^\infty\! {d\omega\over 2\pi}  { e^{i\omega (t_1-t_2)} \over 2|\omega|}\left\{  \begin{array}{cc}    -e^{-|\omega| (x_1-x_2) }      &  x_2<0 \cr   e^{-|\omega||x_1-x_2| }  &  x_2>0
   \end{array} \right\}     }
where we have again taken $x_1>0$ but allow either sign of $x_2$.  This exhibits a sign flip of the correlator as $x_2$ crosses the origin, as characterizes the topological defect.

\subsection{Poles}
\label{poles}

The correlator exhibits poles in $\omega_1$ when the denominator of $X(\omega_1,p_1,p_2)$ in \rf{c20} vanishes.  Rewriting this vanishing condition in terms of $(a,b,c,d)$ given in \rf{c25a} yields 
\eq{c26}{  {\rm poles:}\quad b\omega_1^2 +(a+d)|\omega_1|+c=0~.}
  For $b\neq 0$ the roots are
\eq{c30}{ \omega_\pm = { -(a+d) \pm \sqrt{(a-d)^2+4} \over 2b} }
These roots are related to the existence of normalizable bound state modes in the presence of the defect, as we discuss in section \rff{sec: decaying modes}.  The existence of solutions with positive $\omega_\pm$ correspond to instabilities, i.e. modes that grow exponentially in real time.   Coupling space can thus be divided into stable and unstable regions, as will be discussed in section \rff{mode}.

\section{Vacuum diagrams}\label{sec: vacuum}
\label{vacuum}

\begin{figure}[H]
\centering
\begin{tikzpicture}[baseline={([yshift=-1ex]current bounding box.center)},vertex/.style={anchor=base,circle,fill=black!25,minimum size=18pt,inner sep=2pt},scale=0.7]
    \draw[line width = 0.8mm, blue] (0,-3.4) -- (0,3.4);
    \draw[line width = 0.4mm] (0,-3) arc(-90:90:3);
    \draw[line width = 0.4mm] (0,-3) arc(-90:90:1 and 0.8);
    \draw[line width = 0.4mm] (0,-1.4) arc(-90:0:1 and 0.8);
    \draw[line width = 0.4mm] (0,1.4) arc(90:0:1 and 0.8);
    \draw[line width = 0.4mm] (0,3) arc(90:-90:1 and 0.8);
    \filldraw[black] (1,0.3) circle (1pt);
    \filldraw[black] (1,0) circle (1pt);
    \filldraw[black] (1,-0.3) circle (1pt);
\end{tikzpicture}
\caption{ Defect  contributions to the connected vacuum diagram  for  $\langle e^{-S_d}\rangle$}
\label{vac}
\end{figure}
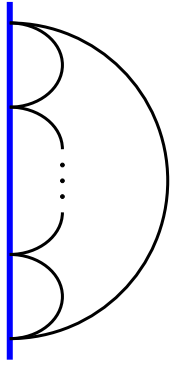

We now consider the sum of vacuum diagrams in the presence of the defect as illustrated in Figure \ref{vac}. 
This can equivalently be thought of as computing the vacuum expectation value of the defect operator,
\eq{c31z} {Z_{\rm vac} =e^W =  \langle e^{-S_d}\rangle~,}
where $W$ represents the connected contributions.     We can evaluate $W$ by summing diagrams, similarly to how we computed the correlator, and arrive at 
\eq{c32}{ W =    -c^B_0 L_d + L_d \sum_{n=0}^\infty {1\over 2(n+1) }  \int_{-\infty}^\infty  {d\omega dp\over (2\pi)^2 } { I_{n}(p,p) \over \omega^2 +p^2 } }
where $I_n(p_1,p_2)$ is the same integral defined in \rf{c8}-\rf{c9} that appeared in the correlator computation.   To get a finite result we have given the defect a large but finite length $L_d$.  We have also included the defect counterterm $c_0^B$.

   To evaluate the sum it is useful to use the identity    
\eq{c33}{ \sum_{n=0}^\infty {M^n \over n+1} = \int_0^1 \! {dz \over I-zM}}
involving the matrix $M$.   We introduce a UV cutoff $\Lambda$ and relate the bare couplings in $S_d$ in terms of renormalized couplings as in  \rf{c19}.    After some algebra we obtain 
\eq{c34}{W &= -c_0^BL_d- L_d  \int_0^1 \! dz  \int_{0 }^\Lambda  {d\omega \over 2\pi }   {1\over \omega}  { \big( -2c_1 +\omega z(c_1 c_3-c_2^2) \big)\omega -\big( c_2^2 z-(2\omega +zc_1) c_3 \big)\omega^2  \over c_1 c_3 \omega z^2-z^2 c_2^2\omega+2z\omega^2 c_3-2zc_1 -4\omega} \cr
& = -c_0^BL_d - L_d \int_0^\Lambda\! {d\omega\over 2\pi}   \ln \left[  {-2c_3\omega^2-(c_1c_3-c_2^2-4)\omega+2c_1 \over 4\omega}  \right] ~.  }
The divergence structure depends on whether the irrelevant $c_3$ and marginal $c_2$ couplings are turned on.   In the simplest case with $c_2=c_3=0$ we have 
\eq{c35}{ W = -c_0^BL_d-\frac{c_1L_d}{4\pi}\ln\frac{2\Lambda}{c_1} - \frac{c_1L_d}{4\pi} ~.}
We therefore take $c_0^B =- {c_1 \over 4\pi} \ln{\Lambda \over \mu}$ where $\mu$ is an RG scale, yielding 
\eq{c36}{  W =- {c_1 L_d\over 4\pi} \left[ 1-\ln\left({c_1\over 2\mu}\right)\right] }
The imaginary part for $c_1<0$ reflects the presence of unstable modes.    
The renormalized defect operator thus depends on the arbitrary RG scale $\mu$  as $\langle e^{-S_d}\rangle \propto \mu^{-{c_1 L_d\over 4\pi} }.$
The fact that the log divergence is linear in $c_1$ implies that the defect anomalous dimension $\gamma_d$ defined in \rf{a7} is one-loop exact when $c_2=c_3=0$.  However, for more general couplings  $\gamma_d$ gets contributions at all loop orders.  For example, setting $c_3=0$ is easily found to imply a log divergent term that is canceled by taking
\eq{c36aa}{ c_0^B =  -{1\over 1+{1\over 4}c_2^2} {c_1 \over 4\pi} \ln (\Lambda/\mu)  + {\rm [linear~divergence]}} 
yielding  $\gamma_d$ in \rf{a8}. With all couplings $c_{1,2,3}$ nonzero one instead finds a divergent series. 

\section{Multiple defects}
\label{fusion_sec}

We now consider two parallel defects separated by a distance $L$.   The partition function, which can be interpreted as computing the inter-defect potential, may be computed in the same manner as for the single defect. 

\subsection{Two defect partition function}

  Writing the defect action as 
\eq{c37}{ S_d = \sum_{a=1}^2\int\! dt \left( c_0^{(a),B} + {1\over 2}c_1^{(a),B} \phi^2(x_a,t)+ c_2^{(a),B}  \phi(x_a,t)\phi'(x_a,t) +  {1\over 2}c_3^{(a),B} \phi'^2(x_a,t) \right) }
with $x_2-x_1=L$ we compute $Z(L)=e^{W(L)}= \langle e^{-S_d}\rangle$ by summing vacuum diagrams.   
For simplicity we will just give the result for $c_2^{(a),B}= c_3^{(a),B}=0$, which is
\eq{c38}{ W(L)=  L_d \left[  \int_0^\Lambda {d\omega \over 2\pi}  \ln \left(  \left( 1+{c_1^{(1)}\over 2\omega}\right) \left( 1+{c_1^{(2)}\over 2\omega}\right)  - {c_1^{(1)} c_1^{(2)} \over 4\omega^2 } e^{-2\omega L}   \right)  - {c_1^{(1)} + c_1^{(2)} \over 4\pi} \ln {\Lambda \over \mu}  \right] }
where we have rewritten the bare couplings in terms of renormalized couplings using the same expressions as before.   The integral is log divergent, but the divergence is canceled by the $c_0^B$ counterterms yielding a finite result as $\Lambda \rt \infty$.  
For small separation we have
\eq{c39}{ W(L) = {(c_1^{(1)}+c_1^{(2)})L_d \over 4\pi }  \left[ 1- \ln \left(  {c_1^{(1)}+c_1^{(2)} \over 2\mu}\right) \right]    +{c_1^{(1)} c_1^{(2)} L_d \over 4\pi} L \ln L + \ldots  }
so that 
\eq{c40}{  Z(L) \approx   Z(0) L^{  c_1^{(1)} c_1^{(2)} LL_d/ 4\pi  }~.  }
For $L=0$ we recover the partition function of a single defect with $c_1= c_1^{(1)}+c_1^{(2)}  $, but the expansion in small $L$ is neither analytic nor a power law.

\subsection{Fusion of two defects}

We now consider the fusion of two parallel defects.  On general grounds, the fusion of two defects with some quadratic couplings $c_i^{(1,2)}$ should result a defect with new quadratic couplings $c_i$.  The expectation is that the couplings $c_i$ are determined by the composition of the matching conditions associated to two fused defects, 
\eq{c40a}{ \left(
                       \begin{array}{cc}
                         a & b \\
                         c & d \\
                       \end{array}
                     \right) = \left(
                       \begin{array}{cc}
                         a^{(2)}  & b^{(2)}  \\
                         c^{(2)}  & d^{(2)}  \\
                       \end{array}
                     \right)\left(
                       \begin{array}{cc}
                         a^{(1)}  & b^{(1)} \\
                         c^{(1)} & d^{(1)} \\
                       \end{array}
                     \right)
}
where we have taken defect $2$ to be to the right of defect $1$.    According to this, the couplings $c_i$ are then extracted from the map between $(a,b,c,d)$ and $c_i$ provided by \rf{c25a}.  This yields a rather complicated relation between the $c_i$ and the $c_i^{(1,2)}$. 

To verify this, we can compute the scalar two-point function in the presence of the two defects separated by a distance $L$, then take $L\rt 0$ and compare to our result for the single defect correlator.    The contribution of the vacuum diagrams is summarized by \rf{c39}.   To define the fused defect via an OPE type procedure, we should strip off the factor of  $L^{  c_1^{(1)} c_1^{(2)} LL_d/ 4\pi  }$.
  The connected diagrams contributing to the correlator may be summed up using the same steps as for the single defect.  The result for general $L$ is complicated and unenlightening.  In appendix \rff{fusion} we write the result for $L\rt 0$.  The result is as expected: the fused couplings $c_i$ are in accord with the rule \rf{c40}.

\section{Mode expansion}
\label{mode}
Consider a free massive or massless $(m=0)$ non-compact scalar theory in  $d=1+1$  Minkowski spacetime
\begin{equation}
S = \int \dr^2x \, \frac{1}{2}\left[ -\left(\partial_t \phi\right)^2 + \left(\partial_x \phi \right)^2 + m^2\phi^2 \right] ~.
\end{equation}
We are interested in a timelike defect described by  the matching conditions at $x=0$ that preserve the time translation symmetry
\eq{eq: gluing condition}{ \left(
           \begin{array}{c}
             \phi_R \\
             \phi'_R \\
           \end{array}
         \right)  =  \left(
                       \begin{array}{cc}
                         a & b \\
                         c & d \\
                       \end{array}
                     \right)
           \left(
           \begin{array}{c}
             \phi_L \\
             \phi'_L \\
           \end{array}
         \right), \quad ad-bc=1 }
where $\phi_{L,R} = \lim_{x\to 0^{\mp}}\phi(t,x)$. Assuming this  matching condition,  the Klein-Gordon (KG) inner product of two solutions,
\begin{equation}
(f,g) \equiv -i \int^\infty_{-\infty} \left(f \dot{g}^* -\dot{f}g^* \right) \, \dr x ~. \label{eq: KG}
\end{equation}
is time independent if and only if $ad-bc=1$, as indicated.

\subsection{Massless scattering modes}

There are two sets of  scattering solutions to the field equation $\partial_\mu\partial^\mu\phi=0$ subject to the matching condition~\eqref{eq: gluing condition},  corresponding to waves incident from the left or right,  %
\begin{equation}
	\varphi_\omega^v= \begin{cases}x<0: & \frac{1}{\sqrt{2\omega}} t_\omega e^{-i \omega v} \\ x>0: & \frac{1}{\sqrt{2\omega}} \left(e^{-i \omega v}+r_\omega^v e^{-i \omega u}\right) \end{cases}, \qquad 	\varphi_\omega^u= \begin{cases}x<0: & \frac{1}{\sqrt{2\omega}} \left(e^{-i \omega u}+r_\omega^u e^{-i \omega v}\right) \\ x>0: & \frac{1}{\sqrt{2\omega}} t_\omega e^{-i \omega u}\end{cases} \label{eq: ab}
\end{equation}
where $u=t-x$, $v=t+x$, and the transmission and reflection coefficients are
\begin{equation}
	t_\omega=\frac{2 i \omega}{b \omega^2+i(a+d) \omega-c}, \quad r_\omega^v=\frac{b \omega^2+i(a-d) \omega+c}{b \omega^2+i(a+d) \omega-c},  \quad r_\omega^u=\frac{b \omega^2-i(a-d) \omega+c}{b \omega^2+i(a+d) \omega-c} ~. \label{eq: aa}
\end{equation}
We normalised the modes with respect to the KG inner product such that
\begin{equation}
    (\varphi^v_\omega,\varphi^v_{\omega'}) = 2\pi \delta(\omega- \omega'),\quad  (\varphi^u_\omega,\varphi^u_{\omega'}) = 2\pi \delta(\omega- \omega'),\quad  (\varphi^u_\omega,\varphi^v_{\omega'}) = 0
\end{equation}
If there are no bound states, the scalar field can be quantised using the scattering solutions as 
\begin{equation}
\phi(t,x) = \int^\infty_0 \frac{\dr\omega}{2\pi} \left(a_\omega^v \varphi_\omega^v + a_\omega^u\varphi_\omega^u + \text{h.c.}\right), \quad \pi(t,x) = \dot{\phi}(t,x) ~.
\end{equation}
Writing
\begin{equation}
	\begin{aligned}
		& a^{v,u}_\omega=\left(\phi, \varphi^{v,u}_\omega\right)=-i \int_{-\infty}^{\infty}\left(\phi(x) \dot{\varphi}^{v,u}_\omega{}^*(x)-\pi(x) \varphi^{v,u}_\omega{}^*(x)\right) \mathrm{d} x \\
		& a^{v,u}_\omega{}^{\dagger}=-\left(\phi, \varphi^{v,u}_\omega{}^*\right)=i \int_{-\infty}^{\infty}\left(\phi(x) \dot{\varphi}^{v,u}_\omega(x)-\pi(x) \varphi^{v,u}_\omega(x)\right) \mathrm{d} x
	\end{aligned}
\end{equation}
and using $[\phi(x),\pi(y)] = i\delta(x-y)$ at equal time, one can derive the canonical quantization conditions 
\begin{equation}
[a_\omega, a_{\omega'}^\dagger] = 2\pi \delta (\omega-\omega') \label{eq: ac} ~.
\end{equation}
For this to be consistent, the classical solutions should satisfy the completeness condition derived from the the equal-time commutator
\begin{equation}
\begin{aligned}
[\phi(x),\pi(y)]&=	\int_0^{\infty}\frac{\mathrm{d} \omega}{2 \pi} \left(\varphi_\omega^v(x) \dot{\varphi}_\omega^{v *}(y)-\varphi_\omega^{v *}(x) \dot{\varphi}_\omega^v(y)\right) +(v \leftrightarrow u)\\
&= i\delta(x-y) ~.
\end{aligned}
\end{equation}
Evaluated on the scattering solutions~\eqref{eq: ab} and \eqref{eq: aa} we find, 
\begin{equation}
\begin{aligned}
[\phi(x),\pi(y)] &= \begin{cases}
 i \delta(x-y) + i\int^\infty_{0} \frac{\dr \omega}{2\pi} r_\omega e^{i\omega|x+y|}~,  & xy>0 \\
 i\int^{\infty}_{0} \frac{\dr \omega}{2\pi} t_\omega\, e^{i\omega|x-y|}~, & xy<0  
\end{cases} ~.
\end{aligned} \label{eq: ad}
\end{equation}
We see that completeness is satisfied provided $r_\omega$ and $t_\omega$ are analytic in the  upper half $\omega$-plane, since then the integrals can be converted into closed contour integrals that vanish by Cauchy's theorem.  On the other hand,  the existence of poles in the upper-half plane implies that the scattering solutions should be accompanied by  bound states for completeness.

\subsection{Bound state  modes} \label{sec: decaying modes}

\begin{figure}[H]
	\centering
	\begin{subfigure}{0.45\textwidth}
		\centering
		\begin{tikzpicture}
			\begin{axis}[
				xlabel={$a$},
				ylabel={$d$},
				xmax=5.4,
				xmin=-5.4,
				xticklabel=\empty,
				yticklabel=\empty,
				axis lines=middle,
				samples=200
				]
				\addplot[blue,thick,domain=-5:-0.1, name path=A] {1/x} node [below,pos=0.2] {$ad=1$};
				\addplot[draw=none,name path=B] {-10};
				\addplot[red,fill opacity=0.2] fill between[of=A and B,soft clip={domain=-8:0}];
				\addplot[blue,thick,domain=0.1:5, name path=C] {1/x} node[above, pos=0.8] {$ad=1$};
				\addplot[yellow,fill opacity=0.2] fill between[of=C and B,soft clip={domain=0:8}];
				\addplot[draw=none,name path=D] {10};
				\addplot[yellow,fill opacity=0.2] fill between[of=D and A,soft clip={domain=-6:0}];
				\addplot[cyan,fill opacity=0.2] fill between[of=C and D,soft clip={domain=0:6}];
				%\node [anchor=north] at (axis cs:-0.3,-0.3) {$O$};
				\node [anchor=north] at (axis cs:-2.4,-5) {$0<\kappa_+<\kappa_-$};
				\node [anchor=north] at (axis cs:2.5,6) {$\kappa_+<\kappa_-<0$};
				\node [anchor=north] at (axis cs:-2.4,6) {$\kappa_+<0<\kappa_-$};
				%	\draw[red!20,dashed] (axis cs:2,-4) -- (axis cs:2,10);
			\end{axis}
		\end{tikzpicture}
		\caption{$b>0$}
		\label{fig:bound state posb>0}
	\end{subfigure}%
		\begin{subfigure}{0.45\textwidth}
		\centering
		\begin{tikzpicture}
			\begin{axis}[
				xlabel={$a$},
				ylabel={$d$},
				xmax=5.4,
				xmin=-5.4,
				xticklabel=\empty,
				yticklabel=\empty,
				axis lines=middle,
				samples=200
				]
				\addplot[blue,thick,domain=-5:-0.1, name path=A] {1/x} node [below,pos=0.2] {$ad=1$};
				\addplot[draw=none,name path=B] {-10};
				\addplot[cyan,fill opacity=0.2] fill between[of=A and B,soft clip={domain=-8:0}];
				\addplot[blue,thick,domain=0.1:5, name path=C] {1/x} node[above, pos=0.8] {$ad=1$};
				\addplot[yellow,fill opacity=0.2] fill between[of=C and B,soft clip={domain=0:8}];
				\addplot[draw=none,name path=D] {10};
				\addplot[yellow,fill opacity=0.2] fill between[of=D and A,soft clip={domain=-6:0}];
				\addplot[red,fill opacity=0.2] fill between[of=C and D,soft clip={domain=0:6}];
				%\node [anchor=north] at (axis cs:-0.3,-0.3) {$O$};
				\node [anchor=north] at (axis cs:-2.4,-5) {$\kappa_-<\kappa_+<0$};
				\node [anchor=north] at (axis cs:2.5,6) {$0<\kappa_-<\kappa_+$};
				\node [anchor=north] at (axis cs:-2.4,6) {$\kappa_-<0<\kappa_+$};
				%	\draw[red!20,dashed] (axis cs:2,-4) -- (axis cs:2,10);
			\end{axis}
		\end{tikzpicture}
		\caption{$b<0$}
		\label{fig:bound state posb<0}
	\end{subfigure}
	\caption{Regions of $(a,b,d)$ space where $\kappa_\pm$ have various signs. One of the signs flips across the blue curves. There are two normalisable decaying modes in the red region, one in yellow region and none in the blue region respectively.}
	\label{fig: bound state frequency2}
\end{figure}
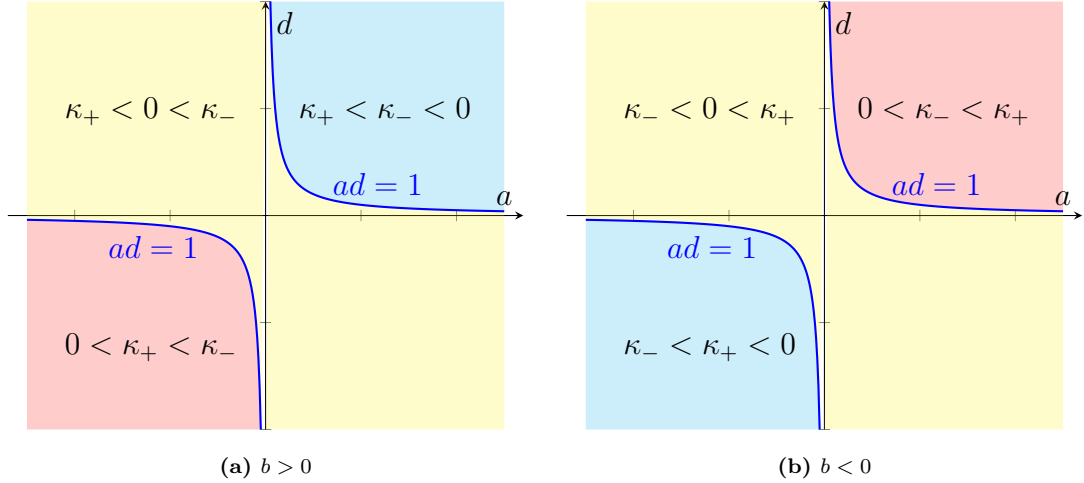

The potential incompleteness of scattering modes leads us to examine the existence of normalizable bound state modes.  These are solutions that decay exponentially as $|x|\to \infty$, and either grow or decay exponentially in $t$ (the two cases being related by time reversal.) 
Plugging  the following  ansatz for modes that grow in $t$,   
\begin{equation}
	\phi=\left\{\begin{array}{ll}
		x<0: & A e^{\kappa v} \\
		x>0: & B  e^{\kappa u}
	\end{array}, \quad \quad Re(\kappa)>0\right.
\end{equation}
where $Re(\kappa)>0$ for normalizability, into the gluing condition~\eqref{eq: gluing condition}, a normalizable mode is seen to exist if at least one solution to 
\eq{qq1}{
	b \kappa^2+(a+d) \kappa+c=0
}
has a positive real part.   Upon writing  $\kappa =-i\omega$ we recognize  \rf{qq1} as  giving the locations of the poles in the  transmission/reflection coefficients \eqref{eq: aa}.
For $b \neq 0$ normalizable solutions exist if 
\begin{equation}
\kappa_+>0 \, \text{ or }\, \kappa_->0, \quad 	\kappa_{\pm}=-\frac{a+d \pm \sqrt{(a-d)^2+4}}{2 b}  \label{eq: bound state frequency} ~.
\end{equation}
For $b=0$ we have $\kappa = - {ac\over 1+a^2 }$   (using $d=1/a$) so that normalizable modes exist for $ac<0$.   
In the previous section, we saw that the scattering modes alone are not complete when $\text{Im}(\omega_+)>0$ or $\text{Im}(\omega_-)>0$. This is precisely the condition for the existence of the  bound states, $\text{Re}(\kappa_+)>0$ or $\text{Re}(\kappa_-)>0$, illustrated in Figure~\ref{fig: bound state frequency2}.  

For such a case, the solutions come in a pair $(\phi_\pm,\bar{\phi}_\pm)$ related by time reversal,  
\begin{equation}
	\begin{aligned}
		\phi_{ \pm} & = \begin{cases}x<0: & N_{ \pm}^{-1} e^{\kappa_{ \pm} v} \\
			x>0: & N_{ \pm}^{-1}\left(a+b \kappa_{ \pm}\right) e^{\kappa_{ \pm} u}\end{cases}, \qquad 		\bar{\phi}_{ \pm} & = \begin{cases}x<0: & \bar{N}_{ \pm}^{-1} e^{-\kappa_{ \pm} u} \\
			x>0: & \bar{N}_{ \pm}^{-1}\left(a+b \kappa_{ \pm}\right) e^{-\kappa_\pm v}\end{cases}
	\end{aligned} \label{eq: bound states}
\end{equation}
where the normalization 
\begin{equation}
	N_{\pm}^{-1}=\bar{N}_{\pm}^{-1}=\frac{1}{\sqrt{1+\left(a+b \kappa_{\pm}\right)^2}} \label{eq: b normalisation}
\end{equation}
is chosen such that the solutions are real-valued and satisfy $(\phi_\pm,\bar{\phi}_\pm) = i$.

The $\phi_\pm$ modes grow exponentially in time,  hence the theory is unstable when such modes exist.   Nonetheless, we can formally quantize the scalar field as
\begin{equation}
	\phi(t, x)=\int_0^{\infty} \frac{\mathrm{d} \omega}{2 \pi}\left(a_\omega \varphi_\omega^L+b_\omega \varphi_\omega^R+\text { h.c. }\right)+c_{+} \phi_{+}+\bar{c}_{+} \bar{\phi}_{+}+c_{-} \phi_{-}+ \bar{c}_{-} \bar{\phi}_{-} \label{eq: ae}
\end{equation}
where $c_\pm$ are hermitian, for $\omega_\pm >0$ and similarly for other cases. The same process used to derive~\eqref{eq: ac} can be recycled to derive the commutation of the modes for the bound states
\begin{equation}
[c_\pm,\bar{c}_\pm] = (\phi_\pm,\bar{\phi}_\pm) = i ~.
\end{equation}
Using this, the contribution from the decaying modes to the equal-time commutator $[\phi(x),\pi(y)]$ exactly cancels the incomplete contribution~\eqref{eq: ad} from the scattering modes and hence restores the completeness 
\begin{equation}
[\phi(x),\pi(y)] = i\delta(x-y) ~.
\end{equation}

\subsection{Spectrum}

The classical normalizable modes grow exponentially in time, and  when such modes exist the Hamiltonian  suffers from a  spectrum unbounded from below. To see this, first note that the bulk Hamiltonian $H=\frac{1}{2}\int^{\infty}_{-\infty} \left(\dot{\phi}^2+(\phi')^2\right)\dr x$ has a   time dependence
\begin{equation}
\frac{\dr H}{\dr t} = \frac{\mathrm{d}}{\mathrm{~d} t}\left(b c \phi_L^{\prime} \phi_L+\frac{a c}{2} \phi_L^2+\frac{b d}{2}\left(\phi_L^{\prime}\right)^2\right)~.
\end{equation}
The Hamiltonian can be made time-independent by subtracting the total derivative above
\begin{equation}
H_{tot} = H - \left(b c \phi_L^{\prime} \phi_L+\frac{a c}{2} \phi_L^2+\frac{b d}{2}\left(\phi_L^{\prime}\right)^2\right)
\end{equation}
corresponding to the defect Hamiltonian. 
The total Hamiltonian restricted to the normalizable modes is
\begin{equation}
\begin{aligned}
H_{tot}&= -\frac{\kappa_+}{2}\left( c_+\bar{c}_+ + \bar{c}_+ c_+ \right) - \frac{\kappa_-}{2}\left( c_-\bar{c}_- + \bar{c}_- c_- \right)  ~.
\end{aligned}
\end{equation}
Renaming $c_+ = \frac{1}{\sqrt{2}}(p+x), \bar{c}_+ = \frac{1}{\sqrt{2}}(p-x)$, $H_{tot}$ is that of the inverted harmonic oscillator, which is known to have continuous spectrum unbounded from below and above, assuming a boundary condition that requires the wave function to vanish at spatial infinity~\cite{BARTON1986322}. The unbounded spectrum implies that whenever there is a bound state, the massless theory with  matching condition~\eqref{eq: gluing condition} is ill-defined. However,  the instability of the classical solutions and the unbounded spectrum can by cured by adding a mass, as we describe in the next subsection,  or by adding bulk interactions, as discussed in Appendix~\ref{app: solitions}.  Alternatively, we can restrict attention to the stable regime shown in Figure~\ref{fig: bound state frequency2} (blue region). 

\subsection{Massive scalar mode expansion}
The massive scalar theory with matching condition~\eqref{eq: gluing condition} has scattering modes given by
\begin{align}
	\varphi_k^L&= \begin{cases}x<0: & \frac{1}{\sqrt{2 \omega}} t_k e^{-i \omega t-i k x} \\ x>0: & \frac{1}{\sqrt{2 \omega}}\left(e^{-i \omega t-i k x}+r_k e^{-i \omega t+i k x}\right)\end{cases}\\
    \varphi_k^R&= \begin{cases}x<0: & \frac{1}{\sqrt{2 \omega}}\left(e^{-i \omega t+i k x}+r_k e^{-i \omega t-i k x}\right) \\ x>0: & \frac{1}{\sqrt{2 \omega}} t_k e^{-i \omega t+i k x}\end{cases}
\end{align}
where $\omega = \sqrt{k^2+m^2}$ and $t_k$ and $r_k$ are the same as $t_\omega$ and $r_\omega$~\eqref{eq: aa} with $\omega \to k$. 

As in the massless case,  bound state modes exist if at least one of $\kappa_\pm$ given in Eq.~\eqref{eq: bound state frequency} has  positive real part. When they exist, those with $\kappa_+^2 > m^2$ or $\kappa_-^2 > m^2$ are unstable, while those with $\kappa_+^2<m^2$ or $\kappa_- <m^2$ are oscillatory. We focus on the oscillatory  modes, which are given by
\begin{equation}
	\begin{aligned}
		\phi_{ \pm} & = \begin{cases}x<0: & N_{ \pm}^{-1} e^{-i \bar{\omega}_{ \pm} t+\kappa_{ \pm} x} \\
			x>0: & N_{ \pm}^{-1}\left(a+b \kappa_{ \pm}\right) e^{-i \bar{\omega}_{ \pm} t-\kappa_{ \pm} x}\end{cases} 
	\end{aligned}
\end{equation}
where 
\begin{equation}
	\kappa_{ \pm}= -\frac{a+d \pm \sqrt{(a-d)^2+4}}{2 b} >0, \quad  \bar{\omega}_{ \pm}=\sqrt{m^2-\kappa_{ \pm}^2} \in \mathbb{R}
\end{equation}
and
\begin{equation}
	N_{+}^{-1}=\sqrt{\frac{\kappa_{+}}{\bar{\omega}_{+}}} \sqrt{\frac{1}{1+\left(a+b \kappa_{+}\right)^2}} \in \mathbb{R}^{+}
\end{equation}
The scalar field is quantized as
\begin{equation}
	\phi(t, x)=\int_0^{\infty} \frac{\mathrm{d} \kappa}{2 \pi}\left(a_\kappa \varphi_\kappa^L+b_\kappa \varphi_\kappa^R+h . c .\right)+\left(c_{+} \phi_{+}+c_{-} \phi_{-}+h . c .\right)
\end{equation}
where
\begin{equation}
	\left[c_{\pm}, c_{\pm}^{\dagger}\right]=\left(\phi_{\pm}, \phi_{\pm}\right)=1 \label{eq: af}
\end{equation}
when both $\kappa_\pm >0$ and similarly for other cases by dropping some bound state modes. The Hamiltonian restricted to the bound states  is
\begin{equation}
\begin{aligned}
H &= \frac{\bar{\omega}_+}{2}\left( c_+^\dagger c_+ + c_+c_+^\dagger \right) + \frac{\bar{\omega}_+}{2}\left( c_-^\dagger c_- + c_-c_-^\dagger \right) 
&= \bar{\omega}_+\left( \bar{c}_+c_+ + \frac{1}{2} \right) + \bar{\omega}_- \left( \bar{c}_-c_- + \frac{1}{2} \right) ~.
\end{aligned}
\end{equation}
Defining the vacuum state as $c_\pm |0\rangle = 0$, one can see that the spectrum is positive.

\section{Ring partition function, g-function,  and entanglement entropy of accelerating defects}
\label{ring}

\subsection{Ring partition function} 

In this section we compute the partition function of a ring shaped defect.  In order to apply this result to the applications  that follows, we consider placing the ring on a cone.  Namely, we consider
\eq{c50}{ ds^2 = dr^2 +r^2 d\theta^2~,\quad \theta \cong \theta + \beta }
with the ring placed at $r=1$.\footnote{Choosing $r=1$ implies no loss of generality since we can always rescale $r$ along with the couplings.} The ring partition function may be written as 
\eq{c51}{ Z(\beta) = \langle 0 | e^{-S_d} |0\rangle }
where we interpret the states in terms of radial quantization and think of $e^{-S_d}$ as an operator on this Hilbert space.  We will consider the simplest case of $c_2=c_3=0$; in this case $e^{-S_d}$ is diagonalized in the field eigenstate basis, since the ``conjugate momentum" $\phi'$ does not appear in the defect action.  Letting $\phib(\theta) = \phi(r=1,\theta)$ we then have 
\eq{c52}{ Z(\beta)= \int {\cal D} \phib \langle 0|\phib\rangle e^{-S_d[\phib]} \langle \phib |0\rangle~.}
We compute $\langle \phib |0\rangle $ by computing the free scalar path integral in the region $r\leq 1$ with boundary condition $\phi(r=1,\theta ) =\phib(\theta)$.  We will only be interested in the $c_1$ dependence of $\log Z(\beta)$, in which case we only need the saddle point approximation to the path integral since the fluctuation determinant does not depend on $c_1$.  

Writing
\eq{c53}{ \phib(\theta) = \phib_0 + \sum_{n\neq 0 } \phib_n e^{2\pi in \theta \over \beta}~,\quad \phib_{-n}=\phib_n^*}
the scalar field solution is 
\eq{c54}{ \phi(r,\theta) =  \phib_0 + \sum_{n=1}^\infty \big( \phib_{-n}e^{2\pi i n \theta \over \beta} +\phib_n e^{-{2\pi i n \theta \over \beta}} \big)r^{2\pi n/\beta} }
  The action interior to the ring is
\eq{c55}{ S_{r<1} & =\int_{r<1} d^2x  {1\over 2} (\p \phi)^2  =  2\pi\sum_{n=1}^\infty n \phib_n \phib_{-n}  ~.}
The contribution of the exterior region $S_{r>1}$ takes the same form, as follows from symmetry under $r\rt 1/r$.   The defect action, including the $c_0^B$ counterterm,  is
\eq{c56}{ S_d &= \int\! d\theta \left( c_0^B+  {1\over 2}c_1 \phib^2\right)  = \beta c_1 \left[ \sum_{n=1}^\infty \phib_n \phib_{-n}+{1\over 2} \phib_0^2\right]+\beta c_0^B ~.   }
The ring partition function is then 
\eq{c57}{ Z(\beta) & = {\cal N}\left[  \prod_n  \int d\phib_n \right] e^{-S_{r<1}-S_{r>1}-S_d} \cr
& =  {\cal N}  e^{-\beta c_0^B} \int_{-\infty}^\infty d\phib_0 e^{-{1\over 2}\beta c_1 \phib_0^2}  \prod_{n\geq 1}  \int d\phib_nd\phib_{-n}  e^{-(4\pi n+ \beta c_1) \phib_n \phib_{-n}   } \cr
& =   {\cal N} { e^{-\beta c_0^B}\over \sqrt{\beta c_1} }   \prod_{n\geq 1} (4\pi n +\beta c_1)^{-1}\cr
& = {\cal N} { e^{-\beta c_0^B}\over \sqrt{\beta c_1} }   \prod_{n\geq 1} \left( 1+{\beta c_1\over 4\pi n} \right)^{-1} ~.}
 where in the final equality we have absorbed a constant factor into ${\cal N}$.  
We thus find
\eq{c58}{ \log Z(\beta) & = \log {\cal N} -{1\over 2} \log{ \beta c_1 \over 4\pi}  -\sum_{n=1}^N \log\left( 1 +{\beta c_1\over 4\pi n} \right)-\beta c_0^B ~,}
where we  imposed a UV cutoff on the mode sum.  To cancel the log divergence we take
\eq{c59}{  c_0^B = -{c_1 \over 4\pi} \log N~.}
We then take  $N\rt \infty$ using the formula\footnote{This follows from the  Weierstrass formula for the Gamma function,
$\displaystyle -\ln \Gamma(z+1) = \sum_{m=1}^\infty \left[ \log (1+ {z\over m} ) - {z\over m} \right] + \gamma z$ 
where $\gamma$ is the Euler-Mascheroni constant,  $\displaystyle  \gamma = \lim_{N\rt \infty} \left[ \sum_{m=1}^N {1\over m} - \log N \right]$.}
\eq{c60}{ \lim_{N\rt \infty} \left[ \sum_{n=1}^N \log\left(1+ {z\over n}\right) - z\log N \right] = -\log \Gamma(1+z)~.}
We finally arrive at
\eq{c61}{ \log Z(\beta) 
& =\log {\cal N}   +{1\over 2} \log{ \beta c_1 \over 4\pi}  + \log \Gamma\left({\beta c_1 \over 4\pi}\right) ~,}
where we recall that ${\cal N}$ does not depend on $c_1$.

\subsection{Defect $g$-function}

Thinking of $Z(\beta)$ as a thermal partition function we can define the associated thermodynamic entropy,\footnote{We use $s$ to avoid confusion with the action $S$.}
\eq{c62}{  s = (1-\beta \p_\beta) \log Z(\beta) ~.}
Keeping just the $c_1$ dependent part, this evaluates to 
\eq{c63}{s & ={1\over 2} \log{ \beta c_1 \over 4\pi}  + \log \Gamma\left({\beta c_1 \over 4\pi}\right) -{\beta c_1 \over 4\pi} \Psi\left({\beta c_1 \over 4\pi}\right) -{1\over 2} }
where $\Psi(x)  = \Gamma'(x)/\Gamma(x)$.  

We now comment on the relation of this result to the g-theorem \cite{Affleck:1991tk,Kutasov:2000qp,Friedan:2003yc}.  We first observe that $s$ is a monotonically decreasing function of $c_1$, which is in accord with the g-theorem.   The connection to the g-theorem, which is usually phrased in terms of a theory defined on a space with boundary, is made via the folding trick under which our ring partition function is related to the scalar field partition function  on  a disk with a $c_1 \phi^2$ interaction on the boundary.  Such partition functions were considered in the context of boundary string field theory \cite{Witten:1992cr,Kutasov:2000qp,Kraus:2000nj,Takayanagi:2000rz}.   It would be natural to extend our computation to allow for general $c_{1,2,3}$ couplings but we were unable to obtain a closed form expression.   The result should decrease along the RG flow according to the argument in \cite{Friedan:2003yc}.   We also note that the g-theorem, defects,  and relations to ring partition functions have been considered in more general contexts in \cite{Casini:2022bsu,Cuomo:2021rkm}. 

\subsection{Entanglement entropy of accelerating defects}

This section involves the following problem.   We consider two defects under going  constant proper acceleration $a$ in opposite directions.   At time $t=0$ they are mutually at rest and  located at positions $x=\pm  a^{-1} $.  The field $\phi$ is assumed to be in the Lorentz invariant quantum state compatible with the Lorentz invariance of the defect trajectories.   Given this setup, we wish to compute the entanglement entropy at $t=0$ between the two half-spaces $x<0$ and $x>0$.     The motivation for posing this precise question is that the answer is easily obtained from the partition function \rf{c61}. 

The Lorentzian trajectories of the defects are
\eq{c45}{ x(\tau) & = \pm  {1\over a}  \cosh(a \tau) ~,\quad
t(\tau)  = {1\over a}  \sinh(a \tau)~}
where $\tau $ is proper time.   Rotating to Euclidean time via $t\rt it$ and $\tau \rt i\tau$ the trajectory turns into a circle of radius $R=1/a$,
\eq{c46}{x(\tau) = {1\over a} \cos (a\tau)~,\quad t(\tau) = {1\over a} \sin(a\tau) }
where $\tau$ is now periodic,  $\tau \cong \tau +2\pi/a$.  This is illustrated in Figure~\ref{fig: defect trajectories}.
\begin{figure}[H]
		\centering
    \begin{subfigure}{0.45\textwidth}
	\centering
	\begin{tikzpicture}[baseline={([yshift=-1ex]current bounding box.center)},vertex/.style={anchor=base,circle,fill=black!25,minimum size=18pt,inner sep=2pt},scale=0.7]
    \draw[line width = 0.4mm, -stealth] (-4,0) -- (4,0);
    \draw[line width = 0.4mm, -stealth] (0,0) -- (0,5);
    \draw[blue, line width = 0.6mm] plot[variable=\x,domain=0:2.5, samples=40] ({2*cosh(\x/2)}, {2*sinh(\x/2)});
    \draw[blue, line width = 0.6mm] plot[variable=\x,domain=0:2.5, samples=40] ({-2*cosh(\x/2)}, {2*sinh(\x/2)});
    \node[below, black] at (1.8,.1) {\scriptsize $\sfrac{1}{a}$};
    \node[below, black] at (-2.5,.1) {\scriptsize $-\sfrac{1}{a}  $};
    \node[right, black] at (0,5) {\footnotesize $t$};
    \node[below, black] at (4,0) {\footnotesize $x$};
    \draw[line width = 0.3mm, stealth-] (-4.5,0) -- (-4.5,1);
    \draw[line width = 0.3mm, -stealth] (-4.5,4) -- (-4.5,5);
    \node[rotate = 90] at (-4.5,2.5) {\scriptsize Lorentzian time};
    \draw[line width = 0.4mm, -stealth] (0,0) -- (3,-2);
    \draw[blue, line width = 0.6mm] plot[variable=\x,domain=0:180, samples=40] ({+2*cos(\x/2) + \x/100},{-1.2*sin(\x/2)});
    \draw[blue, line width = 0.6mm] plot[variable=\x,domain=180:360, samples=40] ({+2*cos(\x/2) - (\x-360)/100},{-1.2*sin(\x/2)});
    \node[below] at (3,-2) {\footnotesize $it$};
    \draw[black, line width = 0.3mm, stealth-] (-4.5,-0.5) -- (-3.75,-1);
    \draw[black, line width = 0.3mm, -stealth] (-1.35,-2.6) -- (-0.6,-3.1);
    \node[rotate = -33.69] at (-2.55,-1.8) {\scriptsize Euclidean time};
\end{tikzpicture}
	\caption{}
	\label{fig:L}
\end{subfigure}
\begin{subfigure}{0.45\textwidth}
	\centering
\begin{tikzpicture}[baseline={([yshift=-1ex]current bounding box.center)},vertex/.style={anchor=base,circle,fill=black!25,minimum size=18pt,inner sep=2pt},scale=2]
	% Triangle
	\draw[line width = 0.4mm] (2,1)-- (0,0) -- (2,-1);
	% Curved arc inside the triangle
	\draw[line width = 0.6mm, blue] (1.4,-0.7) arc[start angle=-26.5, end angle=26.5, radius=1.5cm];
	% Label for the arc
	\node at (1.9,0) {$\beta$};
	% Dotted detail inside the triangle
	\draw[line width = 0.6mm, blue, dashed] (1.4,0.7)  arc[start angle=153.5, end angle=206.5, radius=1.5cm];
\end{tikzpicture}
	\caption{}
        \label{fig:E}
\end{subfigure}
\caption{Figure \ref{fig:L}: The blue curve depicts the Lorentzian trajectory of the accelerating defect and its Euclidean counter part. Figure \ref{fig:E}: The ring defect in Euclidean space that corresponds to the accelerating defect in the Lorentzian spacetime. Here $\beta = 2\pi/a$, and we have included a cone to indicate that we will be varying $\beta$ to compute the entanglement entropy. }
\label{fig: defect trajectories}
\end{figure}

The density matrix $\rho_R[\phi_+,\phi_-]$ describing the $x>0$ half-space has a standard path integral representation, obtained (up to normalization) by doing the path integral on the Euclidean plane with boundary conditions $\phi_+ (\phi_-)$ imposed just above (below) the positive real axis, as shown in figure \rff{accel}. 
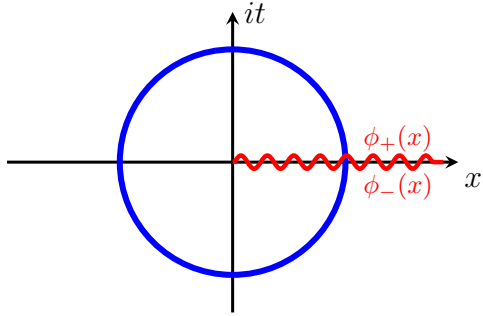
\begin{figure}[H]
\centering
\begin{tikzpicture}[baseline={([yshift=-1ex]current bounding box.center)},vertex/.style={anchor=base,circle,fill=black!25,minimum size=18pt,inner sep=2pt},scale=1]
    \draw[line width = 0.4mm, -stealth] (-3,0) -- (3,0);
    \node[below] at (3.2,0) {$x$};
    \draw[line width = 0.4mm, -stealth] (0,-2) -- (0,2);
    \node[above] at (0.3,1.7) {$it$};
    \draw[line width = 0.8mm, blue] (0,0) circle (1.5);
    \draw[snake it, line width = 0.6mm, red] (0,0) -- (2.8,0);
    \node[red, above] at (2.2,0) {\footnotesize $\phi_+(x)$};
    \node[red, below] at (2.2,0) {\footnotesize $\phi_-(x)$};
\end{tikzpicture}
\caption{Pair of accelerating defects continued to imaginary time where they describe a ring.   The density matrix describing the $x>0$ half-space is obtained by performing the path integral in the presence of the defect with prescribed boundary conditions above and below the cut along the positive real axis.}
\label{accel}
\end{figure}

The entanglement entropy is obtained as
\eq{c47}{ s  = -\Tr (\rho \ln \rho)~.}
As  in the absence of the defect, it's convenient to think  of the usual angular direction $\theta$  on the plane as a time direction generated by a Rindler  Hamiltonian $H_{\rm Rin} \sim  \p_\theta$.  In particular the density matrix becomes
\eq{c48}{ \rho[\phi_+,\phi_-] =N \langle \phi_- |e^{- {2\pi} H_{\rm Rin} }|  \phi_+\rangle }
where $N$ is the normalization factor needed to set $\Tr \rho =1$.
If we define
\eq{c48z}{ Z(\beta ) = \Tr e^{-\beta H_{\rm Rin} }  }
then the entanglement entropy can be obtained as the  entropy associated to the partition function $Z(\beta)$ using the standard thermodynamic expression
\eq{c49}{  s = (1-\beta \p_\beta) \log Z(\beta) \big|_{\beta =2\pi}~.}
Finally, $Z(\beta)$ also has a simple path formulation:  it is given by the path integral on the cone, since the periodicity $\theta \cong \theta +2\pi$ has been replaced by the periodicity $\theta \cong \theta +\beta$ according to \rf{c48}.

For simplicity we will only consider the defect with the $c_1 \phi^2$ coupling.
The entanglement entropy will be a function of the dimensionless combination $c_1/a$.  This being the case we will henceforth set $a=1$, as it can always be restored by taking $c_1 \rt c_1/a$.  The entanglement entropy is then obtained by setting $\beta =2\pi$ in \rf{c63}, which gives 
\eq{c64}{s & ={1\over 2} \log{  c_1 \over 2}  + \log \Gamma\left({ c_1 \over 2}\right) -{ c_1 \over 2} \Psi\left({ c_1 \over 2}\right) -{1\over 2} ~.}
It follows that this entanglement entropy monotonically decreases with increasing $c_1$, yielding an entanglement interpretation of the g-theorem in this example. 

\section{Defect on torus}\label{torusZ} 

In this section we consider the torus partition function with a defect wrapped around one of the cycles.   One motivation is to understand how to relate the computation of the partition via a trace in either of two channels on the torus; this is a bit more subtle than usual due to the $\mu$ dependence of the defect.     The partition function will also allow us to obtain a Cardy type formula for the asymptotic density of states of the defect Hilbert space.  We restrict out attention to the $c_1$ defect, with $c_2=c_3=0$.

\subsection{Defect Hilbert space calculation}\label{sec:dh}
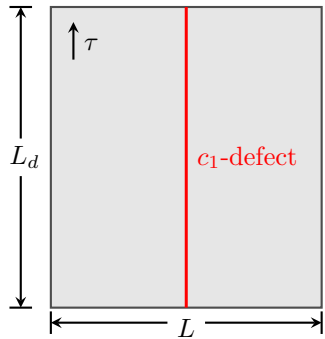
\begin{figure}[H]
    \centering
    \begin{tikzpicture}[baseline={([yshift=-1ex]current bounding box.center)},vertex/.style={anchor=base,circle,fill=black!25,minimum size=18pt,inner sep=2pt},scale=1]
    \filldraw[grey] (-1.8,-2) rectangle ++(3.6,4);
    \draw[thick, dgrey] (-1.8,-2) rectangle ++(3.6,4);
    \draw[line width = 0.4mm, red] (0,-2) -- (0,2);
    \draw[thick, black] (-1.8,-2.35) -- (-1.8,-2.05);
    \draw[thick, black] (1.8,-2.35) -- (1.8,-2.05);
    \draw[thick, black, -stealth] (0.25,-2.2) -- (1.8,-2.2);
    \draw[thick, black, -stealth] (-0.25,-2.2) -- (-1.8,-2.2);
    \node[below] at (0,-2) {\footnotesize $L$};

    \draw[thick, black] (-2.05,-2) -- (-2.35,-2);
    \draw[thick, black] (-2.05,+2) -- (-2.35,+2);
    \draw[thick, black, -stealth] (-2.2,-0.25) -- (-2.2,-2);
    \draw[thick, black, -stealth] (-2.2,+0.25) -- (-2.2,+2);
    \node[left] at (-1.8,0) {\footnotesize $L_d$};
    
    \node[red, right] at (0,0) {\footnotesize $c_1$-defect};

    \draw[thick, black, -stealth] (-1.5,1.3) -- (-1.5,1.8);
    \node[black, right] at (-1.5,1.5) {\footnotesize $\tau$}; 
    \end{tikzpicture}
    \caption{We choose the Euclidean time along $c_1$-defect direction, and the result computed in this channel is denoted as $\Zd(L_d,L,\mu)$.}
    \label{fig:d_o_tt}
\end{figure}

We first compute the torus partition function in the channel where the $c_1$-defect is placed along the Euclidean time direction as shown in Figure \ref{fig:d_o_tt}. We denote the result as $\Zd(L_d,L,\mu)$ where $\mu$ is the RG scale, as will be discussed shortly. Here, the subscript $d$ denotes that the partition function is a trace over $c_1$-defect Hilbert space.
In this channel, the partition function is computing the trace of $e^{-\beta (H - 1/12)}$ over the defect Hilbert space. We will parameterize the spatial direction $x \in [0,L)$ and place the defect at $x = 0$. From \eqref{b15}, we learn that a $c_1$-defect will impose the following twisted boundary condition at $x = 0$
\begin{equation}\label{eq:c1_bc}
    \begin{pmatrix} \phi(L) \\ \phi'(L) \end{pmatrix} = \begin{pmatrix} 1 & 0 \\ -c_1 & 1 \end{pmatrix}\begin{pmatrix} \phi(0) \\ \phi'(0) \end{pmatrix} ~.
\end{equation}
For $c_1 > 0$, the eigenfunctions satisfying the above boundary conditions fall into two classes. The first class is
\begin{equation}
    \psi_n^{I}(x) = \sqrt{2} \sin\left(\frac{2\pi n}{L} x\right) ~, \quad n \in \mathbb{Z}_{>0} ~.
\end{equation}
Notice that these eigenfunctions satisfy the boundary condition \eqref{eq:c1_bc} as $\psi^{I}_n(0) = \psi^{I}_n(L) = 0$ and $\psi^{I\prime}_n(0) = \psi^{I\prime}_n(L)$, and are independent of the coupling $c_1$. 

The second class is
\begin{equation}
    \psi_\omega^{II}(x) = A(\omega) (\cos(\omega x) + \cos(\omega(L_2 - x))) ~,
\end{equation}
where the normalization factor $A(\omega) = \sqrt{\frac{\omega L}{2(\omega L + \sin(\omega L))}}$ and $\omega > 0$ satisfies
\begin{equation}\label{eq:class_II_ev}
    2 \omega \tan\left(\frac{\omega L}{2}\right) = c_1 ~.
\end{equation}
As one can see, the second class depends continuously on the coupling $c_1$, and the modes can be viewed as the flows of the periodic eigenfunctions $\sqrt{2}\cos(\frac{2\pi nx}{L})$ ($n\in \mathbb{Z}_{\geq 0}$) for $c_1 = 0$. Therefore, we can naturally use $\omega_n$ ($n\geq 0$) to denote solutions at positive $c_1$.

\begin{figure}[h]
    \centering
    \includegraphics[width=0.8\linewidth]{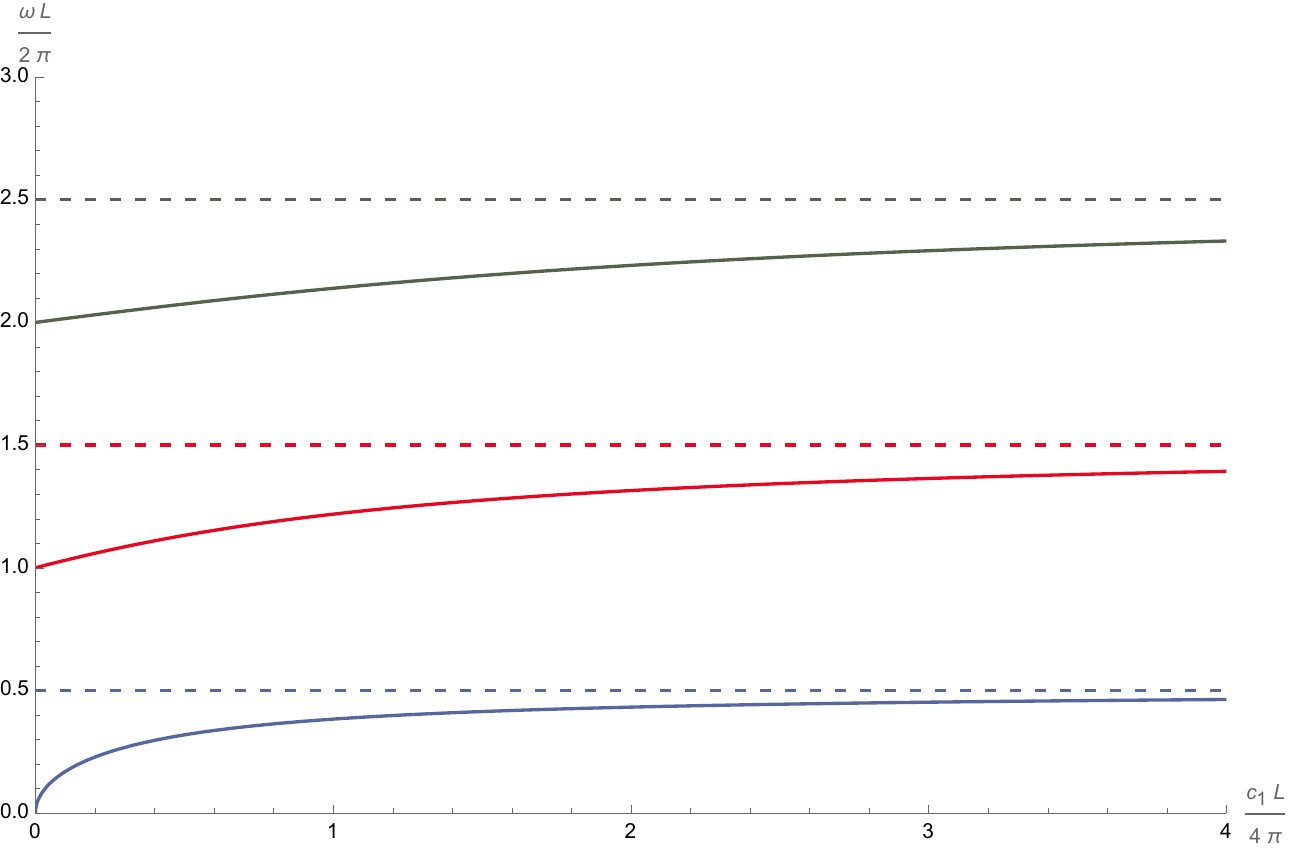}
    \caption{Here, we plot the first three solutions to the equation \eqref{eq:class_II_ev} as a function of the coupling $c_1$. Notice that we have rescaled  $\omega$ and $c_1$ with $\frac{1}{L}$ so that they are dimensionless.}
    \label{fig:flow_of_omega}
\end{figure}

It is notable that the zero mode $\psi = const.$ at $c_1 = 0$ is lifted upon turning on the coupling, which we take to be positive. Furthermore, at small $c_1$, the leading behavior of $\omega_0$ as a function of $c_1$ is distinct from the other $\omega_n$'s, as one can see in Figure \ref{fig:flow_of_omega}.
It is straightforward to check that at small $c_1$ the solutions $\omega_n$ admit an expansion in $c_1$:
\begin{equation}\label{eq:omega_expan}
    \omega_0 = \sqrt{\frac{c_1}{L}} \left(1 - \frac{c_1 L}{24} + O(c_1^2)\right) ~, \quad 
    \omega_n = \frac{2\pi n}{L} + \frac{c_1}{2n \pi} + O(c_1^2) ~, \quad n>0 ~.
\end{equation}
In the limit where $c_1 \rightarrow \infty$, all the solutions $\omega_n$ have the same behavior:
\begin{equation}\label{eq:omlc1}
    \omega_n \rightarrow \frac{2\pi}{L} \left(n + \frac{1}{2}\right) \left(1 - {4\pi \over  c_1L} + O\left(\frac{1}{c_1^2}\right) \right) ~, \quad c_1 \rightarrow \infty ~.
\end{equation}
For finite $c_1$  one can see in Figure \ref{fig:flow_of_omega} that as $n$ gets larger  $\omega_n$ slowly approaches $\left(n + \frac{1}{2}\right) \frac{2\pi}{L}$. We can also write down an expansion for large $n$ at fixed $c_1$:
\begin{equation}\label{eq:omg_expan}
    \omega_n = \frac{2\pi n}{L} + \frac{c_1}{2n\pi} + O\left(\frac{1}{n^2}\right) ~.
\end{equation}

On the other hand, if $c_1 < 0$, the system is pathological because one eigenfunction will have imaginary $\omega$. It's not hard to see this as the perturbative expansion \eqref{eq:omega_expan} holds for negative $c_1$ as well, and $\omega_0$ becomes imaginary when turning on small negative $c_1$. For finite negative $c_1$, the function $2ix\tan\left(\frac{i x L}{2}\right) = - 2x \tanh\left(\frac{\omega L}{2}\right)$ remains negative and monotonically decreasing for $x>0$, thus \eqref{eq:class_II_ev} always has a purely imaginary solution when $c_1 < 0$. Quantization of this mode either leads to non-unitarity or an energy spectrum unbounded from below. Thus, we will only consider the case where $c_1 > 0$.

Once we understand the eigenfunctions, it is straightforward to construct the defect Hilbert space, as the states are formed from the harmonic oscillators with frequencies given by the $\omega_n$'s from the eigenfunctions. Naively assembling the partition functions of each oscillator together, we find 
\begin{equation}\label{eq:dp_unreg}
    \Zd(L_d,L) = \left(\prod_{n=1}^\infty \frac{q^{\frac{n}{2}}}{1 - q^n}\right) \left(\prod_{n=0}^\infty \frac{q^{\frac{\omega_n L}{4\pi}}}{1 - q^{\frac{\omega_n L}{2\pi}}} \right) ~,
\end{equation}
where $q = e^{-2\pi \frac{L_d}{L}}$. Notice that the  factor arising from the vacuum energy of the oscillators:
\begin{equation}
    q^{\sum_{n=1} \frac{n}{2} + \sum_{n=0}\frac{\omega_n L}{4\pi}} ~,
\end{equation}
is divergent and therefore requires regularization. We split it into three parts as
\eq{zpq}{ 
    \frac{1}{2}\sum_{n=1}^\infty n + \frac{1}{2}\sum_{n=0}^\infty \frac{\omega_n L}{2\pi} = \sum_{n=1}^\infty n + \frac{c_1 L}{8\pi^2}\sum_{n=1}^\infty \frac{1}{n} + \frac{L}{4\pi}\left(\omega_0 + \sum_{n=1}^\infty \left(\omega_n - \frac{2\pi n}{L} - \frac{c_1}{2n \pi}\right)\right) ~.
}
The first piece is regularized using familiar $\zeta$-function regularization, and leads to $-\frac{1}{12}$. The second piece diverges logarithmically, and we regulate it by considering an energy cut-off at $\Lambda$; since the momentum is measured in the unit of $\frac{1}{L}$, this will cut the infinite sum at $\Lambda L$. Then we add a counter term along the defect 
\begin{equation}\label{eq: 9a}
    \frac{c_1}{4\pi} \int dt \log\left(\frac{\Lambda}{\mu}\right) ~, 
\end{equation}
where $\mu$ is the RG scale. Taking the limit $\Lambda \rightarrow \infty$, we find
\begin{equation}
    \lim_{\Lambda \rightarrow \infty} \left[-\frac{c_1 L}{8\pi^2}\log\left(\frac{\Lambda}{\mu}\right) + \frac{c_1 L}{8\pi^2} \sum_{n=1}^{\Lambda L} \frac{1}{n}\right] = \frac{c_1 L}{8\pi^2}[\gamma + \log(\mu L)] ~,
\end{equation}
where $\gamma$ is the Euler–Mascheroni constant. The remaining part of \rf{zpq} is finite as one can see from \eqref{eq:omg_expan} and numerically verify. To summarize, the partition function computed  in this channel is 
\begin{equation}\label{eq:dh_part}
\begin{aligned}
    \Zd(L_d,L,\mu) & = e^{-\frac{c_1 L_d}{4\pi}(\gamma + \log(\mu L))} q^{- \frac{1}{12} + \frac{L}{2\pi} E_0(c_1)}  \prod_{n = 0}^\infty \left(1 - q^{\frac{\omega_n L}{2\pi}}\right)^{-1} \prod_{m>0} (1-q^m)^{-1} \\
    & = e^{-\frac{c_1 L_d}{4\pi}(\gamma + \log(\mu L))} \frac{q^{- \frac{1}{24} + \frac{L}{2\pi} E_0(c_1)}}{\eta(q)} \prod_{n = 0}^\infty \frac{1}{1 - q^{\frac{\omega_n L}{2\pi}}} ~,
\end{aligned}
\end{equation}
where the constant $E_0(c_1)$ represents the shifted ground state energy
\begin{equation}
    E_0(c_1) = \frac{\omega_0}{2} + \frac{1}{2} \sum_{n=1}^\infty \left(\omega_n - \frac{2\pi n}{L} - \frac{c_1}{2n \pi}\right) ~.
\end{equation}

As a consistency check, one can also compute the partition function from the path integral by expanding the scale field $\phi(t,x)$ using the eigenfunctions described above. The path integral then reduces to a infinite product of Gaussian integrals, which can then be regulated following the same approach taken for a free scalar field \cite{DiFrancesco:1997nk}. This reproduces the result \eqref{eq:dh_part}, and after a further regularization leads to the same final result \eqref{eq:dh_part}.

To conclude, we want to quickly point it out that the coefficient $-\frac{c_1 L}{4\pi}$ in front of the $-\frac{c_1 L}{4\pi}\log(\mu L)$ term can be viewed as the anomalous dimension of the $c_1$-defect.   This way of computing anomalous dimension can be generalized to the case where both $c_1$ and $c_2$ are activated. In this case, the twisted boundary condition is given by
\begin{equation}\label{eq:c1_bca}
    \begin{pmatrix} \phi(L) \\ \phi'(L) \end{pmatrix} = \begin{pmatrix} \frac{(2-c_2)^2}{4-c_2^2} & 0 \\ -\frac{c_1}{4-c_2^2} &  \frac{(2+c_2)^2}{4-c_2^2} \end{pmatrix}\begin{pmatrix} \phi(0) \\ \phi'(0) \end{pmatrix} ~,
\end{equation}
and the equation for the eigenvalues become
\begin{equation}
    2\omega \tan\left(\frac{\omega L}{2}\right) = c_1 + \frac{c_2^2}{2} \omega \cot\left(\frac{\omega L}{2}\right) ~.
\end{equation}
Notice that the equations can be solved exactly when $c_1 = 0$, and for large $\omega$ they admit a similar expansion as \eqref{eq:omega_expan} (where for simplicity we assume $c_2 > 0$)
\begin{equation}
\begin{aligned}
    \omega_n &= + \frac{2}{L}\left(\arctan\left(\frac{c_2}{2}\right)\right) + \frac{2\pi n}{L} + \frac{c_1}{4+c_2^2} \frac{1}{n\pi + \arctan(c_2/2)} + O(c_1^2) ~, \quad n \geq 0 ~, \\
    \omega'_n &= - \frac{2}{L}\left(\arctan\left(\frac{c_2}{2}\right)\right) + \frac{2\pi n}{L} + \frac{c_1}{4+c_2^2} \frac{1}{n\pi + \arctan(c_2/2)} + O(c_1^2) ~, \quad n > 0 ~.
\end{aligned}
\end{equation}
Then the logarithmic divergence in the zero point energy sum contributes to the partition function as
\begin{equation}
\begin{aligned}
    & \exp\left(-\frac{L_d}{2} \left(\sum_{n=0}^\infty \omega_n + \sum_{n=1}^\infty \omega'_n\right)\right) \\
    \supset & \exp\left(-\frac{L_d}{2} \left(\sum_{n=0}^\Lambda \frac{c_1}{4+c_2^2} \frac{1}{n\pi + \arctan(c_2/2)} + \sum_{n=1}^\Lambda \frac{c_1}{4+c_2^2} \frac{1}{n\pi + \arctan(c_2/2)}\right)\right) \\
    \simeq & \exp\left(- \frac{c_1 L_d}{4\pi} \frac{1}{1+c_2^2/4} \log(\Lambda)\right) ~.
\end{aligned}
\end{equation}
Thus, we see that the anomalous dimension for a $c_1,c_2$-defect is given by $-\frac{c_1 L_d}{4\pi} \frac{1}{1+c_2^2/4}$, matching the calculation in \rf{c36aa}.

\subsection{Torus partition function with the space-like defect}

\begin{figure}
    \centering
    \begin{tikzpicture}[baseline={([yshift=-1ex]current bounding box.center)},vertex/.style={anchor=base,circle,fill=black!25,minimum size=18pt,inner sep=2pt},scale=1]
    \filldraw[grey] (-2,-1.8) rectangle ++(4,3.6);
    \draw[thick, dgrey] (-2,-1.8) rectangle ++(4,3.6);
    \draw[line width = 0.4mm, red] (-2,0) -- (2,0);
    \draw[thick, black] (-2,-2.15) -- (-2,-1.85);
    \draw[thick, black] (2,-2.15) -- (2,-1.85);
    \draw[thick, black, -stealth] (0.25,-2) -- (2,-2);
    \draw[thick, black, -stealth] (-0.25,-2) -- (-2,-2);
    \node[below] at (0,-1.8) {\footnotesize $L_d$};

    \draw[thick, black] (-2.05,-1.8) -- (-2.35,-1.8);
    \draw[thick, black] (-2.05,+1.8) -- (-2.35,+1.8);
    \draw[thick, black, -stealth] (-2.2,-0.25) -- (-2.2,-1.8);
    \draw[thick, black, -stealth] (-2.2,+0.25) -- (-2.2,+1.8);
    \node[left] at (-2,0) {\footnotesize $L$};
    
    \node[red, below] at (0,0) {\footnotesize $c_1$-defect};
    \draw[thick, black, -stealth] (-1.7,1.1) -- (-1.7,1.6);
    \node[black, right] at (-1.7,1.3) {\footnotesize $\tau$}; 
\end{tikzpicture}
    \caption{In the dual channel, we choose the Euclidean time to run transversely to the $c_1$-defect, and the result computed in this channel is denoted as $\Zd(L_d,L,\mu)$.}
    \label{fig:d_o_ts}
\end{figure}
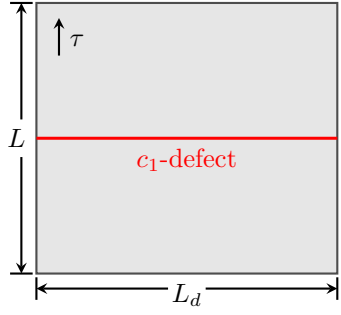

Next, we compute the same torus partition function but now as a trace in the dual channel where Euclidean time is taken run transverse to the defect, as shown in figure \ref{fig:d_o_ts}. We emphasize  that this is precisely the same setup as in Figure \ref{fig:d_o_tt}, and we are simply identifying Euclidean time differently than before. Naturally, we expect to get the same answer as \eqref{eq:dh_part}; nevertheless, for the moment we will use $\Zo$ to denote the corresponding result. The calculation can be computed using either the Hamiltonian approach or the path-integral approach; below  we demonstrate explicitly the Hamiltonian approach and briefly sketch on the path integral computation.

Because the defect is placed along the spatial direction, it should be treated as an operator insertion that does not change the Hilbert space but rather acts on it. Because of this, we take the familiar expansion of the free scalar field, $\phi(x,t) = \sum_n \varphi_n(t)e^{\frac{2\pi i nx}{L_d}}$,  to get the free Hamiltonian (and use the $\zeta$-function to regularize the vacuum energy)
\begin{equation}
    H_{free} = - \frac{\pi}{6L_d} + \frac{1}{2L_d} \hat{\pi}_0^2 + \frac{2\pi}{L_d}\sum_{n > 0} n (\hat{b}_n^\dagger \hat{b}_n + \hat{d}_n^\dagger d_n) ~, 
\end{equation}
as well as the defect operator
\begin{equation}
    e^{-S_d} = \exp\left(- \frac{c_1 L_d}{2}\hat{\varphi}_0^2 - \frac{c_1 L_d}{4\pi} \sum_{n > 0} \frac{1}{n}(\hat{b}_n^\dagger - \hat{d}_n)(\hat{b}_n - \hat{d}_n^\dagger) \right)~.
\end{equation}
Here, the $\hat{b}_n,\hat{b}^\dagger_n$'s and $\hat{d}_n, \hat{d}^\dagger_n$'s represent creation and annihilation operators of harmonic oscillators with normalization $[\hat{b}_n,\hat{b}^\dagger_n] = [\hat{d}_n,\hat{d}^\dagger_n] = 1$, and $\hat{\pi}_0$ and $\hat{\varphi}_0$ represent the momentum and the coordinate of the zero mode. Note that the defect operator effectively introduces a mass term for the zero mode. 

The torus partition function in this channel may be computed by inserting a complete basis of states (formed by the eigenstates of the free Hamiltonian $H_{free}$) in the trace
\begin{equation}
    \Zo(L_d, L, \mu) = \Tr(e^{-S_d} e^{-L H_{free}}) = \sum_{\psi,\phi} \langle \psi| e^{S_d} | \phi \rangle \langle \phi | e^{- L H_{free}} |\psi\rangle ~,
\end{equation}
where again the RG scale $\mu$ will be introduced shortly. It is useful to note that the defect operator $e^{-S_d}$ only mixes the two oscillators labeled by the same $n$, thus the partition function factorizes. The contribution for a given $n$ can be computed by expanding $e^{-\frac{c_1}{4\pi n}(\hat{b}_n^\dagger - \hat{d}_n)(\hat{b}_n - \hat{d}_n^\dagger)}$ and use $[\hat{b}_n^\dagger - \hat{d}_n,\hat{b}_n - \hat{d}_n^\dagger] = 0$. We find
\begin{equation}
    \sum_{\ell,\tilde{\ell}=0}^\infty \langle \ell, \tilde{\ell}| e^{-\frac{c_1 L_d}{4\pi n} (b_n^\dagger - d_n)(b_n - d_n^\dagger)}|\ell,\tilde{\ell}\rangle \tilde{q}^{n\ell+n\tilde{\ell}} = \left[(1-\tilde{q}^n)\left(1 + \frac{c_1 L_d}{4\pi n} - \tilde{q}^n\left(1-\frac{c_1 L_d}{4\pi n}\right)\right)\right]^{-1} ~,
\end{equation}
where $\tilde{q} = e^{-2\pi\frac{L}{L_d}}$. And the contribution from the zero mode is
\begin{equation}\label{eq:zero1}
    \sqrt{\frac{1}{2\pi}} \int d\pi_0 \langle \pi_0| e^{-\frac{c_1 L_d}{2}\hat{\varphi}_0^2} |\pi_0\rangle e^{-\frac{\pi_0^2 L}{2L_d}} = \frac{1}{2\pi} \int d\pi_0 d\varphi_0 \langle \pi_0| e^{-\frac{c_1 L_d}{2}\varphi_0^2} |\varphi_0 \rangle \langle \varphi_0 |\pi_0\rangle e^{-\frac{\pi_0^2 L}{2L_d}} = \frac{1}{\sqrt{c_1 L}} ~.
\end{equation}
Combining everything together, we find the partition function is given by
\begin{equation}\label{eq:tro_bare}
    \frac{1}{\sqrt{c_1 L}} \tilde{q}^{-\frac{1}{12}} \prod_{n=1}^\infty  \frac{1}{(1-\tilde{q}^n)(1 + \frac{c_1 L_d}{4\pi n} - \tilde{q}^n(1-\frac{c_1 L_d}{4\pi n}))} ~,
\end{equation}
where the factor $\tilde{q}^{-\frac{1}{12}}$ comes from the vacuum energy term in $H_{free}$. The above expression contains a similar divergent term, which can be seen by writing
\begin{equation}
\begin{aligned}
    \prod_{n=1}^\infty  \frac{1}{1 + \frac{c_1 L_d}{4\pi n} - \tilde{q}^n(1-\frac{c_1 L_d}{4\pi n})} = & \exp\left[-\sum_{n>0} \log\left(1 + \frac{c_1 L_d}{4\pi n} - \tilde{q}^n\left(1-\frac{c_1 L_d}{4\pi n}\right)\right)\right] \\
    = & \exp\left[-\sum_{n>0} \frac{c_1 L_d}{4\pi n} \left(1 + O\left(\frac{1}{n}\right)\right) \right] ~,
\end{aligned}
\end{equation}
and we will regularize it in the same way as in the previous calculation. Namely, we first introduce the same UV cut-off at energy $\Lambda$. But because the energy is measured in unit of $\frac{1}{L_d}$, this would instead cut the infinite sum at $\Lambda L_d$. Next, we introduce the same counter term
\begin{equation}
    \frac{c_1}{4\pi} \int dt \, \log\left(\frac{\Lambda}{\mu}\right) ~,
\end{equation}
and taking the limit $\Lambda \rightarrow \infty$ then leads to
\begin{equation}
    \lim_{\Lambda \rightarrow \infty} \left[\frac{c_1 L_d}{4\pi}\log\left(\frac{\Lambda}{\mu}\right) - \frac{c_1 L_d}{4\pi} \sum_{n=1}^{\Lambda L_d} \frac{1}{n}\right] = - \frac{c_1 L_d}{4\pi}[\gamma + \log(\mu L_d)] ~.
\end{equation}
The partition function is then given by
\begin{equation}\label{eq:Zo}
     \Zo(L_d,L,\mu) = e^{-\frac{c_1 L_d}{4\pi}\left(\gamma + \log(\mu L_d)\right)}\frac{1}{\sqrt{c_1 L}} \frac{\tilde{q}^{-\frac{1}{24}}}{\eta(\tilde{q})} \prod_{n=1}^\infty \frac{e^{\frac{c_1 L_d}{4\pi n}}}{1 + \frac{c_1 L_d}{4\pi n} - \tilde{q}^n(1-\frac{c_1 L_d}{4\pi n})} ~, \quad \tilde{q} = e^{-2\pi \frac{L}{L_d}} ~.
\end{equation}
As a consistency check, let's consider the vacuum expectation value from \eqref{eq:Zo}, which can be obtained from the coefficient of the leading term in the $\tilde{q}$-expansion:
\begin{equation}\label{eq:vft}
    \langle e^{-S_d} \rangle = e^{-\frac{c_1 L_d}{4\pi}\left(\gamma + \log(\mu L_d)\right)}\frac{1}{\sqrt{c_1 L}} \prod_{n=1}^{\infty} \frac{e^{\frac{c_1 L_d}{4\pi n}}}{1 + \frac{c_1 L_d}{4\pi n}} = \frac{e^{-\frac{c_1 L_d}{4\pi}\log(\mu L_d)}}{\sqrt{c_1 L}} \Gamma\left(1 + \frac{c_1 L_d}{4\pi}\right) ~,
\end{equation}
The path integral calculation can be done by expanding the scalar field $\phi(x,t)$ using periodic functions and treat the defect operator $e^{-S_d}$ as an operator insertion in the Gaussian integral. It is straightforward to check that it reproduces the same unregularized result \eqref{eq:tro_bare}.

\subsection{Matching the two calculations}
As pointed out, since we are computing the same configuration with the same counter terms, we expect the result $\Zd(L_d,L,\mu)$ and $\Zo(L_d,L,\mu)$ to match exactly:
\begin{equation}
    \Zd(L_d,L,\mu) = \Zo(L_d,L,\mu) ~.
\end{equation}
This equality amounts to the claim
\begin{equation}
\label{ppa} 
\begin{aligned}
    & e^{-\frac{c_1 L_d}{4\pi}(\gamma + \log(\mu L))} \frac{q^{- \frac{1}{24} + \frac{L}{2\pi} E_0(c_1)}}{\eta(q)} \prod_{n = 0}^\infty \frac{1}{1 - q^{\frac{\omega_n L}{2\pi}}} \\
    =& e^{-\frac{c_1 L_d}{4\pi}\left(\gamma + \log(\mu L_d)\right)}\frac{1}{\sqrt{c_1 L}} \frac{\tilde{q}^{-\frac{1}{24}}}{\eta(\tilde{q})} \prod_{n=1}^\infty \frac{e^{\frac{c_1 L_d}{4\pi n}}}{1 + \frac{c_1 L_d}{4\pi n} - \tilde{q}^n(1-\frac{c_1 L_d}{4\pi n})} ~.
\end{aligned}
\end{equation}
The RG scale $\mu$ automatically cancels out from both sides, and leads to
\begin{equation}\label{eq:modular_cor}
    \frac{q^{- \frac{1}{24} + \frac{L}{2\pi} E_0(c_1)}}{\eta(q)} \prod_{n = 0}^\infty \frac{1}{1 - q^{\frac{\omega_n L}{2\pi}}} = \left(\frac{L}{L_d}\right)^{\frac{c_1 L_d}{4\pi}} \frac{\tilde{q}^{-\frac{1}{24}}}{\eta(\tilde{q})} \prod_{n=1}^\infty \frac{e^{\frac{c_1 L_d}{4\pi n}}}{1 + \frac{c_1 L_d}{4\pi n} - \tilde{q}^n(1-\frac{c_1 L_d}{4\pi n})} ~.
\end{equation}

\

Let's first check \eqref{eq:modular_cor} at the leading order of $c_1$. Keeping the leading term on both sides, we find
\begin{equation}
\begin{aligned}
    LHS &= \sqrt{\frac{L}{c_1}}\frac{1}{L_d}  \frac{1}{|\eta(q)|^2}\left(1 + O(c_1)\right) ~, \quad q = e^{-2\pi \frac{L_d}{L}} ~, \\
    RHS &= \frac{1}{\sqrt{c_1 L}} \frac{1}{\eta(\tilde{q})^2} \left(1 + O(c_1)\right) ~, \quad \tilde{q} = e^{-2\pi \frac{L}{L_d}} ~.
\end{aligned}
\end{equation}
The modular property of $\eta(q)$ (that is, $\eta(\tilde{q}) = \sqrt{L_d/L} \eta(q)$) implies the two sides agree at the leading order. This agreement has a simple interpretation, that is, at the leading order, the $c_1$-defect provides a mass-like regulator of the zero mode. This regulator is only modular covariant as it soaks up the modular transformation of the partition function of the oscillator modes to ensure the full partition function is modular invariant at this order.

To see the effect of the anomalous dimension $\left(\frac{L}{L_d}\right)^{\frac{c_1 L_d}{4\pi}}$, we have to consider the next order correction. Using
\begin{equation}
    \frac{e^{\frac{c_1 L_d}{4\pi n}}}{1 + \frac{c_1 L_d}{4\pi n} - \tilde{q}^n(1-\frac{c_1 L_d}{4\pi n})} = \frac{1}{1-\tilde{q}^n}\left(1 - \frac{c_1 L_d}{2\pi n} \frac{\tilde{q}^n}{1-\tilde{q}^n} + O(c_1^2) \right) ~,
\end{equation}
we find
\begin{equation}\label{eq:Z_c1_1o}
\begin{aligned}
     RHS = & \frac{1}{\sqrt{c_1 L}} \frac{1}{|\eta(\tilde{q})|^2} \left[1 - \frac{c_1 L_d}{2\pi}\left(\frac{1}{2}\log\left(\frac{L_2}{L} \right) + \sum_{n=1}^\infty \frac{1}{n} \frac{\tilde{q}^n}{1-\tilde{q}^n}\right) + O(c_1^2)\right] \\
     = & \frac{1}{\sqrt{c_1 L}} \frac{1}{|\eta(\tilde{q})|^2} \left[1 - \frac{c_1 L_d}{2\pi}\left(- \frac{\pi}{12}\frac{L}{L_d} + \frac{1}{2}\log\left(\frac{L_d}{L} \right) - \log(\eta(\tilde{q})) \right) + O(c_1^2)\right] ~,
\end{aligned}
\end{equation}
where we have used
\begin{equation}
    \sum_{n=1}^\infty \frac{1}{n} \frac{\tilde{q}^n}{1-\tilde{q}^n} = \sum_{n,m=1}^\infty \frac{1}{n} \tilde{q}^{nm} = - \sum_{m=1}^\infty \log(1 - \tilde{q}^m) = - \frac{\pi}{12} \frac{L}{L_d} - \log(\eta(\tilde{q})) ~.
\end{equation}

To derive the first order correction in $c_1$ on the LHS of \eqref{eq:modular_cor}, first notice that from \eqref{eq:omega_expan} we have
\begin{equation}
    \frac{q^{\frac{\omega_0 L}{4\pi}}}{1 - q^{\frac{\omega_0 L}{2\pi}}} = \frac{1}{L_d} \sqrt{\frac{L}{c_1}} \left(1 + \frac{c_1 L_d}{2\pi} \frac{\pi}{12}\left(\frac{L}{L_d} - \frac{L_d}{L}\right) + O(c_1^2)\right) ~,
\end{equation}
and
\begin{equation}
    \frac{1}{1 - q^{\frac{\omega_n L}{2\pi}}} = \frac{1}{1 - \tilde{q}^n}\left(1 - \frac{c_1 L_d}{2n\pi} \frac{q^n}{1 - q^n} + O(c_1^2)\right) ~, \quad n > 0 ~,
\end{equation}
where $q = e^{-2\pi \frac{L_d}{L}}$. Combining these contributions, we find
\begin{equation}
\begin{aligned}
   LHS &= \frac{1}{L_d} \sqrt{\frac{L}{c_1}} \frac{1}{|\eta(q)|^2}\left[1 - \frac{c_1 L_d}{2\pi} \left( -\frac{\pi}{12} \frac{L}{L_d} - \log(\eta(q))\right) + O(c_1^2)\right] \\
    &= \frac{1}{\sqrt{c_1 L}} \frac{1}{|\eta(\tilde{q})|^2}\left[1 - \frac{c_1 L_d}{2\pi} \left( - \frac{\pi}{12}\frac{L}{L_d} + \frac{1}{2} \log\left(\frac{L_d}{L}\right) - \log(\eta(\tilde{q}))\right) + O(c_1^2)\right] ~,
\end{aligned}
\end{equation}
where we find agreement with \eqref{eq:Z_c1_1o} after using the modular property of the Dedekind $\eta$-function $ \eta(q) = \sqrt{L/L_d} \eta(\tilde{q})$.

\begin{figure}
    \centering
    \begin{subfigure}{0.32\textwidth}
		\centering
    \includegraphics[width=\linewidth]{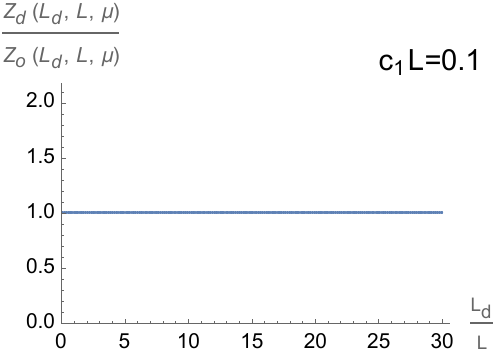}
    \caption{$c_1 L = 0.1$}
    \label{fig:num1}
    \end{subfigure}
    \begin{subfigure}{0.32\textwidth}
		\centering
    \includegraphics[width=\linewidth]{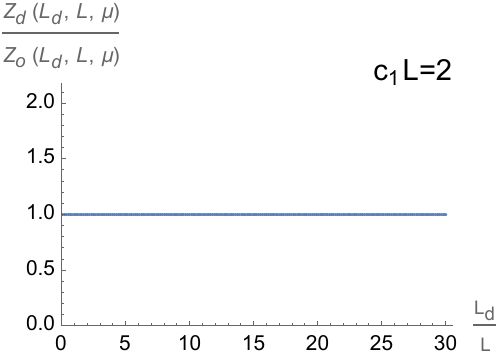}
    \caption{$c_1 L = 2$}
    \label{fig:num2}
    \end{subfigure}
    \begin{subfigure}{0.32\textwidth}
		\centering
    \includegraphics[width=\linewidth]{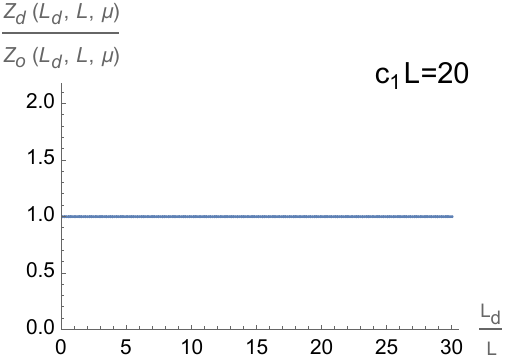}
    \caption{$c_1 L = 20$}
    \label{fig:num3}
    \end{subfigure}
    \caption{Numerical verification of \eqref{eq:modular_cor}, where we fix $c_1 L = 0.1,2,20$ and plot $\frac{Z_d(L_d,L,\mu)}{Z_o(L_d,L,\mu)}$ as a function $\frac{L_d}{L}$ (recall that the ratio does not depends on the RG scale $\mu$).}
    \label{fig:num}
\end{figure}

To conclude, we consider numerical verification of \eqref{eq:modular_cor}. To do so, we  fix the value of $c_1 L$ and plot the ratio (recall that the ratio does not depend on $\mu$)
\begin{equation}
    \frac{\Zo(L_d,L,\mu)}{\Zd(L_d,L,\mu)}
\end{equation}
as a function of $\frac{L_d}{L}$. We find  a nice numerical match of \eqref{eq:modular_cor} as shown in Figure \ref{fig:num}.

\subsection{Cardy formulas}
The modular invariance of a CFT with compact spectrum relates the high and  low temperature limits, allowing one to estimate the asymptotic growth of the density of states, yielding a  Cardy formula \cite{Cardy:1986ie}.  Though we consider a non-conformal defect, we can apply the same reasoning by computing the partition function in the two different channels.  

\subsubsection{Review of the standard Cardy formula}

Let us first recall the derivation in the case of a generic compact CFT with finite central charge $c$. Its partition function on a square torus is defined as
\begin{equation}
    Z(L_d,L) = \int_0^\infty dE \, \rho(E) e^{-L_d (E - \frac{\pi c}{6L})} 
\end{equation}
where $\rho(E)$ is the density of states, and we use $L_d$ to denote the length of the Euclidean time cycle, and $L$ denotes the length of the spatial circle, even though no defect is inserted. Modular invariance implies that $Z(L_d,L) = Z(L,L_d)$, and in the limit $\frac{L_d}{L} \rightarrow 0$, $Z(L,L_d) \simeq e^{\frac{\pi L}{6L_d}}$ and we find 
\begin{equation}
    \int_0^\infty dE \, \rho(E) e^{-L_d E} \simeq e^{- \frac{\pi c L_d}{6L}} e^{\frac{\pi c L}{6L_d}} ~.
\end{equation}
The inverse Laplace transformation $\rho_0(E)$ of the RHS then describes the asymptotic growth of $\rho(E)$ in the sense that 
\begin{equation}
    \frac{\int_0^{E_{max}} dE \, \rho(E)}{\int_0^{E_{max}} dE \, \rho_0(E)} \rightarrow 1 ~, \quad E_{max} \rightarrow \infty ~. 
\end{equation}
The inverse Laplace transformation 
\begin{equation}
    \rho_0(E) = {1 \over 2\pi i} \int_{-i\infty}^{i\infty}   \! dL_d \, e^{E L_d + {\pi c L  \over 6L_d} - \frac{\pi c L_d}{6L}} 
\end{equation}
can be evaluated using the saddle point approximation. Upon changing variable $L_d = \frac{i z}{\sqrt{E}}$, the integral becomes
\begin{equation}
    \rho_0(E) = {1 \over 2\pi \sqrt{E}} \int_{-\infty}^{\infty}   \! dz \, e^{\sqrt{E}(iz - \frac{i\pi c L}{6z}) + \frac{i\pi c}{6\sqrt{E}L} z} ~, 
\end{equation}
and the large $E$ saddle point can be seen from
\begin{equation}
    i z - {i\pi c L \over 6z} = \sqrt{\frac{2\pi c L}{3}} - \sqrt{\frac{6}{\pi c L}}(z-z_0)^2 + \ldots 
\end{equation}
where $z_0 = - i \sqrt{\pi c L \over 6}$. Evaluating the integral around $z_0$ leads to
\begin{equation}
    \rho_0(E) = \left( { cL/2\pi \over 48 E^3}\right)^{1/4}  e^{2\pi \sqrt{c EL/2\pi \over 3} } \left(1 + O\left(\frac{1}{\sqrt{cEL}}\right)\right) ~.
\end{equation}
Notice that we have  neglected the term $e^{\frac{i\pi c}{6\sqrt{E}L} z}$ in the integral for the following reasons. First, while it modifies the location of the saddle point, the leading error is of order $\frac{c^{5/2}}{(EL)^{3/2}}$ which is negligible compared to the leading error of the saddle point approximation. Second, the leading order contribution of $e^{\frac{i\pi c}{6\sqrt{E}L} z}$ at the saddle point is of order $\frac{c^{\frac{3}{2}}}{\sqrt{EL}}$ which is again negligible as it belongs to $O\left(\frac{1}{\sqrt{cEL}}\right)$ (recall that we have assume that $c$ is finite therefore of order $O(1)$).

\subsubsection{Defect Cardy formula}

We can similarly derive a Cardy formula for the spectrum of the Hamiltonian acting on the  Hilbert space of the $c_1$ defect. In this case, the relation \eqref{eq:modular_cor} relates the density of states $\rho_{c_1}(E)$ of the defect Hilbert space to the high temperature limit of the partition function in the following way. First, expanding $\eta(q)$ on the LHS of \eqref{eq:modular_cor} leads to
\begin{equation}
    q^{-\frac{1}{12} + \frac{L}{2\pi} E_0(c_1)} \left[\prod_{n=1}^\infty (1-q^n)^{-1} \prod_{n=0}^\infty(1-q^{\frac{\omega_n L}{2\pi}})^{-1}\right] = \left(\frac{L}{L_d}\right)^{\frac{c_1 L_d}{4\pi}} \frac{\tilde{q}^{-\frac{1}{24}}}{\eta(\tilde{q})} \prod_{n=1}^\infty \frac{e^{\frac{c_1 L_d}{4\pi n}}}{1 + \frac{c_1 L_d}{4\pi n} - \tilde{q}^n(1-\frac{c_1 L_d}{4\pi n})} ~,
\end{equation}
and the two infinite products on the LHS can be expressed using the density of states $\rho_{c_1}(E)$ 
\begin{equation}
    \left[\prod_{n=1}^\infty (1-q^n)^{-1} \prod_{n=0}^\infty(1-q^{\frac{\omega_n L}{2\pi}})^{-1}\right] = \int_0^\infty \rho_{c_1}(E) e^{-EL_d} dE ~,
\end{equation}
where we have chosen the ground state energy to be $E = 0$. This implies that
\begin{equation}
    \int_0^\infty \rho_{c_1}(E) e^{-EL_d} dE = q^{\frac{1}{12} - \frac{L}{2\pi} E_0(c_1)} \left(\frac{L}{L_d}\right)^{\frac{c_1 L_d}{4\pi}} \frac{\tilde{q}^{-\frac{1}{24}}}{\eta(\tilde{q})} \prod_{n=1}^\infty \frac{e^{\frac{c_1 L_d}{4\pi n}}}{1 + \frac{c_1 L_d}{4\pi n} - \tilde{q}^n(1-\frac{c_1 L_d}{4\pi n})} ~, 
\end{equation}
and in the high temperature limit $\frac{L_d}{L} \rightarrow 0$, the RHS is dominated by the contribution from the vacuum state, and we find
\begin{equation}\label{eq:c1d_htl}
\begin{aligned}
    \int_0^{\infty} \rho_{c_1}(E) e^{-L_d E} \approx \frac{1}{\sqrt{c_1 L}} e^{L_d \left(E_0(c_1) - \frac{\pi}{6L}\right)} e^{-\frac{c_1 L_d}{4\pi} \log\left(\frac{L_d}{L}\right)} e^{\frac{\pi L}{6L_d}} e^{\frac{c_1 L_d \gamma}{4\pi}} \Gamma\left(1 + \frac{c_1 L_d}{4\pi}\right) ~, \quad \frac{L_d}{L} \rightarrow 0 ~.
\end{aligned}
\end{equation}

Inverse Laplace transformation of the RHS of \eqref{eq:c1d_htl} then leads to an approximate density of state $\rho_{c_1,0}(E)$ capturing the asymptotic behavior of $\rho_{c_1}$. However, since now we have two dimensionless parameters $EL$ and $c_1 L$, the result of the saddle point approximation will then depends on the relation between $E$ and $c_1$. 

\vspace{.2cm} 
\noindent
{\bf $c_1 L = O(1)$}
\vspace{.2cm} 

Let us first consider the simplest case where $c_1 L = O(1)$, which is to say that $c_1 L $ is held fixed at a nonzero value of order $1$ while we take $EL\gg 1$. To evaluate the saddle point approximation only requires $EL \gg 1$, and it is straightforward to check that the saddle point remains the same as the usual Cardy case, and we find
\begin{equation}\label{eq:C1}
    \rho_{c_1,0}(E) = \frac{1}{\sqrt{c_1 L}} \left( { L/2\pi \over 48 E^3}\right)^{1/4} e^{\sqrt{\frac{\pi}{6EL}} \frac{c_1 L}{8\pi} \log\left(\frac{6EL}{\pi}\right)} e^{2\pi \sqrt{\frac{EL/2\pi}{3}}}\left(1 + O\left(\frac{1}{\sqrt{EL}}\right)\right) ~, \quad EL \gg 1~.
\end{equation}
Notice that the factor $e^{\sqrt{\frac{\pi}{6EL}} \frac{c_1 L}{8\pi} \log\left(\frac{6EL}{\pi}\right)}$ is kept as the next leading term in its expansion is of order $\frac{\log E}{\sqrt{E}}$.

\

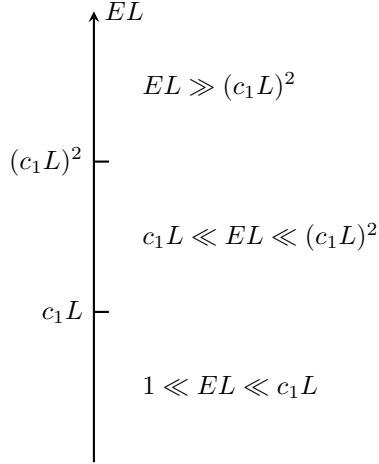
\begin{figure}
    \centering
    \begin{tikzpicture}[baseline={([yshift=-1ex]current bounding box.center)},vertex/.style={anchor=base,circle,fill=black!25,minimum size=18pt,inner sep=2pt},scale=1]
    \draw[black, thick, -stealth] (0,0) -- (0,6);
    \node[black, right] at (0,6) {\footnotesize $EL$};
    \draw[black, thick] (0,4) -- (0.2,4);
    \draw[black, thick] (0,2) -- (0.2,2);
    \node[black, left] at (0,4) {\footnotesize $(c_1L)^2$};
    \node[black, left] at (0,2) {\footnotesize $c_1L$};
    \node[black, right] at (0.5,5) {\footnotesize $EL\gg (c_1 L)^2$};
    \node[black, right] at (0.5,3) {\footnotesize $c_1 L \ll EL \ll (c_1 L)^2$};
    \node[black, right] at (0.5,1) {\footnotesize $1 \ll EL \ll c_1 L$};
    \end{tikzpicture}
    \caption{There are three regimes when $c_1 L \gg 1$. When $EL \gg (c_1 L)^2$, the asymptotic density of states take a similar form \eqref{eq:C11} as in the case $c_1 L = O(1)$. When $1 \ll EL \ll c_1 L$, the asymptotic density of states can be approximated by the asymptotic density of states of a Dirichlet defect as in \eqref{eq:C2}. Finally, the regime where $c_1 L \ll EL \ll (c_1L)^2$ is hard to compute analytically.}
    \label{fig:CardyRegime}
\end{figure}

On the other hand, when $c_1 L \gg 1$, we must carefully discuss the relation between $c_1 L$ and $EL$. As we will demonstrate, there are three regimes as shown in Figure \ref{fig:CardyRegime}.

\vspace{.2cm} 
\noindent
{\bf $EL \gg (c_1 L)^2 \gg 1$}
\vspace{.2cm} 

Let us first consider the first regime $EL \gg (c_1 L)^2 \gg 1$. In the case, we can still take the saddle point to be the same as the previous conventional Cardy case, and the saddle point approximation leads to a similar result as \eqref{eq:C1}
\begin{equation}\label{eq:C11}
    \rho_{c_1,0}(E) = \frac{1}{\sqrt{c_1 L}} \left( { L/2\pi \over 48 E^3}\right)^{1/4} e^{\sqrt{\frac{\pi}{6EL}} \frac{c_1 L}{8\pi} \log\left(\frac{6EL}{\pi}\right)} e^{2\pi \sqrt{\frac{EL/2\pi}{3}}}\left(1 + O\left(\frac{c_1 L}{\sqrt{EL}}\right)\right) ~, \quad EL \gg (c_1 L)^2 \gg 1 ~,
\end{equation}
except now the leading error is of the order $O\left(\frac{c_1 L}{\sqrt{EL}}\right)$.

\vspace{.2cm} 
\noindent
{\bf $c_1 L \gg EL \gg 1$ and comparison to Dirichlet defect}
\vspace{.2cm}

Another interesting regime that can be computed is where $c_1 L \gg EL \gg 1$, which can be analyzed by expanding the RHS of \eqref{eq:c1d_htl} for large $c_1 L$, and inverse Laplace transforming the leading term. The leading order contribution of expanding \eqref{eq:c1d_htl} can be computed by taking $c_1 L \rightarrow +\infty$; alternatively, it can be computed using the fact that $c_1 = +\infty$ corresponds to Dirichlet defect.

To be more concrete, let's consider the massless scalar with boundary conditions $\phi(x=0) = \phi(x=L) = 0$. The allowed mode solutions are then given by
\begin{equation}
    \phi_n(x) = \sin\left(\frac{\pi n}{L} x\right) ~, \quad n = 1,2, \cdots ~,
\end{equation}
with allowed frequency $\omega_n = \frac{\pi n}{2}$. Notice this can also be derived from taking $c_1 \rightarrow \infty$ limit for the two classes of modes discussed in Section \ref{sec:dh}. The torus partition function computing the trace over defect Hilbert space can then be straightforwardly computed as
\begin{equation}
    Z_{Dir}(L_d,L) = \prod_{n=1}^\infty \frac{q^{\frac{1}{4}n}}{1 - q^{\frac{1}{2}}} = q^{-\frac{1}{48}} \prod_{n=1}^\infty (1 - q^{\frac{n}{2}}) = \eta(q^{\frac{1}{2}}) ~, \quad q = e^{-2\pi \frac{L_d}{L}} ~, 
\end{equation}
where we have regularized the zero point energy using $\displaystyle \sum_{n=1}^\infty n = - \frac{1}{12}$. The high temperature limit of the dual channel can then be established using the modular property of the Dedekind eta function, and we find
\begin{equation}\label{eq:htDir}
    \int_0^\infty \rho_{c_1 = \infty}(E) e^{-EL_d} dE \approx \sqrt{\frac{L_d}{2L}} e^{\frac{\pi L}{6L_d}} e^{-\frac{\pi L_d}{24L}} ~, \quad \frac{L_d}{L} \rightarrow 0 ~.
\end{equation}
From this, we conclude that the asymptotic density of states in the regime $c_1 L \gg EL \gg 1$ can be approximated by the inverse Laplace transformation of the RHS of \eqref{eq:htDir} 
\begin{equation}\label{eq:C2}
    \rho_{c_1,0}(E) = \frac{1}{4\sqrt{3}E} e^{2\pi \sqrt{\frac{EL/2\pi}{3}}} \left(1 + O\left(\frac{EL}{c_1 L}\right) + O\left(\frac{1}{\sqrt{EL}}\right)\right)~, \quad c_1 L \gg EL \gg 1 ~.
\end{equation}

To conclude, we want to point out that this analysis implies that the leading order term in the large $c_1$ expansion of the RHS of \eqref{eq:c1d_htl} must agree with the \eqref{eq:htDir}. In the following, we will show that this is indeed the case up to $O(1)$ number in the $c_1 L \rightarrow \infty$ limit as a consistency check. Using the Stirling approximation, it is straightforward to write the RHS of \eqref{eq:c1d_htl} as
\begin{equation}
\begin{aligned}
    & \frac{1}{\sqrt{c_1 L}} e^{L_d \left(E_0(c_1) - \frac{\pi}{6L}\right)} e^{-\frac{c_1 L_d}{4\pi} \log\left(\frac{L_d}{L}\right)} e^{\frac{\pi L}{6L_d}} e^{\frac{c_1 L_d \gamma}{4\pi}} \Gamma\left(1 + \frac{c_1 L_d}{4\pi}\right) \\
    =& \sqrt{\frac{L_d}{2L}} e^{\frac{\pi L}{6L_d}} \exp\left[L_d\left(E_{0}(c_1) - \frac{\pi}{6L} + \frac{c_1}{4\pi}\log\left(\frac{c_1 L}{4\pi}\right)  - \frac{c_1}{4\pi}(1-\gamma) \right)\right]\left(1 + O\left(\frac{1}{c_1 L_d}\right)\right) ~.
\end{aligned}
\end{equation}
To match with the calculation from the Dirichlet defect \eqref{eq:htDir}, we must have 
\begin{equation}\label{eq:E0lc}
    E_0(c_1) = - \frac{c_1}{4\pi}\log\left(\frac{c_1 L}{4\pi}\right) + \frac{c_1}{4\pi}(1-\gamma) + \frac{\pi}{8L} + O\left(\frac{1}{c_1L^2}\right) ~, \quad c_1L \rightarrow \infty ~.
\end{equation}
To study this, let's recall that 
\begin{equation}
    E_0(c_1) = \frac{\omega_0}{2} + \frac{1}{2} \sum_{n=1}^\infty \left(\omega_n - \frac{2\pi n}{L} - \frac{c_1}{2\pi n}\right) ~.
\end{equation}
To estimate its large $c_1$ behavior, we notice that the expansion \eqref{eq:omega_expan} is valid when $n \gg \sqrt{c_1 L}$ while the expansion \eqref{eq:omlc1} is valid when $n \ll c_1 L$. In the region $\sqrt{c_1 L} \ll n \ll c_1 L$, both expansions are simultaneously valid. 

Thus, we consider split the infinite sum at $N = \epsilon c_1 L$ into two pieces
\begin{equation}\label{eq:chop}
     E_0(c_1) = \frac{\omega_0}{2} + \frac{1}{2} \sum_{n=1}^N \left(\omega_n - \frac{2\pi n}{L} - \frac{c_1}{2\pi n}\right)  + \frac{1}{2} \sum_{n=N+1}^\infty \left(\omega_n - \frac{2\pi n}{L} - \frac{c_1}{2\pi n}\right)
\end{equation}
where $\epsilon$ is chosen such that $\sqrt{c_1 L} \ll \epsilon c_1 L \ll c_1 L$. Notice that even though this implies that $\frac{1}{\sqrt{c_1 L}} \ll \epsilon \ll 1$, we still consider $\epsilon$ as being independent of $c_1 L$.

The first two terms in \eqref{eq:chop} we estimate using the expansion \eqref{eq:omlc1}, which leads to
\begin{equation}
\begin{aligned}
    &\frac{\omega_0}{2} + \frac{1}{2}\sum_{n=1}^{\epsilon c_1 L}\left(\omega_n - \frac{2\pi n}{L} - \frac{c_1}{2\pi n}\right) \\
    =& \frac{\pi}{2L} + \frac{1}{2} \sum_{n=1}^{\epsilon c_1 L} \left(-\frac{8\pi^2}{c_1 L^2} n + \frac{\pi}{L}\left(1 - \frac{4\pi}{c_1 L}\right) - \frac{c_1}{2n\pi}\right) + O\left(\frac{1}{c_1 L^2}\right) \\
    =& -\frac{c_1}{4\pi}\log(c_1 L) + c_1 \left( -\frac{1}{4\pi}\log\epsilon - \frac{\gamma}{4\pi} + O(\epsilon) \right) + \left(\frac{\pi}{2L} + O(\epsilon)\right) + O\left(\frac{1}{c_1 L^2}\right) ~.
\end{aligned}
\end{equation}
To estimate the last term, we use the expansion \eqref{eq:omega_expan}, and to compute the $O(c_1)$ and $O(1)$ contribution, one must use a more detailed version of \eqref{eq:omega_expan}
\begin{equation}
\begin{aligned}
    \omega_n - \frac{2\pi n}{L} - \frac{c_1 }{2n\pi} =
        & - \frac{L}{8\pi^3 n^3} c_1^2 \\
        & + \left(-\frac{L^2}{96\pi^3 n^3} + \frac{L^2}{16n^5\pi^5}\right) c_1^3 \\
        & + \left(+\frac{L^3}{96n^5\pi^5} - \frac{5L^3}{128n^7\pi^7}\right)c_1^4 \\
        & + \left(+\frac{L^4}{2560n^5\pi^5} + O\left(\frac{1}{n^7}\right)\right)c_1^5 \\
        & + \left(-\frac{23}{30720n^7\pi^7} + O\left(\frac{1}{n^9}\right)\right)c_1^6 \\
        & + \cdots \\
        & + \left(\alpha_{2m} \frac{L^{2m-1}}{n^{2m+1}} + O\left(\frac{1}{n^{2m+3}}\right)\right) (c_1)^{2m} \\
        & + \left(\frac{2(-1)^m}{(2m+1)(4\pi)^m}\frac{L^{2m}}{n^{2m+1}} + O\left(\frac{1}{n^{2m+3}}\right)\right)(c_1)^{2m+1} \\
        & + \cdots ~. \\
\end{aligned}
\end{equation}
As we will see shortly, the order $O(c_1)$ contribution arises from the term of the form $\displaystyle \sum_{n=\epsilon c_1 L_1 + 1}^\infty \frac{(c_1)^{2m+1}}{n^{2m+1}} $ while the order $O(1)$ contribution arises from the term of the form $\displaystyle \sum_{n=\epsilon c_1 L_1 + 1 }^\infty \frac{(c_1)^{2m}}{n^{2m+1}}$. But because we do not have a generic expression for the coefficients $\alpha_{2m}$ of the latter term, we only managed to compute explicitly the order $O(c_1)$ contribution.

To see this, we notice that
\begin{equation}
\begin{aligned}
    \sum_{n=\epsilon c_1 L_1 + 1}^\infty \frac{1}{n^{p}} = \frac{(-1)^p}{(p-1)!}\psi^{(p-1)}(\epsilon c_1 L_1 + 1)  = \frac{1}{p-1} \frac{1}{(\epsilon c_1 L)^{p-1}}\left(1 + O\left(\frac{1}{\epsilon c_1 L}\right)\right) ~,
\end{aligned}
\end{equation}
thus the order $O(c_1)$ contribution of the second sum is given by
\begin{equation}
\begin{aligned}
    &\frac{1}{2}\sum_{n=\epsilon c_1 L+1}^\infty  \omega_n - \frac{2\pi n}{L} - \frac{c_1 }{2n\pi} \\
    =& \frac{c_1}{4\pi} \sum_{m=1}^\infty \frac{(-1)^m}{2m(2m+1)} \left(\frac{1}{4\pi\epsilon}\right)^{2m} + O(1) \\
    =& \frac{c_1}{4\pi} \left(1 + \log\left(4\pi\epsilon\right) + O(\epsilon)\right) + O(1) ~,
\end{aligned}
\end{equation}
where the last result is acquired by explicitly summing over $m$ (which we omit for simplicity) and then expand near small $\epsilon$.

Combining the two results, we find that the result correctly  matches the expected behavior \eqref{eq:E0lc} up to an order 1 number:
\begin{equation}
    E_0(c_1) = -\frac{c_1}{4\pi} \log\left(\frac{c_1 L}{4\pi}\right) + \frac{c_1}{4\pi} \left(1-\gamma\right) + O(1) ~.
\end{equation}

\section{Spacelike $c_1$ defect and quench}
\label{quench} 

In this section we consider time evolution on the Lorentzian plane in the presence of a spacelike defect. For $t<0$  the scalar field is in its standard (no defect) vacuum state.  A spacelike defect is then inserted at $t=0$, leading to an excited state at $t>0$.    We compute this excited state, along with  the two-point scalar correlator and stress tensor expectation value  therein.   This is a particular example of a quench, described by an instantaneous change in the Hamiltonian.

\subsection{In and out vacuum states for the general defect}
In the presence of a spacelike defect inserted at $t=0$, there are two types of \textit{in} modes which are either left or right moving purely positive frequency waves for $t<0$. To find them, we can plug the ansatz
\begin{equation}
\phi^\omega_{in, L} = \begin{cases}
    t<0 : & N \, e^{-i\omega v}\\ t>0: & N\left( A e^{-i\omega v} + B e^{i\omega u} \right)
\end{cases}, \qquad \phi^\omega_{in, R} = \begin{cases}
    t<0 : & N' \, e^{-i\omega u}\\ t>0: & N'\left( A' e^{-i\omega u} + B' e^{i\omega v} \right)
\end{cases}
\end{equation}
where $u=t-x, v=t+x$, into the matching condition
\begin{equation}
    \begin{pmatrix}
        \phi(t=0^+,x) \\ \dot{\phi}(t=0^+,x)
    \end{pmatrix} = \begin{pmatrix}
        a & b \\ c & d
    \end{pmatrix} \begin{pmatrix}
        \phi(t=0^-,x) \\ \dot{\phi}(t=0^-,x)
    \end{pmatrix}, \qquad ad-bc=1
\end{equation}
and normalise them so their  KG inner products \eqref{eq: KG} are,
\begin{equation}
    (\phi^\omega_{L,in},\phi^{\omega'}_{L,in}) = 2\pi\delta(\omega-\omega') = (\phi^\omega_{R,in},\phi^{\omega'}_{R,in}), \quad \text{all other inner products} = 0 ~.
\end{equation}
Note that the condition $ab-bc=1$ is necessary (and sufficient) for conservation of the KG inner product across the defect. 
The resulting $in$ modes are
\begin{align}
&\phi^\omega_{in, L} = \begin{cases}
    t<0 : & \frac{1}{\sqrt{2\omega}} \, e^{-i\omega v}\\ t>0: & \frac{1}{\sqrt{2\omega}} \left( A_\omega e^{-i\omega v} + B_\omega e^{i\omega u} \right)
\end{cases}\\
&\phi^\omega_{in, R} = \begin{cases}
    t<0 : & \frac{1}{\sqrt{2\omega}}  e^{-i\omega u}\\ t>0: & \frac{1}{\sqrt{2\omega}}\left( A_\omega e^{-i\omega u} + B_\omega e^{i\omega v} \right)
\end{cases} \label{eq: ja}
\end{align}
where
\begin{equation}
\begin{aligned}
    A_\omega &= \frac{1}{2}\left(-\frac{c}{i\omega}+ a+d-i\omega b \right) = t_\omega^{-1} \\
    B_\omega &= \frac{1}{2} \left(\frac{c}{i\omega}+a-d-i\omega b \right) = \frac{r_\omega^v}{t_\omega}
\end{aligned}
\end{equation}
where $r_\omega$ and $t_\omega$ are the reflection and transmission coefficients given in Eq.~\eqref{eq: aa}.  While $A_\omega$ doesn't vanish for the real-valued frequency $\omega$ and $(a,b,c,d)$ satisfying $ad-bc=1$, $B_\omega$ vanishes for the trivial $(a=d=1, b=c=0)$ and $\mathbb{Z}_2$ $(a=d=-1,b=c=0)$ defects, as it is analogous to the reflection coefficients in the presence of the time-like defect.

To quantise the theory we define the \textit{in} vacuum as
\begin{equation}
    a_\omega |0;in\rangle = 0, \quad b_\omega|0;in\rangle = 0, \quad \forall \omega 
\end{equation}
where the only non-trivial commutators between the oscillators are
\begin{equation}
    [a_\omega, a_{\omega'}^\dagger] = 2\pi \delta(\omega-\omega') = [b_\omega, b_{\omega'}^\dagger] ~. \label{eq: jb}
\end{equation}
Then, one can quantise the scalar field using the in modes~\eqref{eq: ja} as
\begin{equation}
    \phi(t,x) = \int^\infty_0 \frac{\dr \omega}{2\pi} \left( a_\omega \phi^\omega_{in,L} + b_\omega \phi^\omega_{in,R} + h.c. \right) ~.
\end{equation}
For $t>0$, which is the time after the defect insertion, one can single out the positive frequency modes in the mode expansion
\begin{equation}
\begin{aligned}
    \phi(t>0,x) &= \int^\infty_0 \frac{\dr \omega}{2\pi} \left[ \frac{e^{-i\omega v}}{\sqrt{2\omega}}\left( A_\omega a_\omega + B_\omega^* b_\omega^\dagger \right) + \frac{e^{-i\omega u}}{\sqrt{2\omega}}\left(B_\omega^* a_\omega^\dagger + A_\omega b_\omega \right) +h.c. \right] \\
    &\equiv \int^\infty_0 \frac{\dr \omega}{2\pi}\left(c_\omega \frac{e^{-i\omega v}}{\sqrt{2\omega}} + d_\omega \frac{e^{-i\omega u}}{\sqrt{2\omega}} + h.c. \right)
\end{aligned}
\end{equation}
where we defined the new oscillators
\begin{equation}
    c_\omega = A_\omega a_\omega + B_\omega^* b_\omega^\dagger ~, \qquad d_\omega = B_\omega^* a_\omega^\dagger + A_\omega b_\omega  ~. \label{eq: jc}
\end{equation}
From the commutators of old oscillators~\eqref{eq: jb}, it follows that the new oscillators are canonically quantised, i.e.
\begin{equation}
    [c_\omega,c^\dagger_{\omega'}] = 2\pi \delta(\omega-\omega') = [d_\omega, d^\dagger_{\omega'}] ~.
\end{equation}
These oscillators define the \textit{out} vacuum state such that
\begin{equation}
    c_\omega |0;out\rangle = 0 ~, \quad d_\omega|0;out\rangle = 0, \quad \forall \omega  ~.
\end{equation}
Due to the appearance of the old creation operators in the new annihilation operators, the $in$ vacuum state is not annihilated by $c_\omega$ and $d_\omega$ and hence  differs from the $out$ vacuum state for $t>0$. Said differently, to an observer at $t>0$, the $in$ vacuum state corresponds to an excited state. To write the $in$ vacuum state in the Fock space basis of the $out$ vacuum w.r.t. the new oscillators we use an ansatz
\begin{equation}
    |0;in\rangle = \bigotimes_{\omega} \left(\sum_{n_\omega,m_\omega} C^\omega_{n_\omega,m_\omega}|n_\omega,m_\omega;out\rangle\right) ~,
\end{equation}

For  ease of notation, let's consider a fixed frequency $\omega$ and drop the subscript $\omega$. The two defining equations of the $in$ vacuum state, i.e. $a_\omega |0;in\rangle = b_\omega|0;in\rangle =0$, provide two relations of the coefficients
\begin{equation}
\begin{aligned}
A_\omega^* \sqrt{m+1}C_{m+1,n} &= B_\omega^* \sqrt{n}C_{m,n-1} ~, \\
A_\omega^*\sqrt{n}C_{m+1,n} &= B_\omega^*\sqrt{m+1}C_{m,n-1} ~.
\end{aligned}
\end{equation}
which are solved as
\begin{equation}
    C_{n,m} = \delta_{n,m}\left(\frac{B_\omega^*}{A_\omega^*} \right)^n C_{0,0} = \delta_{n,m} \left( \frac{b\omega^2 - i(a-d)\omega +c}{b\omega^2 - i(a+d)\omega - c} \right)^nC_{0,0} ~.
\end{equation}
Hence the $in$ vacuum state is
\begin{equation}
    |0;in\rangle = \prod_\omega \exp\left(\frac{1}{2\pi}\frac{b\omega^2 - i(a-d)\omega +c}{b\omega^2 - i(a+d)\omega - c} c_\omega^\dagger d_\omega^\dagger \right) C^\omega_{0,0}|0;out\rangle  ~.
\end{equation}
where $C^\omega_{0,0}$ are normalisation constants which should be fixed  so that $\langle 0;in| 0;in\rangle = 1$.
For example, for the $c_1$ defect $(a=d=1, b=0, c=c_1)$
\begin{equation}
    |0;in\rangle = \prod_\omega \exp\left(-\frac{1}{2\pi}\frac{c_1}{2i\omega+c_1}c_\omega^\dagger d_\omega^\dagger \right) C^\omega_{0,0}|0;out\rangle  ~.
\end{equation}

\subsection{Correlation functions} \label{Sec: corr} 
The scalar two-point correlation function is a solution of the field equation subject to the matching condition. When both points are below the defect, i.e. $t_{1,2} <0$, the correlation function in the in vacuum state is the standard free theory correlation function 
\begin{equation}
	\langle in | \phi(t_1,x_1)\phi(t_2,x_2 ) |in\rangle = -\frac{1}{4\pi } \log \left[ m^2 \left((x_1-x_2)^2 -(t_1-t_2-i\epsilon)^2\right) \right] ~, \label{eq: bc}
\end{equation}
where the $i\epsilon$ prescription is chosen so that the correlator is analytic in the lower half $t_1-t_2$ plane, as appropriate for this operator ordering. 
We will be working in the massless theory but have introduced $m^2$ in the above as an IR regulator. 
Due to the discontinuity across the defect, the two-point function behaves differently depending on whether one or both points is (are) placed above or below the defect. There are six different cases:
\begin{equation}
\begin{aligned}
\RN{1}:  \quad & t_1>t_2>0 ~, \qquad \RN{1}' :\quad  && t_2>t_1>0 ~, \\
\RN{2}: \quad & t_1>0>t_2 ~, \qquad \RN{2}': \quad && t_2>0>t_1 ~, \\
\RN{3}: \quad & 0>t_1>t_2 ~, \qquad \RN{3}': \quad && 0>t_2 >t_1 ~.
\end{aligned}
\end{equation}
Let's denote the two-point function at Region $X = \RN{1}, \cdots, \RN{3}'$ by
\begin{equation}
f^X(t_1,x_1;t_2,x_2) \equiv \langle 0;in | \phi(t_1,x_1) \phi(t_2,x_2) | 0;in \rangle ~.
\end{equation}
When at least one point is above the defect, one can use $f^{\RN{3}}$ or $f^{\RN{3}'}$ and the matching conditions to fix the ``initial data" to the immediate future of the defect and use that to solve the field equation. Namely, we look for  solutions of the initial value problem
\begin{equation}
	\begin{aligned}
		\varphi(x, 0)= & \varphi_0(x) ~, \quad \dot{\varphi}(x, 0)=v_0(x) ~, \\
		& \left(\partial_t^2-\partial_x^2\right) \varphi=0
	\end{aligned}
\end{equation}
where the initial data $\varphi_0(x)$ and $v_0(x)$ are obtained by $f^{\RN{3}}$ or $f^{\RN{3}'}$~\eqref{eq: bc} and the matching condition. The solution of d'Alembert~\cite{Stone_Goldbart_2009} is
\begin{equation}
	\varphi(t, x)=\frac{\varphi_0(x+t)+\varphi_0(x-t)}{2}+\frac{1}{2} \int_{x-t}^{x+t} v_0(s) \mathrm{d} s ~. \label{eq: bd}
\end{equation}

In Region $\RN{2}$ ($t_1>0>t_2$), the initial data are
\begin{equation}
\begin{aligned}
\varphi_0(x_1) &= f^{\RN{3}}(0,x_1;t_2,x_2) = -\frac{1}{4\pi}\log[m^2(x_{12} - t_2 -i\epsilon)(x_{12} + t_2 + i\epsilon)] ~,\\
v_0(x_1) &= \partial_{t_1}f^{\RN{3}}(0,x_1;t_2,x_2) +c_1f^{\RN{3}}(0,x_1;t_2,x_2) ~,
\end{aligned}
\end{equation}
where $x_{12} \equiv x_1 - x_2$.

Plugging into d'Alembert's solution~\eqref{eq: bd}, the correlation function is
\begin{equation}
	\begin{aligned}
		&f^{\mathrm{II}}\left(t_1, x_1 ; t_2, x_2\right)\\
        &\quad =  -\frac{1}{4 \pi} \log \left[m^2(u_2-u_1+i\epsilon)(v_1-v_2-i\epsilon)\right]+\frac{c_1}{4 \pi}\left(u_1+v_1\right) \\
		&\qquad - \frac{c_1}{8 \pi}\bigg[ (v_1-v_2)\log \left[m(v_1-v_2-i\epsilon)\right] + (u_1+v_2)\log \left[m(-u_1-v_2-i\epsilon)\right]  \\
		&\qquad \qquad \qquad + (v_1+u_2)\log \left[m (v_1+u_2+i\epsilon)\right] -(u_2-u_1)\log \left[m(u_2-u_1+i\epsilon)\right] \bigg] ~.
	\end{aligned} \label{eq: bea}
\end{equation}

Following the same procedure at Region~$\RN{1}$ using $f^{\RN{2}}$ obtained above, the correlation function is
\begin{equation}
	\begin{aligned}
		&f^{\mathrm{I}}\left(t_1, x_1 ; t_2, x_2\right)\\
        & \quad =  -\frac{1}{4 \pi} \log \left[m^2\left(u_2-u_1+i\epsilon\right)\left(v_1-v_2-i\epsilon\right)\right] +\frac{c_1}{4 \pi}\left(u_1+v_1+u_2+v_2\right) \\
        &\qquad -\frac{c_1}{8\pi} \bigg[\left(v_1+u_2\right) \log \left[m^2(v_1+u_2)^2+\epsilon^2\right] + \left(u_1+v_2\right) \log \left[m^2(u_1+v_2)^2+\epsilon^2\right]\bigg] \\
		&\qquad + \frac{3c_1^2}{16 \pi}\left(u_1+v_1\right)\left(u_2+v_2\right) \\
		&\qquad + \frac{c_1^2}{32\pi}\bigg[\left(v_1-v_2\right)^2 \log \left[m^2(v_1-v_2)^2+\epsilon^2\right] +\left(u_1-u_2\right)^2 \log \left[m^2(u_1-u_2)^2+\epsilon^2\right]  \\
		& \qquad \qquad \qquad -\left(v_1+u_2\right)^2 \log \left[m^2(v_1+u_2)^2 + \epsilon^2\right]  -\left(u_1+v_2\right)^2 \log \left[m^2(u_1+v_2)^2+\epsilon^2\right] \bigg] ~.
	\end{aligned} \label{eq: beb}
\end{equation}

One can also compute these correlation functions in  momentum space, starting from Region~\RN{3} and consecutively applying the matching condition and Fourier transforming back to the position space.

Using the correlation function, one can compute the energy momentum tensor in the $in$-vacuum state.  After subtracting off the value in the absence of the defect we have
\begin{equation}
	\begin{aligned}
		\langle 0;in| T_{uu}(u)| 0;in \rangle -\left.\langle 0;in | T_{uu}(u)|0;in\rangle\right|_{c_1=0}  &= \lim\limits_{u_1,u_2\to u} \partial_{u_1}\partial_{u_2}f^{\RN{1}}(t_1,x_1;t_2,x_2)  \\
		&=  \frac{ c_1^2}{16\pi}\log\frac{\Lambda}{m} ~,\\
		\langle 0;in| T_{vv}(v) | 0;in \rangle -\left.\langle 0;in | T_{vv}(v)|0;in \rangle\right|_{c_1=0}  &= \lim\limits_{v_1,v_2\to v}  \partial_{v_1}\partial_{v_2}f^{\RN{1}}(t_1,x_1;t_2,x_2) \\
		&=  \frac{ c_1^2}{16\pi}\log\frac{\Lambda}{m} ~.
	\end{aligned} \label{eq: ia}
\end{equation}
where we regulated the UV divergence through the point splitting, $|u_1-u_2|=\Lambda^{-1}=|v_1-v_2|$. Using these, the energy density and the flux are  
\begin{equation}
	\begin{aligned}
		\langle 0;in | T_{tt}(u,v)|0;in \rangle -\left.\langle \text{in}| T_{tt}(u,v) |0;in \rangle\right|_{c_1=0} &=\left[ \langle 0;in| T_{uu}(u) |0;in\rangle + \langle 0;in| T_{vv}(v) |0;in\rangle \right]^{c_1}_{c_1=0} \\
		&= \frac{ c_1^2}{8\pi}\log\frac{\Lambda}{m} ~, \\
		\langle 0;in| T_{tx}(u,v) |0;in\rangle &=\left[ - \langle 0;in| T_{uu}(u) |0;in \rangle + \langle 0;in| T_{vv}(v) |0;in\rangle \right] \\
		&= 0 ~.
	\end{aligned}
\end{equation}
$T_{uu}$ and $T_{vv}$ for $t>0$~\eqref{eq: ia} have no $O(c_1)$ term, because such a term comes from a single interaction vertex on the defect given by $c_1\phi(p)\phi(-p)$ which contains both left and right moving particles while $T_{uu}$ ($T_{vv})$  contain only  right(left)  moving  particles. On the other hand, the $O(c_1^2)$ term comes from two interaction vertices on the defect, allowing two like-moving fields to contract against each other, leaving two like-moving fields to contribute to $T_{uu}$ and $T_{vv}$.   All higher order vertices are suppressed in the limit $\Lambda \to \infty$.

It is also worth pointing out that for the defect in Euclidean signature we have $\langle T_{\mu\nu}\rangle =0$ at any point not on the defect, as can be easily shown as follows.   Conservation implies that $\langle T_{zz}\rangle$ can only depend on $z=x+it$.  Taking the defect to extend along $t$, we must have that $\langle T_{zz}\rangle$ is independent of $t$ by time translation invariance.  This implies that $\langle T_{zz}\rangle$  is a constant.  But since it must also vanish at infinity, it must vanish at all points off of the defect. The same holds for  $\langle T_{\zb\zb}\rangle$.  Note that it is the last condition  of vanishing at infinity that does not hold for the spacelike quench defect, as this defect creates an excited state that does not disperse at late times. 

\section{Comments}
\label{comments} 

In this paper we studied various aspects of codimension-one defects in free scalar field theory.   Even in this simplified setting there are many possible extensions that could be pursued.  For example: non-parallel defects; defects with cusps and their associated anomalous dimensions;  moving and dynamical defects, and so forth.  Going beyond free field theory there are a large number of directions, many of which have been studied in the existing literature in one form or another.  For example, one could add dynamical degrees of freedom on the defect; replace the bulk theory with an interacting CFT such as Liouville theory; couple the defect to gravity, etc.    For many of these cases, our simplified free field theory treatment should serve as a useful starting point for perturbation theory, where applicable.

\section*{Acknowledgments}

We thank Thomas Dumitrescu and Pierluigi Niro for discussions. S.K. thanks Richard Myers for the multiple helpful discussions. P.K. is supported in part by the National Science Foundation grant PHY-2209700. Z.S. is supported by the Simons Collaboration on Global Categorical Symmetries.

\appendix

\section{Higher transverse derivative terms}
\label{higher}

In this appendix we illustrate how defect terms with higher transverse derivatives can be traded away for terms with at most first transverse derivatives.    As a concrete example consider
\eq{q1}{ S= {1\over 2} \int\! d^2x (\p \phi)^2 + \int\! dt  \big({1\over 2}c_1^B \phi^2 + c_2^B \phi \phi' +{1\over 2} c_3^B \phi'^2 +  c_4^B \phi \phi''  \big)~.}
The expectation is that we can use the equations of motion to replace $\phi''\rt - \ddot{\phi}$, which in momentum space will correspond to a shift of $c_1^B$  proportional to $\omega^2 $.   This quick argument does not  take into account the renormalization of couplings, so it is worthwhile to flesh this out explicitly.

The two-point function is computed in the same manner as before.  The expression \rf{c7} for $I_0(\omega_1,p_1,p_2) $ is now 
\eq{q2}{ I_0(\omega_1,p_1,p_2) =-c_1^B+ic_2^B(p_1-p_2)-c_3^B p_1p_2+c_4^B(p_1^2+p_2^2)~. }
The relationship between the bare and renormalized couplings  is now found to be 
\eq{q3}{ c^B_1& = {\left(  {\pi^2  \over 3}\Lambdah^3 - \omega_1^2 \Lambdah\right) c_4^2 \over (1+\Lambdah c_4)^2}  +{c_1 \over  (1+\Lambdah c_4)^2}+ {\Lambdah c_2^2 \over  (1+\Lambdah c_4)^2 (1-c_3 \Lambdah) }\cr
c^B_2& = {c_2 \over (1+\Lambdah c_4) (1-c_3 \Lambdah) } \cr
c^B_3 & = {c_3 \over 1-c_3 \Lambdah} \cr
c^B_4 & = {c_4 \over 1+c_4 \Lambdah} }
We then work out the function $X(\omega_1,p_1,p_2) $ appearing in \rf{c5} as 
\eq{q4}{ X_{num}& =  -c_1 + {1\over 2} |\omega_1| (c_1 c_3-c_2^2) + {1\over 2} |\omega_1|^3  c_4^2 - {1\over 4}\omega_1^4 c_3 c_4^2 \cr
&  \quad + i c_2(1+{1\over 2} |\omega_1| c_4) (p_1+p_2) \cr
&\quad  +c_4(1+{1\over 2} |\omega_1|  c_4)(1-{1\over 2}|\omega_1| c_3)(p_1^2+p_2^2)\cr
& \quad + \Big( c_3 + {1\over 2 |\omega_1|} ( c_1 c_3-c_2^2+2 c_3 c_4 \omega_1^2)\Big) p_1 p_2 \cr
&\quad   +{i\over 2 |\omega_1|}  c_2 c_4(p_1+p_2)p_1p_2\cr
&\quad  + {1\over 2|\omega_1|}  c_4^2 (1- {1\over 2}|\omega_1|  c_3)p_1^2 p_2^2 }
and 
\eq{q5}{ X_{den} = 1+  {1 \over 2|\omega_1| } \big(c_1+2 c_4 \omega_1^2- c_3 \omega_1^2\big)   -{1\over 4} \big((c_1+2 c_4 \omega_1^2) c_3-c_2^2\big)  ~.}
We are interested in correlation functions of fields inserted away from the defect. Such correlators are unaffected by the shift
 \eq{q10}{ X(\omega_1,p_1,p_2) \rt X(\omega_1,p_1,p_2)  + {\sum_{m, n\geq 0}}'  a_{mn} (p_1^2+\omega_1^2)^m(p_2^2+\omega_1^2)^n }
where $\sum'$ instructs to omit the $m=n=0$ term, since this just adds to the position space correlator terms which are delta function localized on the defect.  Therefore, in \rf{q4} we are free to make the replacements $p_1^2 \rt -\omega_1^2$ and $p_2^2\rt -\omega_1^2$, since this just changes $X$ by terms of the form in \rf{q10}.   Doing so gives
\eq{q11}{  X_{num} =  -\ct_1 +{|\omega_1| \over 2} (\ct_1 c_3-c_2^2) +ic_2(p_1+p_2) +\big( {1 \over 2|\omega_1| } (\ct_1 c_3-c_2^2)  +c_3\big)p_1p_2 
 }
with 
\eq{q12}{ \ct_1 = c_1+2 c_4 \omega_1^2~.}
Examining \rf{q5} and \rf{q11} we see that all $c_4 $ dependence has been absorbed into the definition of $\ct_1$.   This confirms our expectation that the $c_4$ coupling can be traded for an $\omega_1$ dependent shift of $c_1$.  The same mechanism holds for all higher transverse derivative defect terms.

\section{Fusion of two defects}
\label{fusion}
 
 We write the two-defect  correlator in the form \rf{c5}, and write 
 \eq{c41}{ X(\omega_1,p_1,p_2)  = {X_{\rm num} \over X_{\rm den}  }~. }
We find, in the coincident limit  $L=0$, 
\eq{c42}{ X_{\rm num}& = {1\over 2}|\omega_1| \Big[({c_1^{(1)}}+{c_1^{(2)}})({c_3^{(1)}}+{c_3^{(2)}})-({c_2^{(1)}}+{c_2^{(2)}})^2   \Big] \cr
& -{1\over 4}\Big[ {c_1^{(1)}}({c_2^{(2)}}+2)^2 +{c_1^{(2)}}({c_2^{(1)}}-2)^2 -{c_1^{(1)}} {c_1^{(2)}} ({c_3^{(1)}}+{c_3^{(2)}}) \Big]   \cr
& +i\Big[{c_2^{(1)}}+{c_2^{(2)}} +{1\over 2}({c_1^{(1)}} {c_3^{(2)}}-{c_1^{(2)}} {c_3^{(1)}})-{1\over 4}({c_1^{(2)}} {c_2^{(1)}} {c_3^{(2)}}+{c_1^{(1)}} {c_2^{(2)}} {c_3^{(1)}})\cr
& \quad +{1\over 4}{c_2^{(1)}} {c_2^{(2)}}({c_2^{(1)}}+{c_2^{(2)}})   \Big]  (p_1+p_2) \cr
& +{1\over 2|\omega_1|}  \Bigg[ ({c_1^{(1)}}+{c_1^{(2)}})({c_3^{(1)}}+{c_3^{(2)}}) -({c_2^{(1)}}+{c_2^{(2)}})^2\Bigg] p_1 p_2 \cr
&+{1\over 4} \Big[ {c_3^{(1)}}({c_2^{(2)}}-2)^2+{c_3^{(2)}}({c_2^{(1)}}+2)^2 -{c_3^{(1)}} {c_3^{(2)}}({c_1^{(1)}}+{c_1^{(2)}}) \Big]p_1 p_2  }
and 
\eq{c43}{ &X_{\rm den} =  {1\over 8 |\omega_1|} \left(-c_1^{(1)} {c_1^{(2)}} \left( {c_3^{(1)}}+{c_3^{(2)}} \right) +
c_1^{(1)} \left( {c_2^{(2)}}+2 \right) ^{2} +{c_1^{(2)}}
 \left( {c_2^{(1)}}-2 \right) ^{2} \right)
 \cr&
 +{1\over 16}\left( 
 ( c_1^{(2)}c_3^{(1)}c_3^{(2)}-{{c_2^{(2)}}}^{2}{c_3^{(1)}}-4{c_3^{(1)}}-8{c_3^{(2)}} ) c_1^{(1)}-( {{c_2^{(1)}}}^{2}{c_3^{(2)}}+8{c_3^{(1)}}+4{c_3^{(2)}}) {c_1^{(2)}} \right) \cr
 &
+\left(1+{c_2^{(1)}c_2^{(2)} \over 4} \right)^2+ {({c_2^{(1)}}+{c_2^{(2)}})^2 \over 4}   
+ {|\omega_1|\over 8 } \left( c_3^{(1)}c_3^{(2)} ({c_1^{(1)}}+{c_1^{(2)}})  -{c_3^{(1)}}  ({c_2^{(2)}}-2 )^{2}
  -c_3^{(2)}
 (c_2^{(1)}+2)^2 \right) ~. \cr}
In the above the couplings are the renormalized versions, defined in terms of the bare couplings by the relations \rf{c19}.
The couplings of the fused defect are then read off by equating the expression for $X$ above with that of a single defect
\eq{c44}{ X(\omega_1,p_1,p_2) =    {-c_1 +{1\over 2} (c_1 c_3-c_2^2)|\omega_1|  +ic_2(p_1+p_2) +\big( {1\over 2|\omega_1| } (c_1 c_3-c_2^2)  +c_3\big)p_1p_2   \over 1+{1\over 2|\omega_1|} (c_1- c_3 \omega_1^2)   -{1\over 4} (c_1 c_3-c_2^2)    }~.}
As noted in the main text, this results in  expressions for the fused couplings $c_i$ that agree with those computed by  composing the matching relations.

\section{Stabilization by interactions} \label{app: solitions}

The unstable  modes bound to the defect are analogous to the small $k$ modes of a tachyonic  $m^2<0$ free scalar, with $\omega = \sqrt{k^2 + m^2} $.  In that case, the scalar potential is unbounded from below but can be stabilized by adding an interaction term such as $\frac{\lambda}{4}\phi^4$. In this subsection, we add a $\frac{\lambda}{4}\phi^4$ term to the free massless scalars in the presence of a defect described by  the matching conditions, and discuss the solutions and fluctuations around them.

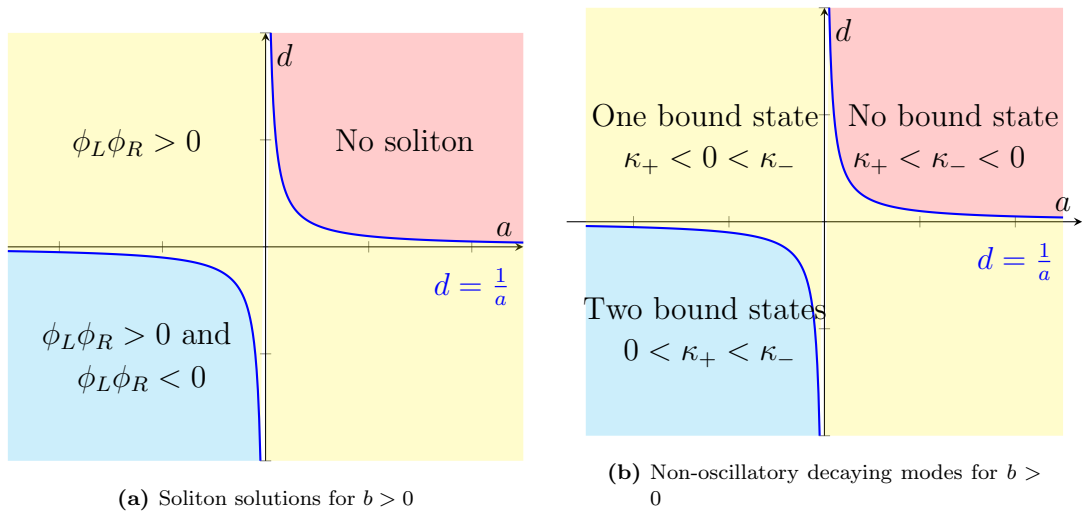
\begin{figure}[H]
	\centering
	\begin{subfigure}{0.45\textwidth}
		\centering
		\begin{tikzpicture}
			\begin{axis}[
				xlabel={$a$},
				ylabel={$d$},
				xmax=5,
				xmin=-5,
				xticklabel=\empty,
				yticklabel=\empty,
				axis lines=middle,
				samples=200
				]
				\addplot[blue,thick,domain=-5:-0.1, name path=A] {1/x};
				\addplot[draw=none,name path=B] {-10};
				\addplot[cyan,fill opacity=0.2] fill between[of=A and B,soft clip={domain=-8:0}];
				\addplot[blue,thick,domain=0.1:5, name path=C] {1/x};
				\addplot[yellow,fill opacity=0.2] fill between[of=C and B,soft clip={domain=0:8}];
				\addplot[draw=none,name path=D] {10};
				\addplot[yellow,fill opacity=0.2] fill between[of=D and A,soft clip={domain=-6:0}];
				\addplot[red,fill opacity=0.2] fill between[of=C and D,soft clip={domain=0:6}];
				%\node [anchor=north] at (axis cs:-0.3,-0.3) {$O$};
				\node [anchor=north] at (axis cs:-2.5,-3) {$\phi_L\phi_R>0$ and};
				\node [anchor=north] at (axis cs:-2.4,-5) {$\phi_L\phi_R<0$};
				\node [anchor=north] at (axis cs:2.7,6) {No soliton};
				\node [anchor=north] at (axis cs:-2.5,6) {$\phi_L\phi_R>0$};
				\node [anchor=north,color=blue] at (axis cs:4,-0.5) {$d=\frac{1}{a}$};
				%	\draw[red!20,dashed] (axis cs:2,-4) -- (axis cs:2,10);
			\end{axis}
		\end{tikzpicture}
		\caption{Soliton solutions for $b>0$}
		\label{fig:soliton solns posb}
	\end{subfigure}%
	\begin{subfigure}{0.45\textwidth}
		\centering
		\begin{tikzpicture}
			\begin{axis}[
				xlabel={$a$},
				ylabel={$d$},
				xmax=5.4,
				xmin=-5.4,
				xticklabel=\empty,
				yticklabel=\empty,
				axis lines=middle,
				samples=200
				]
				\addplot[blue,thick,domain=-5:-0.1, name path=A] {1/x};
				\addplot[draw=none,name path=B] {-10};
				\addplot[cyan,fill opacity=0.2] fill between[of=A and B,soft clip={domain=-8:0}];
				\addplot[blue,thick,domain=0.1:5, name path=C] {1/x};
				\addplot[yellow,fill opacity=0.2] fill between[of=C and B,soft clip={domain=0:8}];
				\addplot[draw=none,name path=D] {10};
				\addplot[yellow,fill opacity=0.2] fill between[of=D and A,soft clip={domain=-6:0}];
				\addplot[red,fill opacity=0.2] fill between[of=C and D,soft clip={domain=0:6}];
				%\node [anchor=north] at (axis cs:-0.3,-0.3) {$O$};
				\node [anchor=north] at (axis cs:-2.5,-3) {Two bound states};
				\node [anchor=north] at (axis cs:-2.4,-5) {$0<\kappa_+<\kappa_-$};
				\node [anchor=north] at (axis cs:2.7,6) {No bound state};
				\node [anchor=north] at (axis cs:2.4,4) {$\kappa_+<\kappa_-<0$};
				\node [anchor=north] at (axis cs:-2.5,6) {One bound state};
				\node [anchor=north] at (axis cs:-2.4,4) {$\kappa_+<0<\kappa_-$};
				\node [anchor=north,color=blue] at (axis cs:4,-0.5) {$d=\frac{1}{a}$};
				%	\draw[red!20,dashed] (axis cs:2,-4) -- (axis cs:2,10);
			\end{axis}
		\end{tikzpicture}
		\caption{Non-oscillatory decaying modes for $b>0$}
		\label{fig:bound state posb}
	\end{subfigure}
	\caption{The parameter space of unstable bound state modes  and  solitons in the $(a,d)$ domain.} 
	\label{fig:bound and solitons posb}
\end{figure}

With the interaction term added, the field equation implies
\begin{equation}
	\phi^{\prime \prime}=V^{\prime}(\phi)=\lambda \phi^3, \quad \rightarrow \quad \phi^{\prime}= \pm \sqrt{\frac{\lambda}{2}} \phi^2
\end{equation}
where the sign can be fixed in terms of the boundary values
\begin{equation}
\phi(x=\pm\infty) = 0, \quad \phi(x=0^-) = \phi_L, \quad \phi(x=0^+) = \phi_R
\end{equation}
where $\phi_{L,R}$ are determined by the matching conditions~\eqref{eq: gluing condition}. The solution is
\begin{equation}
	\phi= \begin{cases}x<0: & \frac{1}{\phi_L^{-1} -\operatorname{sgn}\left(\phi_L\right) \sqrt{\frac{\lambda}{2}} x} \\ x>0: & \frac{1}{\phi_R^{-1}+\operatorname{sgn}\left(\phi_R\right) \sqrt{\frac{\lambda}{2}} x}\end{cases} ~.
\end{equation}
Plugging them into the matching condition~\eqref{eq: gluing condition}, one gets
\begin{equation}
	\begin{aligned}
		\phi_R & =a \phi_L+b \cdot \operatorname{sign}\left(\phi_L\right) \sqrt{\frac{\lambda}{2}} \phi_L^2 ~, \\
		\phi_L & =d \phi_R+b \cdot \operatorname{sign}\left(\phi_R\right) \sqrt{\frac{\lambda}{2}} \phi_R^2 ~.
	\end{aligned} \label{eq: ag}
\end{equation}
The existence of solutions, which we refer to as solitons, depend on the matching matrix. For $b>0$,
\begin{itemize}
	\item $0<\frac{1}{d}<a$: No soliton (also no non-oscillatory bound states around $\phi=0$)
	\item $0<a<\frac{1}{d}$ or $\frac{1}{d}<a<0$: One pair of $\phi_L\phi_R>0$ solitons
	\item $a<\frac{1}{d}<0$: One $\phi_L\phi_R>0$ solition \textit{and} one $\phi_L\phi_R<0$ soliton
\end{itemize}
This is illustrated in Figure~\ref{fig:soliton solns posb}. The number of  solitonic solutions matches the number of bound state modes around $\phi=0$ as can be seen from Figure~\ref{fig:bound state posb}. Thus, in the interacting theory, whenever there is a decaying mode around $\phi=0$, the solitonic solution is also available. 

To see which solution is energetically preferred, we use that the Hamiltonian in the solitonic background is
\begin{equation}
	H_{\mathrm{tot}}=H+Y
\end{equation}
where
\begin{equation}
	\begin{aligned}
		H & =\int_{-\infty}^{\infty}\left(\frac{1}{2} \phi^{\prime 2}+\frac{\lambda}{4} \phi^4\right) \mathrm{d} x \\
		& =\int_{-\infty}^{\infty} \frac{\lambda}{2} \phi^4 \mathrm{~d} x \\
		& =\frac{1}{3} \sqrt{\frac{\lambda}{2}}\left(\left|\phi_L\right|^3+\left|\phi_R\right|^3\right)
	\end{aligned}
\end{equation}
and $Y$ is
\begin{equation}
	\begin{aligned}
		Y & =b c \phi_L^{\prime} \phi_L+\frac{a c}{2} \phi_L^2+\frac{b d}{2}\left(\phi_L^{\prime}\right)^2 \\
		& =\phi_L^2\left(\frac{a c}{2}+b c \sqrt{\frac{\lambda}{2}}\left|\phi_L\right|+\frac{\lambda}{4} b d \phi_L^2\right) ~.
	\end{aligned}
\end{equation}
Plugging the solutions $\phi_{L,R}$ of the matching equations~\eqref{eq: ag}, we find that the maximum energy is negative in all regions on the $(a,d)$ plane where there is a soliton. Hence, we conclude that the solitonic solutions are energetically  preferred to the trivial $\phi=0$ solution. In principle, we should also consider fluctuations around the solitons and check for stability, but we have not done so.

\newpage
\bibliographystyle{bibstyle2017}
\bibliography{defectrefs}

\providecommand{\href}[2]{#2}\begingroup\begin{thebibliography}{10}

\bibitem{Goldberger:2001tn}
W.~D. Goldberger and M.~B. Wise, {\it {Renormalization group flows for brane couplings}},  \href{http://dx.doi.org/10.1103/PhysRevD.65.025011}{{\sf Phys. Rev. D} {\sf {65} }{\sf (2002) }{\sf 025011}}, \href{http://arxiv.org/abs/hep-th/0104170}{{\ttfamily arXiv:hep-th/0104170}}.

\bibitem{Michel:2014lva}
B.~Michel, E.~Mintun, J.~Polchinski, A.~Puhm, and P.~Saad, {\it {Remarks on brane and antibrane dynamics}},  \href{http://dx.doi.org/10.1007/JHEP09(2015)021}{{\sf JHEP} {\sf {09} }{\sf (2015) }{\sf 021}}, \href{http://arxiv.org/abs/1412.5702}{{\ttfamily arXiv:1412.5702 [hep-th]}}.

\bibitem{Witten:1992cr}
E.~Witten, {\it {Some computations in background independent off-shell string theory}},  \href{http://dx.doi.org/10.1103/PhysRevD.47.3405}{{\sf Phys. Rev. D} {\sf {47} }{\sf (1993) }{\sf 3405--3410}}, \href{http://arxiv.org/abs/hep-th/9210065}{{\ttfamily arXiv:hep-th/9210065}}.

\bibitem{Kutasov:2000qp}
D.~Kutasov, M.~Marino, and G.~W. Moore, {\it {Some exact results on tachyon condensation in string field theory}},  \href{http://dx.doi.org/10.1088/1126-6708/2000/10/045}{{\sf JHEP} {\sf {10} }{\sf (2000) }{\sf 045}}, \href{http://arxiv.org/abs/hep-th/0009148}{{\ttfamily arXiv:hep-th/0009148}}.

\bibitem{Affleck:1991tk}
I.~Affleck and A.~W.~W. Ludwig, {\it {Universal noninteger 'ground state degeneracy' in critical quantum systems}},  \href{http://dx.doi.org/10.1103/PhysRevLett.67.161}{{\sf Phys. Rev. Lett.} {\sf {67} }{\sf (1991) }{\sf 161--164}}.

\bibitem{Friedan:2003yc}
D.~Friedan and A.~Konechny, {\it {On the boundary entropy of one-dimensional quantum systems at low temperature}},  \href{http://dx.doi.org/10.1103/PhysRevLett.93.030402}{{\sf Phys. Rev. Lett.} {\sf {93} }{\sf (2004) }{\sf 030402}}, \href{http://arxiv.org/abs/hep-th/0312197}{{\ttfamily arXiv:hep-th/0312197}}.

\bibitem{Bachas:2001vj}
C.~Bachas, J.~de~Boer, R.~Dijkgraaf, and H.~Ooguri, {\it {Permeable conformal walls and holography}},  \href{http://dx.doi.org/10.1088/1126-6708/2002/06/027}{{\sf JHEP} {\sf {06} }{\sf (2002) }{\sf 027}}, \href{http://arxiv.org/abs/hep-th/0111210}{{\ttfamily arXiv:hep-th/0111210}}.

\bibitem{Calabrese:2006rx}
P.~Calabrese and J.~L. Cardy, {\it {Time-dependence of correlation functions following a quantum quench}},  \href{http://dx.doi.org/10.1103/PhysRevLett.96.136801}{{\sf Phys. Rev. Lett.} {\sf {96} }{\sf (2006) }{\sf 136801}}, \href{http://arxiv.org/abs/cond-mat/0601225}{{\ttfamily arXiv:cond-mat/0601225}}.

\bibitem{Raviv-Moshe:2023yvq}
A.~Raviv-Moshe and S.~Zhong, {\it {Phases of surface defects in Scalar Field Theories}},  \href{http://dx.doi.org/10.1007/JHEP08(2023)143}{{\sf JHEP} {\sf {08} }{\sf (2023) }{\sf 143}}, \href{http://arxiv.org/abs/2305.11370}{{\ttfamily arXiv:2305.11370 [hep-th]}}.

\bibitem{Giombi:2023dqs}
S.~Giombi and B.~Liu, {\it {Notes on a surface defect in the O(N) model}},  \href{http://dx.doi.org/10.1007/JHEP12(2023)004}{{\sf JHEP} {\sf {12} }{\sf (2023) }{\sf 004}}, \href{http://arxiv.org/abs/2305.11402}{{\ttfamily arXiv:2305.11402 [hep-th]}}.

\bibitem{Trepanier:2023tvb}
M.~Tr\'epanier, {\it {Surface defects in the O(N) model}},  \href{http://dx.doi.org/10.1007/JHEP09(2023)074}{{\sf JHEP} {\sf {09} }{\sf (2023) }{\sf 074}}, \href{http://arxiv.org/abs/2305.10486}{{\ttfamily arXiv:2305.10486 [hep-th]}}.

\bibitem{Bashmakov:2024suh}
V.~Bashmakov and J.~Sisti, {\it {Exploring Defects with Degrees of Freedom in Free Scalar CFTs}},  \href{http://arxiv.org/abs/2410.01716}{{\ttfamily arXiv:2410.01716 [hep-th]}}.

\bibitem{Casini:2022bsu}
H.~Casini, I.~Salazar~Landea, and G.~Torroba, {\it {Entropic g Theorem in General Spacetime Dimensions}},  \href{http://dx.doi.org/10.1103/PhysRevLett.130.111603}{{\sf Phys. Rev. Lett.} {\sf {130} }{\sf no.~11, }{\sf (2023) }{\sf 111603}}, \href{http://arxiv.org/abs/2212.10575}{{\ttfamily arXiv:2212.10575 [hep-th]}}.

\bibitem{Cuomo:2021rkm}
G.~Cuomo, Z.~Komargodski, and A.~Raviv-Moshe, {\it {Renormalization Group Flows on Line Defects}},  \href{http://dx.doi.org/10.1103/PhysRevLett.128.021603}{{\sf Phys. Rev. Lett.} {\sf {128} }{\sf no.~2, }{\sf (2022) }{\sf 021603}}, \href{http://arxiv.org/abs/2108.01117}{{\ttfamily arXiv:2108.01117 [hep-th]}}.

\bibitem{Diatlyk:2024zkk}
O.~Diatlyk, H.~Khanchandani, F.~K. Popov, and Y.~Wang, {\it {Defect fusion and Casimir energy in higher dimensions}},  \href{http://dx.doi.org/10.1007/JHEP09(2024)006}{{\sf JHEP} {\sf {09} }{\sf (2024) }{\sf 006}}, \href{http://arxiv.org/abs/2404.05815}{{\ttfamily arXiv:2404.05815 [hep-th]}}.

\bibitem{Kravchuk:2024qoh}
P.~Kravchuk, A.~Radcliffe, and R.~Sinha, {\it {Effective theory for fusion of conformal defects}},  \href{http://arxiv.org/abs/2406.04561}{{\ttfamily arXiv:2406.04561 [hep-th]}}.

\bibitem{Diatlyk:2024qpr}
O.~Diatlyk, H.~Khanchandani, F.~K. Popov, and Y.~Wang, {\it {Effective Field Theory of Conformal Boundaries}},  \href{http://dx.doi.org/10.1103/PhysRevLett.133.261601}{{\sf Phys. Rev. Lett.} {\sf {133} }{\sf no.~26, }{\sf (2024) }{\sf 261601}}, \href{http://arxiv.org/abs/2406.01550}{{\ttfamily arXiv:2406.01550 [hep-th]}}.

\bibitem{Shachar:2024ubf}
T.~Shachar, R.~Sinha, and M.~Smolkin, {\it {The defect b-theorem under bulk RG flows}},  \href{http://dx.doi.org/10.1007/JHEP09(2024)057}{{\sf JHEP} {\sf {09} }{\sf (2024) }{\sf 057}}, \href{http://arxiv.org/abs/2404.18403}{{\ttfamily arXiv:2404.18403 [hep-th]}}.

\bibitem{Cuomo:2024psk}
G.~Cuomo, Y.-C. He, and Z.~Komargodski, {\it {Impurities with a cusp: general theory and 3d Ising}},  \href{http://dx.doi.org/10.1007/JHEP11(2024)061}{{\sf JHEP} {\sf {11} }{\sf (2024) }{\sf 061}}, \href{http://arxiv.org/abs/2406.10186}{{\ttfamily arXiv:2406.10186 [hep-th]}}.

\bibitem{Shachar:2024cwk}
T.~Shachar, {\it {On Intersecting Conformal Defects}},  \href{http://arxiv.org/abs/2411.14543}{{\ttfamily arXiv:2411.14543 [hep-th]}}.

\bibitem{Fredenhagen:2006dn}
S.~Fredenhagen, M.~R. Gaberdiel, and C.~A. Keller, {\it {Bulk induced boundary perturbations}},  \href{http://dx.doi.org/10.1088/1751-8113/40/1/F03}{{\sf J. Phys. A} {\sf {40} }{\sf (2007) }{\sf F17}}, \href{http://arxiv.org/abs/hep-th/0609034}{{\ttfamily arXiv:hep-th/0609034}}.

\bibitem{Pannell:2023pwz}
W.~H. Pannell and A.~Stergiou, {\it {Line defect RG flows in the \ensuremath{\varepsilon} expansion}},  \href{http://dx.doi.org/10.1007/JHEP06(2023)186}{{\sf JHEP} {\sf {06} }{\sf (2023) }{\sf 186}}, \href{http://arxiv.org/abs/2302.14069}{{\ttfamily arXiv:2302.14069 [hep-th]}}.

\bibitem{Brunner:2007ur}
I.~Brunner and D.~Roggenkamp, {\it {Defects and bulk perturbations of boundary Landau-Ginzburg orbifolds}},  \href{http://dx.doi.org/10.1088/1126-6708/2008/04/001}{{\sf JHEP} {\sf {04} }{\sf (2008) }{\sf 001}}, \href{http://arxiv.org/abs/0712.0188}{{\ttfamily arXiv:0712.0188 [hep-th]}}.

\bibitem{Gaiotto:2012np}
D.~Gaiotto, {\it {Domain Walls for Two-Dimensional Renormalization Group Flows}},  \href{http://dx.doi.org/10.1007/JHEP12(2012)103}{{\sf JHEP} {\sf {12} }{\sf (2012) }{\sf 103}}, \href{http://arxiv.org/abs/1201.0767}{{\ttfamily arXiv:1201.0767 [hep-th]}}.

\bibitem{Brunner:2015vva}
I.~Brunner and C.~Schmidt-Colinet, {\it {Reflection and transmission of conformal perturbation defects}},  \href{http://dx.doi.org/10.1088/1751-8113/49/19/195401}{{\sf J. Phys. A} {\sf {49} }{\sf no.~19, }{\sf (2016) }{\sf 195401}}, \href{http://arxiv.org/abs/1508.04350}{{\ttfamily arXiv:1508.04350 [hep-th]}}.

\bibitem{Shimamori:2024yms}
S.~Shimamori, {\it {Conformal field theory with composite defect}},  \href{http://dx.doi.org/10.1007/JHEP08(2024)131}{{\sf JHEP} {\sf {08} }{\sf (2024) }{\sf 131}}, \href{http://arxiv.org/abs/2404.08411}{{\ttfamily arXiv:2404.08411 [hep-th]}}.

\bibitem{Ge:2024hei}
D.~Ge, T.~Nishioka, and S.~Shimamori, {\it {Localized RG flows on composite defects and $ \mathcal{C} $-theorem}},  \href{http://dx.doi.org/10.1007/JHEP02(2025)012}{{\sf JHEP} {\sf {02} }{\sf (2025) }{\sf 012}}, \href{http://arxiv.org/abs/2408.04428}{{\ttfamily arXiv:2408.04428 [hep-th]}}.

\bibitem{Giombi:2022vnz}
S.~Giombi, E.~Helfenberger, and H.~Khanchandani, {\it {Line defects in fermionic CFTs}},  \href{http://dx.doi.org/10.1007/JHEP08(2023)224}{{\sf JHEP} {\sf {08} }{\sf (2023) }{\sf 224}}, \href{http://arxiv.org/abs/2211.11073}{{\ttfamily arXiv:2211.11073 [hep-th]}}.

\bibitem{Barrat:2023ivo}
J.~Barrat, P.~Liendo, and P.~van Vliet, {\it {Line defect correlators in fermionic CFTs}},  \href{http://arxiv.org/abs/2304.13588}{{\ttfamily arXiv:2304.13588 [hep-th]}}.

\bibitem{Konechny:2015qla}
A.~Konechny, {\it {Fusion of conformal interfaces and bulk induced boundary RG flows}},  \href{http://dx.doi.org/10.1007/JHEP12(2015)114}{{\sf JHEP} {\sf {12} }{\sf (2015) }{\sf 114}}, \href{http://arxiv.org/abs/1509.07787}{{\ttfamily arXiv:1509.07787 [hep-th]}}.

\bibitem{Lauria:2020emq}
E.~Lauria, P.~Liendo, B.~C. Van~Rees, and X.~Zhao, {\it {Line and surface defects for the free scalar field}},  \href{http://dx.doi.org/10.1007/JHEP01(2021)060}{{\sf JHEP} {\sf {01} }{\sf (2021) }{\sf 060}}, \href{http://arxiv.org/abs/2005.02413}{{\ttfamily arXiv:2005.02413 [hep-th]}}.

\bibitem{DiPietro:2020fya}
L.~Di~Pietro, E.~Lauria, and P.~Niro, {\it {3d large $N$ vector models at the boundary}},  \href{http://dx.doi.org/10.21468/SciPostPhys.11.3.050}{{\sf SciPost Phys.} {\sf {11} }{\sf no.~3, }{\sf (2021) }{\sf 050}}, \href{http://arxiv.org/abs/2012.07733}{{\ttfamily arXiv:2012.07733 [hep-th]}}.

\bibitem{Behan:2020nsf}
C.~Behan, L.~Di~Pietro, E.~Lauria, and B.~C. Van~Rees, {\it {Bootstrapping boundary-localized interactions}},  \href{http://dx.doi.org/10.1007/JHEP12(2020)182}{{\sf JHEP} {\sf {12} }{\sf (2020) }{\sf 182}}, \href{http://arxiv.org/abs/2009.03336}{{\ttfamily arXiv:2009.03336 [hep-th]}}.

\bibitem{Behan:2021tcn}
C.~Behan, L.~Di~Pietro, E.~Lauria, and B.~C. van Rees, {\it {Bootstrapping boundary-localized interactions II. Minimal models at the boundary}},  \href{http://dx.doi.org/10.1007/JHEP03(2022)146}{{\sf JHEP} {\sf {03} }{\sf (2022) }{\sf 146}}, \href{http://arxiv.org/abs/2111.04747}{{\ttfamily arXiv:2111.04747 [hep-th]}}.

\bibitem{Castiglioni:2022yes}
L.~Castiglioni, S.~Penati, M.~Tenser, and D.~Trancanelli, {\it {Interpolating Wilson loops and enriched RG flows}},  \href{http://dx.doi.org/10.1007/JHEP08(2023)106}{{\sf JHEP} {\sf {08} }{\sf (2023) }{\sf 106}}, \href{http://arxiv.org/abs/2211.16501}{{\ttfamily arXiv:2211.16501 [hep-th]}}.

\bibitem{Castiglioni:2023uus}
L.~Castiglioni, S.~Penati, M.~Tenser, and D.~Trancanelli, {\it {Wilson loops and defect RG flows in ABJM}},  \href{http://dx.doi.org/10.1007/JHEP06(2023)157}{{\sf JHEP} {\sf {06} }{\sf (2023) }{\sf 157}}, \href{http://arxiv.org/abs/2305.01647}{{\ttfamily arXiv:2305.01647 [hep-th]}}.

\bibitem{DiPietro:2023gzi}
L.~Di~Pietro, E.~Lauria, and P.~Niro, {\it {Conformal boundary conditions for a 4d scalar field}},  \href{http://dx.doi.org/10.21468/SciPostPhys.16.4.090}{{\sf SciPost Phys.} {\sf {16} }{\sf no.~4, }{\sf (2024) }{\sf 090}}, \href{http://arxiv.org/abs/2312.11633}{{\ttfamily arXiv:2312.11633 [hep-th]}}.

\bibitem{Nagar:2024mjz}
I.~Nagar, A.~Sever, and D.-l. Zhong, {\it {Planar RG flows on line defects}},  \href{http://dx.doi.org/10.1007/JHEP06(2024)110}{{\sf JHEP} {\sf {06} }{\sf (2024) }{\sf 110}}, \href{http://arxiv.org/abs/2404.07290}{{\ttfamily arXiv:2404.07290 [hep-th]}}.

\bibitem{Sun:2025ihw}
X.~Sun and S.-K. Jian, {\it {Boundary operator expansion and extraordinary phase transition in the tricritical O(N) model}},  \href{http://arxiv.org/abs/2501.06287}{{\ttfamily arXiv:2501.06287 [cond-mat.str-el]}}.

\bibitem{BARTON1986322}
G.~Barton, {\it Quantum mechanics of the inverted oscillator potential},  \href{http://dx.doi.org/https://doi.org/10.1016/0003-4916(86)90142-9}{{\sf Annals of Physics} {\sf {166} }{\sf no.~2, }{\sf (1986) }{\sf 322--363}}. \url{https://www.sciencedirect.com/science/article/pii/0003491686901429}.

\bibitem{Kraus:2000nj}
P.~Kraus and F.~Larsen, {\it {Boundary string field theory of the D anti-D system}},  \href{http://dx.doi.org/10.1103/PhysRevD.63.106004}{{\sf Phys. Rev. D} {\sf {63} }{\sf (2001) }{\sf 106004}}, \href{http://arxiv.org/abs/hep-th/0012198}{{\ttfamily arXiv:hep-th/0012198}}.

\bibitem{Takayanagi:2000rz}
T.~Takayanagi, S.~Terashima, and T.~Uesugi, {\it {Brane - anti-brane action from boundary string field theory}},  \href{http://dx.doi.org/10.1088/1126-6708/2001/03/019}{{\sf JHEP} {\sf {03} }{\sf (2001) }{\sf 019}}, \href{http://arxiv.org/abs/hep-th/0012210}{{\ttfamily arXiv:hep-th/0012210}}.

\bibitem{DiFrancesco:1997nk}
P.~Di~Francesco, P.~Mathieu, and D.~Senechal, \href{http://dx.doi.org/10.1007/978-1-4612-2256-9}{{\it {Conformal Field Theory}}, }.
\newblock Graduate Texts in Contemporary Physics. Springer-Verlag, New York, 1997.

\bibitem{Cardy:1986ie}
J.~L. Cardy, {\it {Operator Content of Two-Dimensional Conformally Invariant Theories}},  \href{http://dx.doi.org/10.1016/0550-3213(86)90552-3}{{\sf Nucl. Phys. B} {\sf {270} }{\sf (1986) }{\sf 186--204}}.

\bibitem{Stone_Goldbart_2009}
M.~Stone and P.~Goldbart, {\it Mathematics for physics: A guided tour for graduate students}, .
\newblock Cambridge University Press, 2009.

\end{thebibliography}\endgroup

\end{document}